\documentclass[aps, prl, twocolumn, superscriptaddress, amsmath, showpacs, tightenlines, footinbib, longbibliography]{revtex4-1}
\usepackage[colorlinks, breaklinks, citecolor=blue, linkcolor=blue,  urlcolor={blue}]{hyperref}
\usepackage{breakurl}
\usepackage{amsfonts}
\usepackage{amsmath}
\usepackage{amssymb}
\usepackage{mathrsfs}
\usepackage{epsfig}
\usepackage{graphicx}
\usepackage{mathtools}
\usepackage{bm}
\usepackage{natbib}

\usepackage{colortbl}
\definecolor{mygray}{gray}{.9}
\definecolor{darkblue}{rgb}{1,1,.70}
\definecolor{lightblue}{rgb}{1,1,.90}

\begin{document}
\title{Nonreciprocal Superradiant Phase Transitions and Multicriticality {\color{black}in a Cavity QED System}}
\author{Gui-Lei Zhu}
\affiliation{Department of Physics, Zhejiang Sci-Tech University, Hangzhou 310018, China}
\affiliation{Theoretical Quantum Physics Laboratory, Cluster for Pioneering Research, RIKEN, Wakoshi, Saitama 351-0198, Japan}

\author{Chang-Sheng Hu}
\affiliation{Department of Physics, Anhui Normal University, Wuhu 241000, China}

\author{Hui Wang}
\affiliation{Theoretical Quantum Physics Laboratory, Cluster for Pioneering Research, RIKEN, Wakoshi, Saitama 351-0198, Japan}

\author{Wei Qin}
\affiliation{Theoretical Quantum Physics Laboratory, Cluster for Pioneering Research, RIKEN, Wakoshi, Saitama 351-0198, Japan}
\affiliation{Center for Joint Quantum Studies and Department of Physics, School of Science, Tianjin University, Tianjin 300350, China}

\author{Xin-You L\"{u}}
\email{xinyoulu@hust.edu.cn}
\affiliation{School of Physics, Huazhong University of Science and Technology and Wuhan Institute of Quantum Technology, Wuhan 430074, China}

\author{Franco Nori}
\email{fnori@riken.jp}
\affiliation{Theoretical Quantum Physics Laboratory, Cluster for Pioneering Research, RIKEN, Wakoshi, Saitama 351-0198, Japan}
\affiliation{Quantum Computing Center, RIKEN, Wakoshi, Saitama 351-0198, Japan}
\affiliation{Department of Physics, The University of Michigan, Ann Arbor, MI, 48109-1040, USA}

\date{\today}

\begin{abstract}
We demonstrate the emergence of nonreciprocal superradiant phase transitions and novel multicriticality in a cavity quantum electrodynamics (QED) system, where a two-level atom interacts with two counter-propagating modes of a whispering-gallery-mode (WGM) microcavity. The cavity rotates at a certain angular velocity, and is directionally squeezed by a unidirectional parametric pumping $\chi^{(2)}$ nonlinearity.
The combination of cavity rotation and directional squeezing leads to nonreciprocal first- and second-order superradiant phase transitions. These transitions do not require ultrastrong atom-field couplings and can be easily controlled by the external pump field. Through a full quantum description of the system Hamiltonian, we identify two types of multicritical points in the phase diagram, both of which exhibit controllable nonreciprocity. These results open a new door for all-optical manipulation of superradiant transitions and multicritical behaviors in light-matter systems, with potential applications in engineering various integrated nonreciprocal quantum devices.
\end{abstract}
\maketitle
Phase transitions and critical phenomena are at the heart of understanding the nature of the matter in condensed matter physics and material science\,\cite{domb2000phase,sachdev1999quantum}. One of the most intriguing topics in light-matter systems is the superradiant phase transition\,\cite{Dicke1954,HEPP1973360,Wang1973}, where increasing the atom-field coupling through a critical value induces a transition from the normal phase (NP) to the superradiant phase (SP)\,\cite{EmaryClive2003,Lambert2004,Dimer2007,Baumann2010,Baden2014,Nagy2010,Xin-You2018001,Kirton2019,Youjiang2021,Jinchen2022}. This superradiant transition typically occurs in the thermodynamic limit, where the number of atoms $N$ approaches infinity. 
In the quantum Rabi model\,\cite{Rabi1937,Braak2011,xie2017quantum} with $N=1$, a similar transition can occur, but it requires both ultrastrong light-matter coupling and an extremely large atomic frequency\,\cite{Hwang2015,Maoxin2017,Xin-You2018,Hwang2018,ZhangYu-Yu2021,Fallas2022,liu2022process,Zhengshibiao2022}. The realization of the phase transition in such single-atom models has been successfully demonstrated in quantum simulation platforms, including nuclear magnetic resonance (NMR) quantum simulators\,\cite{Georgescu2014,chen2021experimental}, driven atoms in trapped ions\,\cite{Puebla2017,cai2021observation} and superconducting qubits\,\cite{zheng2022emergent}. 

In open systems, the presence of photon loss can have a significant impact on the critical behavior {\color{black} of the Hamiltonian part of the system}\,\cite{Kirton2017,Shammah2018}. It can induce multicritical phenomena\,\cite{Soriente2018,ZhuCJ2020,Lin2022Rui,Youjiang2019,Han-Jie2020}, or even completely suppress its criticality\,\cite{Larson2017}.
In addition, this class of driven-dissipative systems allows for the demonstration of richer physics, such as the breakdown of photon blockade\,\cite{Carmichael2015,Fink2017,Reiter2020}, stable superradiant lasers\,\cite{Meiser2009,bohnet2012steady,Norcia2016,norcia2016superradiance}, time crystals\,\cite{Gong2018,mattes2023entangled,ferioli2023non}, and atomic synchronization\,\cite{Minghui2014,masson2022universality}. However, thus far, the realization of superradiant transitions has not simultaneously combined the features necessary for exquisite controllability, such as strong nonreciprocity, tuneability, and compact integration. 

Optical nonreciprocity\,\cite{Zin2011,peng2014parity,chang2014parity,fan2012all, Qi-Tao2017,Manipatruni2009,shen2016experimental,Wang2013, Ramezani2018,zhang2018thermal,Keyu2018,Tang2022} {\color{black} is characterized by the asymmetric behavior of optical signals as they travel through an optical system in opposite directions.} This phenomenon plays a crucial role in optical information processing and quantum networks\,\cite{bennett2000quantum,Buluta_2011,kimble2008quantum}. Notably, a theory of nonreciprocal phase transitions in non-equilibrium systems has been proposed, suggesting that asymmetric couplings of multiple species can give rise to time-dependent phases\,\cite{fruchart2021non}. Recently, this concept has been applied to the Dicke model with two spin species\,\cite{chiacchio2023}. Yet, the technique used to achieve nonreciprocal phase transition in these schemes is not readily applicable to achieving {\it controllable nonreciprocal transitions in a light-matter system.} 
In general, achieving the superradiant transition requires tunability of the dipole coupling of the radiation field to atoms. To this end, many pioneering approaches have been proposed, including stimulated Raman transitions\,\cite{Ferri2021,klinder2015dynamical,zhiqiang2017nonequilibrium} and quantum simulators\,\cite{feng2015exploring,chen2021experimental,zheng2022emergent}. However, breaking the reciprocity of the system, in these approaches, remains challenging. The {\it exquisite control} of superradiant transitions and multicriticality {\it through external fields} {\color{black}is an intriguing topic, {\color{black}may inspire new applications, such as} on-chip unidirectional superradiant lasers\,\cite{Meiser2009,bohnet2012steady,Norcia2016,norcia2016superradiance} and integrated high-precision quantum sensing\,\cite{Garbe2020,Chu2021,YingZu-Jian2022,he2023criticality}.}

In this work, inspired by recent experimental advances\,\cite{aoki2006observation,alton2011strong,Junge2013,shomroni2014all,shomroni2014all,scheucher2016quantum,bechler2018passive,Will2021,maayani2018flying} and related works\,\cite{bliokh2015transverse,bliokh2015quantum}, we propose an experimentally feasible approach for all-optical control of superradiant phase transitions in a cavity QED system. Our method focuses on a rotating dual-coupling Jaynes-Cummings (JC) model, where the clockwise and counterclockwise resonator modes are simultaneously coupled to a two-level atom. Generally, the JC model does not exhibit superradiant transitions in the presence of cavity dissipation\,\cite{Larson2017}. Here, we revive such transitions by introducing a classical field to subtly parametrically pump one of the cavity modes. This driven-dissipative setup enables steady-state superradiant transitions to occur in the experimentally friendly cavity QED system, which {\it do not require ultrastrong coupling strengths}, and the extremely large atomic detuning can be easily achieved by {\it tuning the external pump field.} {\color{black}This all-optical control has some main advantages compared to its magnetic and electronic counterparts, such as compactness, ease of integration, and lower power consumption (see Sec. S7.B in Supplementary Material \cite{SM}).}

Interestingly, the combination of cavity rotation\,\cite{maayani2018flying,Jing2018,Huang2018,Jiao2020} and directional squeezing causes the critical points of phase transitions to shift in opposite directions. As a result, the system exhibits {\it nonreciprocal first-order} and {\it second-order} superradiant phase transitions. Moreover, we observe a rich phase diagram featuring controllable {\it tricritical point} and {\it multicritical points}, all displaying nonreciprocity. 
 Our work fundamentally combines the theories of phase transitions and multicriticality with nonreciprocal physics, and could provide valuable resources for quantum metrology\,\cite{Hotter2024}.
\begin{figure}
\includegraphics[width=7.4cm]{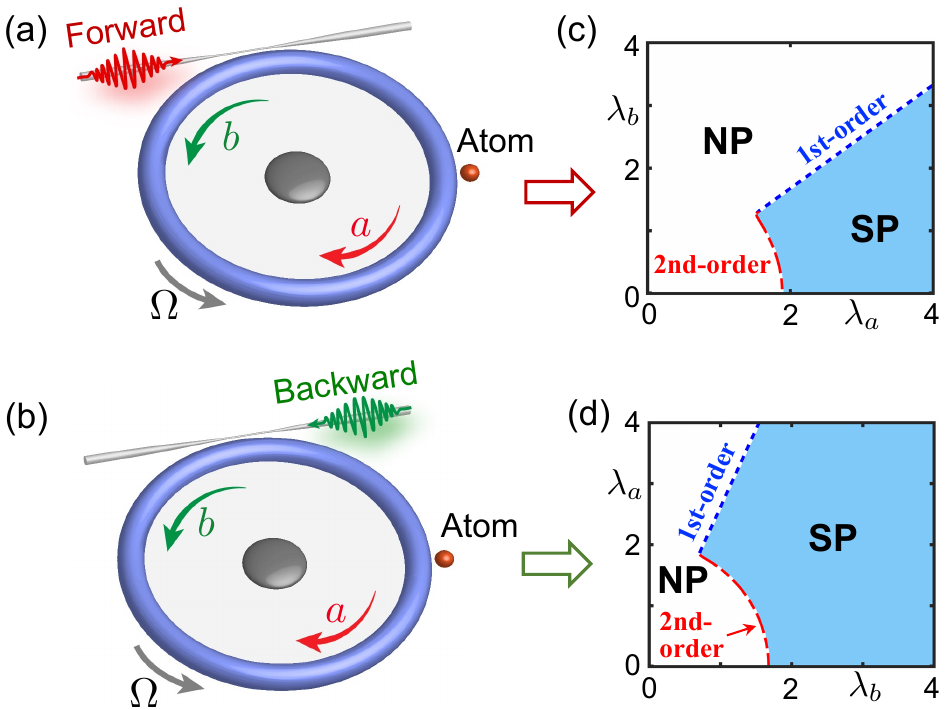}
\caption{Schematic illustration of the rotating dual-coupling Jaynes-Cummings (JC) model in the forward (a) and backward (b) pumps. The whispering-gallery-mode (WGM) resonator with $\chi^{(2)}$ nonlinearity embedded (not shown here), supports counterclockwise and clockwise cavity modes labeled as $a$ and $b$ respectively; and both modes interact with a two-level atom. The resonator rotates counterclockwise at an angular velocity $\Omega$. Panels (c,d) show the boundaries of the normal phase (NP) and superradiant phase (SP) for forward and backward pumps, respectively. The parameters used here are: $\Delta=2,\kappa/\Delta=0.05,G/\kappa=1.5$, $J=0$, and $\Delta_F/\Delta=0.5$ in (c), and $\Delta_F/\Delta=-0.5$ in (d).}
\label{fig1}
\end{figure}

{\it Model.---}Here, we consider the model of a two-level atom interacting with two counter-propagating modes of a whispering-gallery-mode (WGM) resonator, as depicted in Figs.\,\ref{fig1}(a,b). The resonator is made of materials with second-order ($\chi^{(2)}$) nonlinearity\,\cite{lu2019periodically,lu2020toward}. A classical field at frequency $\omega_p$ is input from either the forward or backward to drive the nonlinearity, which generates directional squeezing cavity mode through an optical parametric amplification (OPA) process\,\cite{scully1999quantum, agarwal2012quantum,Huang2009, Xin-You2015, Qin2018, Tang2022,QinWei2022} {\color{black}(see Sec. S9.B in\,\cite{SM})}. The resonator rotates counterclockwise with an angular velocity, denoted as $\Omega$. {\color{black}This rotation can be experimentally achieved by mounting the resonator on a turbine\,\cite{maayani2018flying} (see Sec. S9.C in\,\cite{SM}).} Therefore, the two cavity modes undergo Sagnac-Fizeau shifts concerning their static resonance frequency, represented by $\omega_0$\,\cite{malykin2000sagnac,maayani2018flying}, i.e., $\omega_0\rightarrow \omega_0\pm \Delta_{F}$, with $\Delta_F\approx{nR\Omega\omega_0(1-n^{-2})}/{c}$. Here $n$ and $R$ present the refractive index and radius of the resonator, and $c$ is the speed of light. Note that a positive Sagnac-Fizeau shift ($\Delta_F>0$) corresponds to the forward pump case, while a negative shift ($\Delta_F<0$) corresponds to the backward pump.

In the frame rotating at $\omega_p/2$, the dual-coupling JC Hamiltonian with the forward pump is given by ($\hbar=1$)
\begin{align}\label{H_original}
&H=H_0+H_{\rm int},\\
&H_0=(\Delta\!+\!\Delta_F)a^{\dagger}a\!+\!(\Delta\!-\!\Delta_F)b^{\dagger}b\!
+\!\frac{\Delta_q}{2}\sigma_z\!+\!G(a^{\dagger 2}\!+\!a^2),\nonumber
\\
&H_{\rm int}=\big[(g_aa+g_bb)\sigma_+\!+\!Ja^{\dagger}b+{\rm H.c.}\big],\nonumber
\end{align}
where $a$ and $b$ are the annihilation operators of the counterclockwise and clockwise cavity modes, respectively, and $\sigma_{\pm}=(\sigma_x\pm i\sigma_y)/2$ are the atomic Pauli matrices. The detunings are defined as $\Delta=\omega_{0}-\omega_p/2$ and $\Delta_q=\omega_q-\omega_p/2$. The cavity modes $a$ and $b$ are coupled to the atom with strengths $g_a$ and $g_b$, respectively, and $G(a^{\dag 2}+a^2)$ is the two-photon drive term generated from the OPA process, with $G$ representing the pump strength. In our analytical calculations, the focus is on the forward pump case, and the backward Hamiltonian can be obtained by replacing the two-photon term with $G(b^{\dag 2}+b^2)$. The parameter $J$ represents the hopping amplitude between two cavity modes, and for simplicity, we assume $J=0$ in the following analytical derivations, as it does not affect the main results (see Sec. S8 in \,\cite{SM}). 

Taking into account both the cavity and atomic dissipation, the system evolution is described by the master equation,
\begin{align}\label{drho}
\frac{d\rho}{dt}=-{i}[H,\rho]+\kappa\mathcal{D}[a]+\kappa\mathcal{D}[b]+\gamma\mathcal{D}[\sigma_-],
\end{align}
where $\rho$ is the densitiy matrix of the system, $\mathcal{D}[o]=2o\rho o^{\dagger}-(o^{\dagger}o\rho+\rho o^{\dagger}o)$ is the Lindblad superoperator. Here, we have assumed equal decay rates for the two cavity modes ($\kappa_a=\kappa_b=\kappa$). To ensure the conservation of the total pseudo-angular momentum, we focus mainly on the effect of cavity dissipation and set the atomic decay rate $\gamma=0$. 

The presence of dissipation, as discussed in\,\cite{Larson2017}, can completely suppress the superradiant transition in the JC model. Interestingly, in our model, the two-photon drive is capable of resurrecting this transition. From a symmetry perspective, the two-photon term changes the system from a continuous $U(1)$ symmetry to a discontinuous $Z_2$ symmetry. This $Z_2$ symmetry is defined by $[H,\Pi]=0$ with the parity operator $\Pi=\exp\{i\pi N\}$, where $N=a^{\dag}a+b^{\dag}b+(\sigma_z+1)/2$ is the total number of excitations in the system\,\cite{Hwang2015}.
We define the ratios between the atomic detuning and cavity detuning as $\eta_{\pm}=\Delta_q/(\Delta\pm\Delta_F)$, and $\mu=\Delta_q/G$. Note that the limit of $\eta_{\pm}\rightarrow \infty$ is physically similar to the infinite-frequency limit in the standard Rabi model ($\omega_q/\omega_0\rightarrow\infty$). Fortunately, in our model, this limit can be easily achieved by tuning the pump frequency $\omega_q$. For convenience, we introduce the dimensionless atom-field couplings as $\lambda_{a,b}=2g_{a,b}/\sqrt{\Delta_q(\Delta\pm\Delta_F)}$. {\color{black}These couplings are determined by the detunings $\Delta_q$ and $\Delta$, rather than the atomic frequency $\omega_q$ and resonator frequency $\omega_0$.  Thus, our approach relaxes the couplings required for the superradiant phase transitions from the ultrastrong coupling regime to the strong coupling regime (see Sec. S7.A in \,\cite{SM}).} In the large atomic detuning limit, $\eta_{\pm}\rightarrow \infty$ and $\mu\rightarrow \infty$, when the couplings $\lambda_a, \lambda_b$ increase beyond their critical values, the system undergoes a Rabi-like phase transition from the NP to the SP, indicating the breaking of $Z_2$ symmetry. 

{\it Nonreciprocal superradiant phase transitions.---}
According to Eq.\,(\ref{drho}), we obtain a set of Heisenberg equations of motion for operators (see Sec.\,S1 in \cite{SM}). We define the renormalized occupation of the cavity modes as $\langle a\rangle=\alpha\sqrt{\eta_+}$ and $\langle b\rangle=\beta\sqrt{\eta_-}$, with $\alpha=\alpha_{\rm re}+i\alpha_{\rm im}$, $\beta=\beta_{\rm re}+i\beta_{\rm im}$. In the long-time limit, the system reaches a steady state and we can obtain the {mean-field solutions for cavity occupations $\alpha_{\rm re},\alpha_{\rm im}$ and $\beta_{\rm re}$, $\beta_{\rm im}$}\,\cite{SM}. {\color{black}The validity of the mean-field approach is discussed in Sec. S6 in \,\cite{SM}}. When $\alpha_{\rm re},\alpha_{\rm im},\beta_{\rm re},\beta_{\rm im}=0$, the system is in the NP, while non-zero values of $\alpha_{\rm re},\alpha_{\rm im},\beta_{\rm re},\beta_{\rm im}$ indicate the SP. Figures\,\ref{fig1}(c,d) display the phase diagrams of $\alpha_{\rm re},\alpha_{\rm im}$ and $\beta_{\rm re}$, $\beta_{\rm im}$, for the forward and backward pumps, respectively. When the pump strength is fixed at $G/\kappa=1.5$, the first- (and second-) order transition boundary differs between the two pump directions, leading to the nonreciprocity of the superradiant transitions (see below). Here, the first- and second-order transition boundaries are, respectively, characterized by discontinuous jumps and continuous changes in the cavity occupations (see Fig.\,S1 in\,\cite{SM}). 

\begin{figure}
	\includegraphics[width=8.2cm]{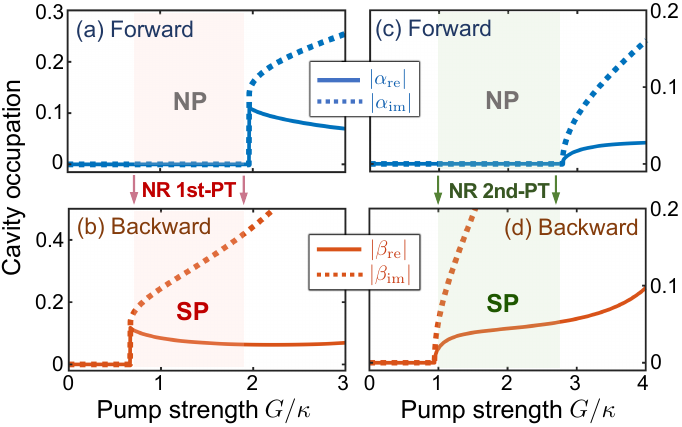}
	\caption{The {\color{black}cavity occupations} $\alpha$ and $\beta$ versus the pump strength $G/\kappa$, for both the forward and backward pumps. The solid (or dotted) curve denotes the real (or imaginary) part of cavity occupations. In panels (a,b), where $\lambda=1.5$, the shaded areas indicate the nonreciprocal first-order phase transition (NR 1st-PT). In panels (c,d), where $\lambda=1.36$, the shaded areas indicate the nonreciprocal second-order phase transition (NR 2nd-PT). Here, $\Delta_F/\Delta=\pm0.5$ for the forward and backward pumps, respectively, $\lambda_a=\lambda_b=\lambda$, and other parameters are the same as in Fig.\,\ref{fig1}.}
	\label{fig2}
\end{figure}

\begin{figure*}
	\includegraphics[width=17cm]{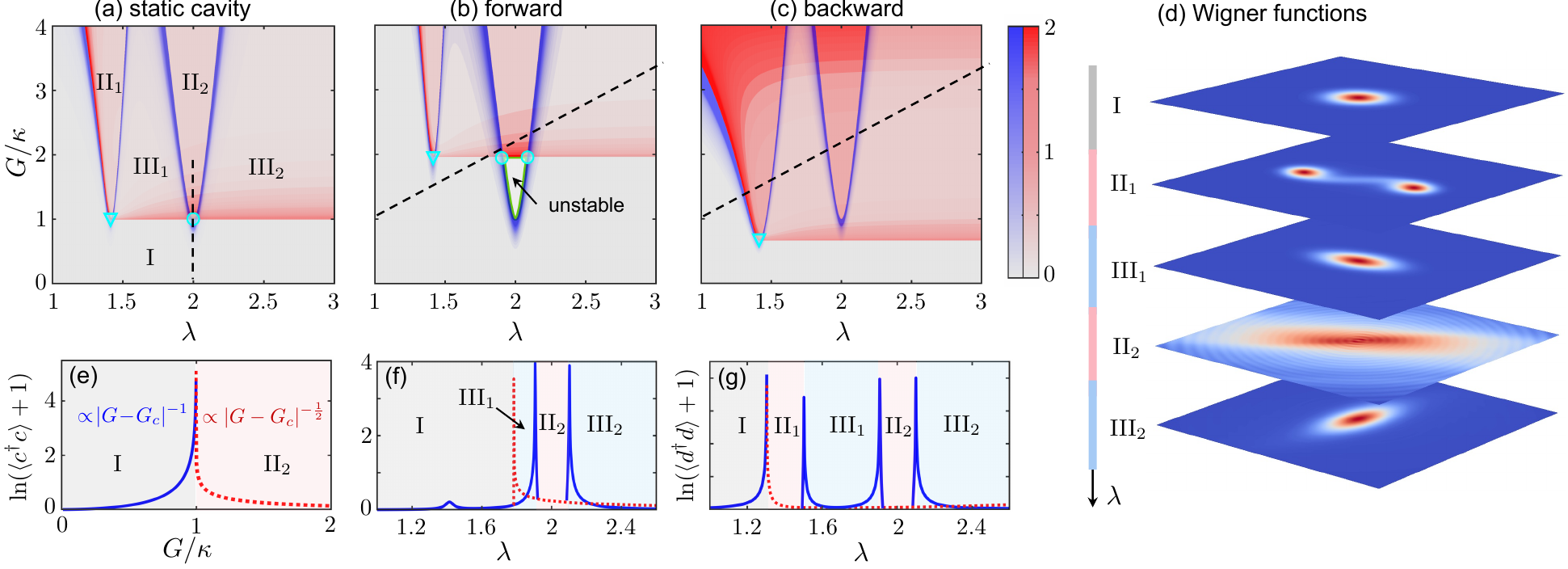}
	\caption{(a)-(c) Phase diagram of the steady-state photon number fluctuations, with $\langle c^{\dagger}c\rangle$ for the forward [panel (b)], $\langle d^{\dagger}d\rangle$ for the backward [panel (c)], and $\langle c^{\dagger}c\rangle$ (or $\langle d^{\dagger}d\rangle$) for the static microcavity [panel (a)]. {\color{black}The blue (red) bar represents fluctuations on top of the mean-field solutions in the normal (superradiant) phase.} Three regimes, denoted as I, II, and III, are identified based on the number of stable coexisting solutions. Subscripts 1 and 2 are employed to distinguish regimes with different Wigner distributions. The tricritical points are shown by blue triangles, and the multicritical points are marked by blue circles. The green curve in panel (b) indicates the unstable region (see details in Sec.\,S5 of\,\cite{SM}). (d) The Wigner functions in different regimes. Here, we consider the coordinates of $\lambda$ and $G/\kappa$ as ${\rm I} (1.0,1.1), {\rm II}_1 (1.42,2.6), {\rm III}_1 (1.8,1.98), {\rm II}_2 (2.04,2.24)$ and ${\rm III}_2 (2.2,2.42)$, $\Delta_q/\Delta=1000$ and atomic decay $\gamma=\kappa$. (e) The critical lines at $\lambda=2$ [the black-dotted line in panel (a)] as a function of $G$. (f,g) Cut plots of ${\rm ln}(\langle c^{\dag}c\rangle$+1) [black-dotted line in (b)] and ${\rm ln}(\langle d^{\dag}d\rangle$+1) [black-dotted line in panel (c)] as a function of $\lambda$. In (e-g), the blue solid and red dotted lines denote cavity fluctuations in the NP and SP, respectively. We consider $\Delta_F=0$ in panels (a,e), $\Delta_F/\Delta=0.5$ in panels (b,d,f) and $\Delta_F/\Delta=-0.5$ in panels (c,g).}
	\label{fig3}
\end{figure*}
{\color{black}To enable all-optical manipulation of superradiant phase transitions using an external field, we simplify the system by adjusting $\lambda_a=\lambda_b=\lambda$. In this configuration, with a fixed coupling $\lambda$, increasing the pump strength $G$ beyond a critical value can induce either a first- or second-order phase transition. The respective critical pump strengths can be analytically calculated as (see Sec. S3 in\,\cite{SM}), 
\begin{small}
\begin{align}\label{G1st}
G_{c}^{\rm 1st}&=\frac{1}{4\kappa\Delta_-}\big[\!-\!u
\!+\!\sqrt{(u\!+\!2\Delta_-\kappa^2)^2\!+\!4\Delta_+^2\Delta_-^2\kappa^2}~\big],\\
G_{c}^{\rm 2nd}\!&=\!\frac{\big[\Delta^4p^2\!\!+\!(2\kappa^2\!\!-\!\Delta_F^2p)^2\!\!
+\!\Delta^2\big(qp^2\!+\!4\kappa^2\big)\big]^{{1}/{2}}}{\big[{16\kappa^2\!+\!\Delta_-^2(p-2)^2}\big]^{1/2}},\label{G2nd}
\end{align}
\end{small}where $\Delta_{\pm}=\Delta\pm \Delta_F$, $u=\Delta_+(\Delta_-^2+\kappa^2)$, $q= \kappa^2\!-\!2\Delta_F ^2$  and $p=\lambda^2-2$. Notably, the first-order critical pump strength $G_{c}^{\rm 1st}$ is irrelevant to the coupling $\lambda$. By setting $G_{c}^{\rm 1st}=G_{c}^{\rm 2nd}$, the tricritical point can be determined at $\lambda_{\rm tric}\approx\sqrt{2}$ (see Fig.\,S2(b) in \cite{SM} for details). 

Figure \ref{fig2} shows the cavity occupations $\alpha$ and $\beta$ versus the pump strength $G$. In the left panel, we consider the case of $\lambda=1.5$. When the system is driven by a forward pump [Fig.\,\ref{fig2}(a)], the cavity occupations exhibit a discontinuous jump from zero to finite values, when $G$ exceeds a critical value $G_{\rm c,f}^{\rm 1st}=1.96\kappa$ (located near the right edge of the shaded region). This is evidence of a first-order phase transition. However, when driven by a backward pump [Fig.\,\ref{fig2}(b)], the critical value is shifted to $G_{\rm c,b}^{\rm 1st}=0.66\kappa$, the left edge of the shaded region. Between these two critical pump strengths (within the shaded areas), if driven forward, the system is in the NP, while if driven backward, the system is in the SP. This phenomenon is called {\it nonreciprocal first-order superradiant phase transition}. 

In the right panels of Fig.\,\ref{fig2}, we fix $\lambda=1.36$ and apply the forward and backward pumps, respectively. In this case, the cavity occupations exhibit continuous increase from zero to finite values at the critical pump strengths $G^{\rm 2nd}_{\rm c,f}=2.78\kappa$ and  $G^{\rm 2nd}_{\rm c,b}=0.94\kappa$, respectively, as shown in Figs.\,\ref{fig2}(c,d). These results indicate a {\it nonreciprocal second-order superradiant phase transition}. Physically, the nonreciprocity originates from the opposite Sagnac shifts induced by the rotation of the cavity. This leads to a shift of the critical point for the forward (or backward) pump towards larger (or smaller) values (see Fig. S2(a) in \cite{SM}).

{\it Nonreciprocal multicriticality.---}
 To explore the steady-state fluctuations of the system, we adopt a semiclassical approach by displacing the cavity fields as $a\rightarrow \langle a\rangle+c$, $b\rightarrow \langle b\rangle+d$, where $c$ and $d$ are the annihilation operators that describe cavity fluctuations. In the large atomic detuning limit, we can obtain the effective low-energy Hamiltonian  (see Sec.\,S2 in\,\cite{SM})
\begin{align}\label{Heff}
\!\!\!\!\!\!\!\!\!H_{\rm eff}=\,&\Lambda_1c^{\dagger}c+\Lambda_2d^{\dag}d+\big[Gc^2+\Lambda_3cd^{\dag}+\Lambda_4e^{2i\phi}c^2\nonumber\\
&+\Lambda_5e^{2i\phi}d^2\!+\Lambda_6e^{2i\phi}cd+{\rm H.c.}\big],
\end{align}
where the parameters $\Lambda_n$ ($n=1,2,...,6$), $\phi$ are determined by the Sagnac shift $\Delta_F$ and the mean-field solutions $\alpha$ and $\beta$. In particular, the trivial solutions of $\alpha=\beta=0$ correspond to the low-energy Hamiltonian in the normal phase. To analyze the nature of the nonreciprocal transitions in the open system, we combine Eq.\,(\ref{Heff}) with the master equation (\ref{drho}) and obtain the steady-state solution of quantum fluctuations (see Sec.\,S4 in\,\cite{SM}). 

Figures\,{\ref{fig3}}(a-c) present the steady-state phase diagrams of photon number fluctuations. For a static cavity (i.e., $\Delta_F=0$) [Fig.\,\ref{fig3}(a)], we observe two types of multicritical behaviors. There is a {\it tricritical point} denoted by the blue triangle where the first- and second-order critical lines meet, which is of the same type as in Refs.\,\cite{Soriente2018, ZhuCJ2020,Lin2022Rui}. The other is the {\it multicritical point} denoted by the blue circle, where regimes I, II, and III coexist. Specially, these two critical lines, exhibiting different asymptotic behaviors\,\cite{Nagy2011}, can morph into a divergent curve [see Fig.\,\ref{fig3}(e)]. Remarkably, this multicritical point shows nonreciprocal characteristics: in the forward pump, the multicritical point splits into two [see Fig.\,\ref{fig3}(b)], while in the backward pump and with the same atom-field coupling, the multicritical point disappears [see Fig.\,\ref{fig3}(c)]. Moreover, the cut plots of the phase diagram also show the nonreciprocity of the photon number fluctuations [see Figs.\,\ref{fig3}(f,g)]. 

To distinguish between regimes I, II, and III, in Fig.\,\ref{fig3}(d), we numerically calculate the Wigner functions of the cavity field using Hamiltonian in Eq.\,(\ref{H_original})\,\cite{JOHANSSON20121760,JOHANSSON20131234}, considering the detuning $\Delta_q/\Delta=10^3$ and the nonzero atomic decay $\gamma=\kappa$. In the NP (regime I), the cavity field is almost in a vacuum state. 
In regime ${\rm II}_1$, the Wigner function shows two peaks, reflecting the $Z_2$ symmetry breaking of the system. In regimes ${\rm III_1}$ and ${\rm III_2}$, the cavity fields are both squeezed, but with an orthogonal squeezing direction.

{\it Possible implementations.---}
{\color{black}The proposed protocols could be implemented using either cold cesium atoms falling onto the surface of a WGM microdisk{\color{black}\,\cite{aoki2006observation,alton2011strong}} or a single trapped $^{85}$Rb atom interacts with a WGM microresonator\,\cite{Will2021} (see more details in Sec.~S9.A of \cite{SM}).} 
We consider the feasible parameters of the microresonator to be $Q=6.0\times 10^{9}$, $R=1.1$ {\rm mm}, $n=1.4$, $\lambda_0=1550 \,{\rm nm}$ and $\Omega=6.6$ {\rm kHz}\,\cite{maayani2018flying}, which yields a Sagnac shift of $\Delta_F=3.2$ MHz and the intrinsic loss rate of $\kappa_i=32$ kHz. The large atomic detuning required for the transition can be achieved by tuning the pump frequency $\omega_p\approx 2\omega_0$. We consider $\eta_+=100$, resulting in $\Delta=6.4$ MHz and $\Delta_q=960$ MHz. Previous studies have shown that a detuning value of $\omega_q/\omega_0=50$ is sufficient to observe the superradiant transitions in NMR simulator\,\cite{chen2021experimental} and trapped ion systems\,\cite{Puebla2017,cai2021observation}. Based on the above parameters, the nonreciprocal first- and second-order superradiant transitions are predicted to occur at the atom-field couplings $g_a\approx 68$ MHz and $g_b\approx 39$ MHz, which are experimentally feasible by adjusting the distance between the atom and the surface\,\cite{Will2021}. Additionally, the squeezed cavity mode can be generated using a periodically poled thin film of lithium niobate media\,\cite{lu2019periodically,lu2020toward}. Here, we consider the pump strength $G/\kappa=2$, the external decay $\kappa_{\rm ex}^p=2\kappa=640$ kHz, and the coupling rate of the parametric nonlinear process ($g/\kappa=0.001$\,\cite{lu2020toward}), which yields the feasible critical pump power $P_c\approx 2.1$ {nW}. 

{\it Conclusions.---}
In summary, we have proposed a method for all-optical manipulating superradiant phase transitions and multicritical phenomena in an open dual-coupling JC model. The nonreciprocal first- and second-order superradiant transitions can be easily achieved by tuning the external pump field. Furthermore, the model allows for exquisite manipulation of the tricritical and multicritical points, both exhibiting controllable nonreciprocity. {\color{black}This general approach can be extended to {\color{black} the case of $N$ particles ($N\gg 1$)}, {specifically in the context of} the dual-coupling Tavis-Cummings model (see Sec. S10 in \cite{SM})}. 
{\color{black}We anticipate that this work will stimulate further theoretical studies and experimental explorations of a broader range of physical phenomena, such as superradiant cooling~\cite{Minghui2016} and atomic synchronization\,\cite{Minghui2014,masson2022universality}, and could find applications in modern quantum technology\,\cite{Braunstein2005}. }


{\it Acknowledgments.---} We acknowledge valuable discussions with Ye-Xiong Zeng, Ran Huang, and Zi-Yong Ge. We also appreciate the valuable comments from Ye-Hong Chen, Zongping Gong, Keyu Xia, Bo Wang, and Zeliang Xiang. X.-Y.L. is supported by the National Key Research and Development Program of China (Grant No. 2021YFA1400700). F.N. is supported in part by:
Nippon Telegraph and Telephone Corporation (NTT) Research,
the Japan Science and Technology Agency (JST)
[via the Quantum Leap Flagship Program (Q-LEAP), and the Moonshot R\&D Grant Number JPMJMS2061],
the Asian Office of Aerospace Research and Development (AOARD) (via Grant No. FA2386-20-1-4069),
and the Office of Naval Research (ONR) (via Grant No. N62909-23-1-2074). The computation is completed in the HPC Platform of Huazhong University of Science and Technology.
%

\newpage

\onecolumngrid

\setcounter{equation}{0} \setcounter{figure}{0}
\setcounter{table}{0}
\setcounter{page}{1}\setcounter{secnumdepth}{3} \makeatletter
\renewcommand{\theequation}{S\arabic{equation}}
\renewcommand{\thefigure}{S\arabic{figure}}
\renewcommand{\bibnumfmt}[1]{[S#1]}
\renewcommand\thesection{S\arabic{section}}
\clearpage

\begin{center}
{\large \bf Supplementary Material for ``Nonreciprocal Superradiant Phase Transitions and Multicriticality in a Cavity QED System''}
\end{center}

\begin{center}
Gui-Lei Zhu$^{1,2}$, Chang-Sheng Hu$^{3}$, Hui Wang$^2$,Wei Qin$^{2,4}$,
Xin-You L\"{u}$^{5,^*}$ and Franco Nori$^{2,6,7,^{\dagger}}$
\end{center}

\begin{minipage}[]{18cm}
\small{\it
\centering $^{1}$Department of Physics, Zhejiang Sci-Tech University, Hangzhou 310018, China\\
\centering $^{2}$Theoretical Quantum Physics Laboratory, Cluster for Pioneering Research, RIKEN, Wakoshi, Saitama 351-0198, Japan \\
\centering $^{3}$Department of Physics, Anhui Normal University, Wuhu 241000, China\\
\centering $^{4}$Center for Joint Quantum Studies and Department of Physics,
School of Science, Tianjin University, Tianjin 300350, China\\
\centering $^{5}$School of Physics, Huazhong University of Science and Technology
and Wuhan Institute of Quantum Technology, Wuhan 430074, China\\
\centering $^{6}$Quantum Computing Center, RIKEN, Wakoshi, Saitama 351-0198, Japan \\
\centering $^{7}$Department of Physics, The University of Michigan, Ann Arbor, MI, 48109-1040, USA\\}

\end{minipage}

\vspace{8mm}
\tableofcontents

\section{Mean-field solutions of $\alpha$ and $\beta$}\label{section1}
We recall that the system Hamiltonian in the forward pump reads,
\begin{align}
H=&(\Delta+\Delta_F)a^{\dagger}a+\frac{\Delta_q}{2}\sigma_z+g_a(a\sigma_++a^{\dagger}\sigma_-)+G(a^{\dagger 2}+a^2)\nonumber
\\
&+(\Delta-\Delta_F)b^{\dagger}b+g_b(b\sigma_++b^{\dagger}\sigma_-)+J(a^{\dagger}b+b^{\dagger}a),\label{H_original}
\end{align}
where $a$ and $b$ are the annihilation operators for the counterclockwise and clockwise cavity modes, respectively. Also, $\sigma_{\pm}=(\sigma_x+i\sigma_y)/2$, and $\sigma_{x,y,z}$ are Pauli matrices of the two-level atom. Here, $\Delta_F$ is the Sagnac shift caused by the rotation of the cavity. The counterclockwise and clockwise cavity modes are coupled to the atom with the strengths $g_a$ and $g_b$, respectively.  By employing the master equation (2) shown in the main text, we can derive the Heisenberg equations of motion for operators,
\begin{align}\label{Heisenberg equations-1}
\frac{da}{dt}&=-i(\Delta+\Delta_F)a-\kappa a-2iGa^{\dag}-iJb-\frac{i}{2}g_a\sigma_x-\frac{1}{2}g_a\sigma_y,\\
\frac{db}{dt}&=-i(\Delta-\Delta_F)b-\kappa b-iJa-\frac{i}{2}g_b\sigma_x-\frac{1}{2}g_b\sigma_y,\\
\frac{d\sigma_x}{dt}&=-\Delta_q\sigma_y+ig_a(a-a^{\dag})\sigma_z+ig_b(b-b^{\dag})\sigma_z,\\
\frac{d\sigma_y}{dt}&=\Delta_q \sigma_x-g_a(a+a^{\dag})\sigma_z-g_b(b+b^{\dag})\sigma_z,\\
\frac{d\sigma_z}{dt}&=ig_a(a^{\dag}-a)\sigma_x+ig_b(b^{\dag}-b)\sigma_x+g_a(a+a^{\dag})\sigma_y+g_b(b+b^{\dag})\sigma_y.\label{Heisenberg equations-6}
\end{align} 
In these equations, we assumed that the decay rates of cavity modes $a$ and $b$ are equal, namely, $\kappa_a=\kappa_b=\kappa$. Additionally, we neglected the atomic loss in the system. 
Similar to the standard Rabi model, in the steady-state limit, we set the following expectation values: $\langle\sigma_x\rangle=X$, $\langle\sigma_y\rangle=Y$, $\langle \sigma_z\rangle =Z$, $\langle a\rangle=\alpha\sqrt{\eta_+}$, $\langle b\rangle=\beta\sqrt{\eta_-}$, where 
\begin{align}
\eta_+=\frac{\Delta_q}{\Delta+\Delta_F}, \,\,\,\eta_-=\frac{\Delta_q}{\Delta-\Delta_F}.
\end{align} 
The mean-field approximation holds in the infinite detuning limit, where $\eta_{\pm}\rightarrow\infty$. {\color{black}The validity of the mean-field approximation is discussed in Sec.\,\ref{section6}.}
Under the spin-conservation law, $X^2+Y^2+Z^2=1$, the following relations are obtained:
\begin{align}\label{Y1}
C&=\Gamma_1\alpha_{\rm re}^2+\Gamma_2\alpha_{\rm re}\alpha_{\rm im}+\Gamma_3\alpha_{\rm im}^2,\\\label{Y2}
C&=K(\alpha_{\rm re}^2+\alpha_{\rm im}^2),
\end{align}
where $C=1-Z^2$, 
\begin{align}\Gamma_{1,3}=\frac{[16(\Delta+\Delta_F\pm2G)^2+\kappa^2]}{(\Delta+\Delta_F)^2\lambda_a^2}, \,\,\,\Gamma_2=\frac{128G\kappa}{(\Delta+\Delta_F)^2\lambda_a^2}, \,\,\,K=\frac{16[(\Delta-\Delta_F)^2+\kappa^2]\lambda_a^2Z^2}{16\kappa^2+(\Delta-\Delta_F)^2(4+\lambda_b^2Z)^2}.
\end{align} 
In these equations, we have separated the real and imaginary parts of $\alpha$ and $\beta$ as $\alpha=\alpha_{\rm re}+i\alpha_{\rm im}$ and $\beta=\beta_{\rm re}+i\beta_{\rm im}$. The constraint parameter $Z$ satisfies
 \begin{align}\label{Zpm}
\!\!\!\!Z_{\pm}=4\frac{\!-\!\Delta_+h_1
-\Delta_-h_2
\pm\sqrt{\big[h_1(2G-\kappa)-\kappa h_2\big]\big[h_1(2G+\kappa)+\kappa h_2\big]} }{\Delta_+h_1\lambda_a^2+2\Delta_+\Delta_-(\Delta_+\Delta_-+\kappa^2)\lambda_a ^2\lambda_b^2+\Delta_-h_2\lambda_b^2},
\end{align}
where we have denoted $\Delta_{\pm}=\Delta\pm\Delta_F$, $h_1=\Delta_+(\Delta_-^2+\kappa^2)\lambda_a ^2$ and $h_2=\Delta_-(\Delta_+^2-4G^2+\kappa^2)\lambda_b^2$ for simplification purposes. 

The mean-field solutions of $\alpha$ and $\beta$ in the steady state can be obtained by solving Eqs.\,(\ref{Y1}) and (\ref{Y2}). The solutions are given by
\begin{align}\label{are}
\alpha_{\rm re}&=\pm\sqrt{C}\sqrt{\frac{\Gamma_2^2-2(\Gamma_1-\Gamma_3)(\Gamma_3-K)+\Gamma_2\sqrt{\Gamma_2^2+4(\Gamma_1-K)(K-\Gamma_3)}}{2(\Gamma_2^2+(\Gamma_1-\Gamma_3)^2)K}},\\\label{aim}
\alpha_{\rm im}&=\frac{\Gamma_2-\sqrt{\Gamma_2^2+4(\Gamma_1-K)(K-\Gamma_3)}}{2(K-\Gamma_3)}\alpha_{\rm re},\\\label{bre}
\beta_{\rm re}&=-4\frac{(\Delta+\Delta_F+2G)\alpha_{\rm re}+\kappa\alpha_{\rm im}}{(\Delta+\Delta_F)\lambda_a\lambda_bZ}-\frac{\lambda_a}{\lambda_b}\alpha_{\rm re},\\\label{bim}
\beta_{\rm im}&=-4\frac{(\Delta+\Delta_F-2G)\alpha_{\rm im}-\kappa\alpha_{\rm re}}{(\Delta+\Delta_F)\lambda_a\lambda_bZ}-\frac{\lambda_a}{\lambda_b}\alpha_{\rm im}.
\end{align}
where we have defined the rescaled atom-field coupling strengths as $\lambda_a=2g_a/\sqrt{(\Delta+\Delta_F)\Delta_q}$ and $\lambda_b=2g_b/\sqrt{(\Delta-\Delta_F)\Delta_q}$. 

\section{Effective Hamiltonian in the infinite-detuning limit ($\Delta_q/(\Delta\pm\Delta_F)\rightarrow \infty$)}\label{section2}
In Sec.\,\ref{section1}, we used the mean-field approximation to obtain the steady-state solutions of $\alpha$ and $\beta$, which neglected quantum fluctuations. In this section, we provide a full quantum description of the Hamiltonian.

\subsection{Low-energy Hamiltonian in the normal phase}
In the large detuning limit, where the atomic energy scale is much larger than the field part, we can decouple the atomic subspaces and focus on the lowest atomic subspace. This allows us to eliminate the atomic part and obtain the effective low-energy Hamiltonian.   To achieve this, we rewrite Hamiltonian (\ref{H_original}) as
\begin{align}
H&=H_c+g_aV_a+g_bV_b,
\end{align}
with
\begin{align}
H_c=(\Delta+\Delta_F)a^{\dagger}a+\frac{\Delta_q}{2}\sigma_z+G(a^{\dagger 2}+a^2)
+(\Delta-\Delta_F)b^{\dagger}b+J(a^{\dag}b+b^{\dag}a),\nonumber
\end{align}
and
\begin{align}
V_a=(a\sigma_+\!+\!a^{\dagger}\sigma_-), \,\,\,\,\,\,\,\,\,V_b=(b\sigma_+\!+\!b^{\dagger}\sigma_-).\nonumber
\end{align}
The Hamiltonian $H_c$ has two decoupled spin subspaces $\mathcal{H}_{\downarrow}$ and $\mathcal{H}_{\uparrow}$, where $\lvert \uparrow\rangle$ and $\lvert \downarrow\rangle$ are the eigenstates of $\sigma_z$. However, the Hamiltonian $g_aV_a+g_bV_b$ introduces the interactions between these two subspaces. To eliminate the interaction terms, we apply the Schrieffer-Woff transformation with the unitary operator $S_{\rm np}=g_a/\Delta_q(a^{\dag}\sigma_--a\sigma_+)+g_b/\Delta_q(b^{\dag}\sigma_--b\sigma_+)$ to the master equation given in Eq. (2) of the main text. We keep terms up to $g_{a,b}^2/\Delta_q$ and neglect higher-order terms in the large detuning limit, i.e., $\Delta_q/(\Delta\pm\Delta_F)\rightarrow \infty$ and $\Delta_q/G\rightarrow \infty$. Under the projection of the $\mathcal{H}_{\downarrow}$ space, we obtain the effective master equation in the normal phase,
\begin{align}\label{mastereq:np}
\dot{\rho}_{\rm np}=-i[H_{\rm np},\rho_{\rm np}]+\kappa(2a\rho_{\rm np}a^{\dag}-a^{\dag}a\rho_{\rm np}-\rho_{\rm np}a^{\dag}a)+\kappa(2b\rho_{\rm np}b^{\dag}-b^{\dag}b\rho_{\rm np}-\rho_{\rm np}b^{\dag}b),
\end{align}
where $\rho_{\rm np}=\langle\downarrow|e^{-S_{\rm np}}\rho e^{S_{\rm np}}|\downarrow\rangle$, and the effective low-energy Hamiltonian in the normal phase becomes,
\begin{align}\label{Hnpeff}
H_{\rm np}&\equiv\langle\downarrow|e^{-S_{\rm np}}He^{S_{\rm np}}|\downarrow\rangle\nonumber\\
&=\Delta_aa^{\dag}a+\Delta_bb^{\dag}b+G(a^{\dag 2}+a^2)+J_s(ab^{\dag}+a^{\dag}b)-\frac{\Delta_q}{2},
\end{align}
where 
$\Delta_a=(\Delta+\Delta_F)(1-{\lambda_a^2}/{4}), 
\Delta_b=(\Delta-\Delta_F)(1-{\lambda_b^2}/{4}), 
J_s=J-\!\lambda_a\lambda_b\sqrt{(\Delta+\Delta_F)(\Delta-\Delta_F)}/{4}$
and the renormalized atom-field couplings $\lambda_a={2g_a}/{\sqrt{(\Delta+\Delta_F)\Delta_q}}$ and $\lambda_b={2g_b}/{\sqrt{(\Delta-\Delta_F)\Delta_q}}$. 

\subsection{Low-energy Hamiltonian in the superradiant phase}

To address the superradiant phase where the cavity fields are macroscopically occupied, we apply displacement transformations to cavity modes. Specifically, we have
\begin{align} D^{\dag}[\alpha\sqrt{\eta_+}]aD[\alpha\sqrt{\eta_+}]=c+\alpha\sqrt{\eta_+},\,\,\, D^{\dag}[\beta\sqrt{\eta_-}]bD[\beta\sqrt{\eta_-}]=d+\beta\sqrt{\eta_-}\end{align} with $D[\alpha\sqrt{\eta_+}]=\exp[\sqrt{\eta_+}(\alpha a^{\dag}-\alpha^* a)]$ and $D[\beta\sqrt{\eta_-}]=\exp[\sqrt{\eta_-}(\beta b^{\dag}-\beta^* b)]$
. Notably, the displacements $\alpha$ and $\beta$ correspond to the mean-field solutions discussed in Sec.\,\ref{section1}. Here, we use $c$ and $d$ to denote the fluctuation operators of the cavity modes. By applying the displacement transformation, the master equation becomes
\begin{align}
\dot{\tilde{\rho}}=-i[H(\alpha,\beta),\tilde{\rho}]+\kappa(2c\tilde{\rho}c^{\dag}-c^{\dag}c\tilde{\rho}-\tilde{\rho}c^{\dag}c)+\kappa(2d\tilde{\rho}d^{\dag}-d^{\dag}d\tilde{\rho}-\tilde{\rho}d^{\dag}d), 
\end{align}
where $\tilde{\rho}=D^{\dag}[\alpha\sqrt{\eta_+}]D^{\dag}[\beta\sqrt{\eta_-}]\rho D[\alpha\sqrt{\eta_+}]D[\beta\sqrt{\eta_-}]$ and $H(\alpha,\beta)$ reads,
\begin{align}\label{Hab}
H(\alpha,\beta)=&\Delta_+ c^{\dag}c+\frac{\Delta_q}{2}\sigma_z+\Delta_-d^{\dag}d+\frac{1}{2}\lambda_a\sqrt{\Delta_q\Delta_+}(c\sigma_++c^{\dag}\sigma_-)+\frac{1}{2}\lambda_b\sqrt{\Delta_q\Delta_-}(d^{\dag}\sigma_-+d\sigma_+)\!\nonumber\\
&+G(c^{\dag  2}+c^2)+J(c^{\dag}d+d^{\dag}c)+\frac{\Delta_q\lambda_a}{2}(\alpha\sigma_++\alpha^*\sigma_-)+\frac{\Delta_q\lambda_b}{2}(\beta\sigma_++\beta^*\sigma_-),
\end{align}
where $\Delta_{\pm}=\Delta\pm\Delta_F$, $\alpha=\alpha_{\rm re}+i\alpha_{\rm im}, \beta=\beta_{\rm re}+i\beta_{\rm im}$ and we have omitted the linear terms and constant terms in the Hamiltonian.

Next, we focus on the atomic part of Hamiltonian, which includes terms such as ${\Delta_q}\sigma_z/{2}+{\Delta_q\lambda_a}(\alpha\sigma_++\alpha^*\sigma_-)/{2}+{\Delta_q\lambda_b}(\beta\sigma_++\beta^*\sigma_-)/{2}$. The eigenvalues of this atomic part are $\pm{\tilde{\Delta}_q}/{2}$, with
\begin{align}\tilde{\Delta}_q=\Delta_q\sqrt{1+\lambda_a^2(\alpha_{\rm re}^2+\alpha_{\rm im}^2)+2\lambda_a\lambda_b(\alpha_{\rm im}\beta_{\rm im}+\alpha_{\rm re}\beta_{\rm re})+\lambda_b^2(\beta_{\rm im}^2+\beta_{\rm re}^2)},\end{align} and the corresponding eigenstates are given by
\begin{align}\label{lowest-energy state}
\lvert\tilde{\downarrow}\rangle&=\cos\theta e^{i\gamma}\lvert\uparrow\rangle+\sin\theta e^{i\phi}\lvert\downarrow\rangle,\\
\lvert\tilde{\uparrow}\rangle&=-\sin\theta e^{-i\phi}\lvert\uparrow\rangle+\cos\theta e^{-i\gamma}\lvert\downarrow\rangle,
\end{align}
where \begin{align}
e^{i\phi}=\frac{\lambda_a(\alpha_{\rm im}+i\alpha_{\rm re})+\lambda_b(\beta_{\rm im}+i\beta_{\rm re})}{\sqrt{\lambda_a^2(\alpha_{\rm re}^2+\alpha_{\rm im}^2)+2\lambda_a\lambda_b(\alpha_{\rm im}\beta_{\rm im}+\alpha_{\rm re}\beta_{\rm re})+\lambda_b^2(\beta_{\rm re}^2+\beta_{\rm im}^2)}}, 
\end{align}
 and $e^{i\gamma}=-i$. The angle $\theta$ satisfies
\begin{align}
\tan 2\theta=-\sqrt{\lambda_a^2(\alpha_{\rm re}^2+\alpha_{\rm im}^2)+2\lambda_a\lambda_b(\alpha_{\rm im}\beta_{\rm im}+\alpha_{\rm re}\beta_{\rm re})+\lambda_b^2(\beta_{\rm re}^2+\beta_{\rm im}^2)}.
\end{align}
 Now, we define the Pauli matrices in the $\{\lvert\tilde{\downarrow}\rangle,\lvert\tilde{\uparrow}\rangle\}$ basis as $\tau_{x}=\lvert\tilde{\uparrow}\rangle\langle\tilde{\downarrow}\rvert+\lvert\tilde{\downarrow}\rangle\langle\tilde{\uparrow}\rvert$, $\tau_y=-i(|\tilde{\uparrow}\rangle\langle\tilde{\downarrow}\rvert-\lvert\tilde{\downarrow}\rangle\langle\tilde{\uparrow}\rvert)$ and $\tau_z=\lvert\tilde{\uparrow}\rangle\langle\tilde{\uparrow}\rvert-\lvert\tilde{\downarrow}\rangle\langle\tilde{\downarrow}\rvert$. Then, $\sigma_{\pm,z}$ in terms of  $\tau_{\pm,z}$ can be expressed as
\begin{align}\label{sigmap}
\sigma_+&=-\sin\theta\cos\theta e^{i(\phi-\gamma)}\tau_z+\cos^2\theta e^{-2i\gamma}\tau_--\sin^2\theta e^{2i\phi}\tau_+,\\\label{sigmaz}
\sigma_z&=-\cos2\theta \tau_z-\sin2\theta e^{i(\phi+\gamma)}\tau_+-\sin2\theta e^{-i(\phi+\gamma)}\tau_-.
\end{align}
By substituting the expression for $\sigma_{\pm}$ and $\sigma_z$ from Eqs.\,{(\ref{sigmap}) and (\ref{sigmaz})} into the Hamiltonian equation\,(\ref{Hab}),we obtain the transformed Hamiltonian as follows:
\begin{align}\label{Hab2}
H(\alpha,\beta)&=\Delta_+ c^{\dag}c+\Delta_-d^{\dag}d+\frac{\tilde{\Delta}_q}{2}\tau_z+G(c^{\dag 2}+c^2)+J(c^{\dag}d+d^{\dag}c)\nonumber\\
&-\frac{1}{2}\lambda_a\sqrt{\Delta_q\Delta_+}\sin^2\theta(c\tau_+e^{2i\phi}+c^{\dag}\tau_-e^{-2i\phi})-\frac{1}{2}\lambda_a\sqrt{\Delta_q\Delta_+}\cos^2\theta(c\tau_-+c^{\dag}\tau_+)\nonumber\\
&-\frac{1}{2}\lambda_b\sqrt{\Delta_q\Delta_-}\sin^2\theta(d\tau_+e^{2i\phi}+d^{\dag}\tau_-e^{-2i\phi})-\frac{1}{2}\lambda_b\sqrt{\Delta_q\Delta_-}\cos^2\theta(d\tau_-+d^{\dag}\tau_+).
\end{align}
Next, we apply the unitary transformation $H'_{\rm sp}=e^{-S_{\rm sp}}H(\alpha,\beta)e^{S_{\rm sp}}$ with the operator $S_{\rm sp}$ given by:
\begin{align}
S_{\rm sp}=&\frac{\lambda_a\sqrt{\Delta_q(\Delta+\Delta_F)}}{2\bar{\Delta}_q}\left[\sin^2\theta(c\tau_+e^{2i\phi}-c^{\dag}\tau_-e^{-2i\phi})+\cos^2(c^{\dag}\tau_+-c\tau_-)\right]\nonumber\\
&+\frac{\lambda_b\sqrt{\Delta_q(\Delta-\Delta_F)}}{2\bar{\Delta}_q}\left[\sin^2\theta(d\tau_+e^{2i\phi}-d^{\dag}\tau_-e^{-2i\phi})+\cos^2\theta(d^{\dag}\tau_+-d\tau_-)\right].
\end{align}
Projecting $H'_{\rm sp}$ onto $|\tilde{\downarrow}\rangle$ basis, we obtain the effective Hamiltonian in the superradiant phase
\begin{align}\label{Heff}
H_{\rm sp}&=\Lambda_1c^{\dagger}c+\Lambda_2d^{\dag}d\!+\!G(c^2\!+\!c^{\dag 2})\!+\!\Lambda_3(cd^{\dag}\!+\!c^{\dag}d)
+\Lambda_4(e^{2i\phi}c^2\!+\!e^{-2i\phi}c^{\dag 2})\!\nonumber\\
&+\!\Lambda_5(e^{2i\phi}d^2\!+\!e^{-2i\phi}d^{\dag 2})+\Lambda_6(e^{2i\phi}cd\!+\!e^{-2i\phi}c^{\dag}d^{\dag}),
\end{align}
where
\begin{align}
&\Lambda_1=(\Delta+\Delta_F)\left(1-\frac{\lambda_a^2\Delta_q(\sin^4\theta+\cos^4\theta)}{4\tilde{\Delta}_q}\right),\,\,\,\,
\Lambda_2=(\Delta-\Delta_F)\left(1-\frac{\lambda_b^2\Delta_q(\sin^4\theta+\cos^4\theta)}{4\tilde{\Delta}_q}\right),\nonumber\\
&\Lambda_3=J-\frac{\sqrt{(\Delta+\Delta_F)(\Delta-\Delta_F)}\lambda_a\lambda_b\Delta_q(\sin^4\theta+\cos^4\theta)}{4\tilde{\Delta}_q},\,\,\,\,
\Lambda_4=-\frac{(\Delta+\Delta_F)\lambda_a^2\Delta_q\sin^2\theta\cos^2\theta}{4\tilde{\Delta}_q},\nonumber\\
&\Lambda_5=-\frac{(\Delta-\Delta_F)\lambda_b^2\Delta_q\sin^2\theta\cos^2\theta}{4\tilde{\Delta}_q},\,\,\,\,
\Lambda_6=-\frac{\sqrt{(\Delta+\Delta_F)(\Delta-\Delta_F)}\lambda_a\lambda_b\Delta_q\sin^2\theta\cos^2\theta}{2\tilde{\Delta}_q}.
\end{align}

\section{Phase transition boundaries}\label{section3}
\subsection{First-order phase transition boundary}
In this section, we derive the first-order phase transition boundary for the superradiant phase transition. Starting from Eq.\,(\ref{Zpm}), we impose a constraint that the square root term must be greater than or equal to zero, resulting in the condition:
\begin{align}\label{constraint}
h_1^2(4G^2-\kappa^2)-h_2^2\kappa^2\ge2h_1h_2,
\end{align}
where $h_1=\Delta_+(\Delta_-^2+\kappa^2)\lambda_a ^2$ and $h_2=\Delta_-(\Delta_+^2-4G^2+\kappa^2)\lambda_b^2$.
We can set this condition with an equal sign to obtain the critical ratio between the atom-field couplings $\lambda_{ac}$ and $\lambda_{bc}$ as follows:
\begin{align}\label{chic}
\chi_c=\frac{\lambda_{bc}}{\lambda_{ac}}=\frac{\sqrt{2G-\kappa}\sqrt{(\Delta+\Delta_F)((\Delta-\Delta_F)^2+\kappa^2)}}{\sqrt{(\Delta-\Delta_F)\kappa((\Delta+\Delta_F)^2-4G^2+\kappa^2)}}.
\end{align}
This ratio $\chi_c$ defines the first-order superradiant phase transition boundary in the $\lambda_a$--$\lambda_b$ phase space.
To ensure that $\chi_c$ is a real number, the pump strength $G$ must satisfy the condition:
\begin{align}
\frac{\kappa}{2}<G<\frac{\sqrt{(\Delta+\Delta_F)^2+\kappa^2}}{2}.
\end{align}
From the condition in Eq.\,(\ref{constraint}), we can derive the critical pump strength for the first-order phase transition, denoted as $G_{\rm c}^{\rm 1st}$,
\begin{align}
G_{\rm c}^{\rm 1st}=\frac{1}{4\Delta_-\kappa \lambda_b^2}\left\{-\Delta_+(\Delta_-^2+\kappa^2)\lambda_a^2+\sqrt{\Big[\Delta_+(\Delta_-^2+\kappa^2)\lambda_a^2+2\Delta_-\kappa^2\lambda_b^2\Big]^2+4\Delta_+^2\Delta_-^2\kappa^2\lambda_b^4}\right\}.
\end{align}
In the special case where $\lambda_a=\lambda_b=\lambda$, the above equation simplifies to 
\begin{align}\label{Gc1st}
G_{\rm c}^{\rm 1st}=\frac{1}{4\Delta_-\kappa}\left\{-\Delta_+(\Delta_-^2+\kappa^2)+\sqrt{\Big[\Delta_+(\Delta_-^2+\kappa^2)+2\Delta_-\kappa^2\Big]^2+4\Delta_+^2\Delta_-^2\kappa^2}\right\}, 
\end{align}
which is Eq.\,(3) in the main text.

\subsection{Second-order phase transition boundary}
In order to explore the second-order phase transition boundary of the system, we start with Eq.\,(\ref{mastereq:np}) and derive the dynamic equation for the first-order bosonic moments
\begin{align}
\dot{\textbf{L}}_{\rm np}=\Sigma_{\rm np}{\textbf L}_{\rm np},
\end{align} 
where ${\textbf L}_{\rm np}=[{\langle a\rangle}, {\langle a^{\dag}\rangle}, {\langle b\rangle}, {\langle b^{\dag}\rangle}]^{\rm T}$
and \begin{eqnarray}
\!\!\!\Sigma_{\rm np}=\left(\begin{array}{cccc}\!-i\Delta_a\!-\!\kappa&\!-2iG&\!-iJ_s&\!0\\
\!2iG &i\Delta_a\!-\!\kappa &\!0&\!iJ_s\\
\!-iJ_s&\!0&\!-i\Delta_b-\kappa &\!0\\
\!0&\!iJ_s&\!0&\!i\Delta_b-\kappa
\end{array}\right),
 \end{eqnarray}
where $\Delta_a, \Delta_b$ and $J_s$ have been defined in Eq.\,(\ref{Hnpeff}).
Diagonalizing the matrix $\Sigma_{\rm np}$ and setting the real part of the eigenvalues to zero, we obtain that for a certain $G$, the two critical atom-field couplings $\lambda_{ac}$ and $\lambda_{bc}$ satisfy:
 \begin{eqnarray}\label{2ndboundary}
\!\!\lambda_{bc}={\frac{\Big\{4q_1\Delta_--\Delta_+(\Delta_+\Delta_-\!+\!\kappa^2)\lambda_{ac}^2\!-\!2\Big[4\kappa^2q_1(\Delta_F\Delta_+\lambda_{ac}^2-q_1)\!+\!\Delta_+^2\Big(\Delta_-^2G^2+(G^2-\Delta_F^2)\kappa^2\Big)\lambda_{ac}^4\Big]^{1/2}~\Big\}^{1/2}}{q_1\Delta_-}},
 \end{eqnarray} 
where $q_1=\Delta_+^2-4G^2+\kappa^2$.
Equation \,(\ref{2ndboundary}) gives the second-order phase transition boundary in the $\lambda_a$--$\lambda_b$ phase space.

In Fig.\,\ref{figS1} we plot the order parameters $\alpha_{\rm re}, \alpha_{\rm im}, \beta_{\rm re}, \beta_{\rm im}$ for the forward [panels (a)-(d)] and backward [panels (e)-(h)] pump cases. For each fixed pump direction, the order parameters exhibit similar features and share the same boundary. The system undergoes phase transitions from the normal phase (NP) to the superradiant phase (SP) through two distinct paths: one is through the first-order phase transition boundary (indicated by the gray dashed line) where the order parameters undergo an abrupt and discontinuous change from zero to a finite value (as given by Eq.\,(\ref{chic})); the other path is through the second-order phase transition boundary (magenta dash-dotted curve), where the order parameters vary continuously from zero to a finite value as $\lambda_a$ or $\lambda_b$ increases [as given by Eq.\,(\ref{2ndboundary})]. The first- and second-order phase transition boundaries meet at the tricritical point (see below). It is worth noting that the phase diagram of the order parameters exhibits {\it different} boundaries for {\it different} pump directions at a fixed pump strength. This {\it nonreciprocal} nature of the phase transition highlights the dependence of the system's behavior on the pump direction. 
\begin{figure}
\includegraphics[width=17.4cm]{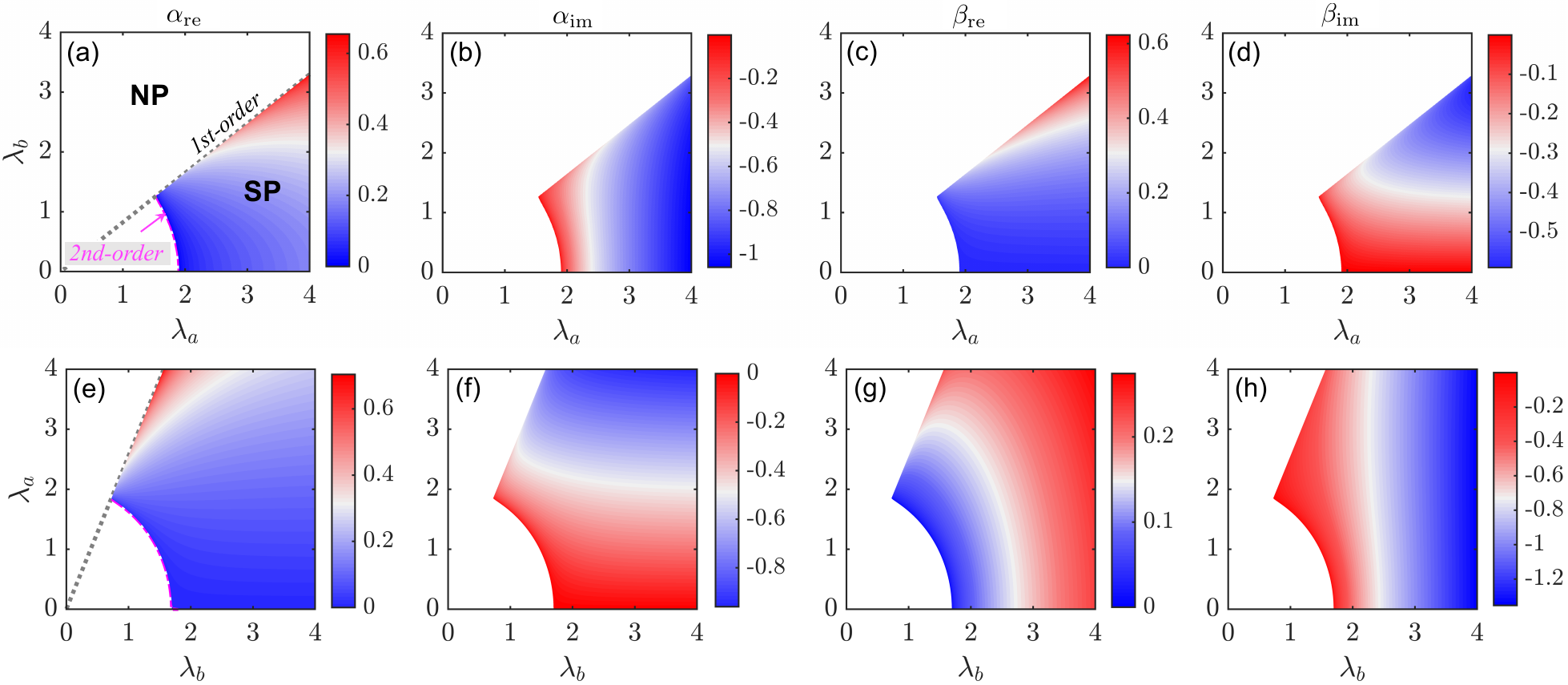}
\caption{Order parameters $\alpha_{\rm re}, \alpha_{\rm im}, \beta_{\rm re}, \beta_{\rm im}$ as functions of the atom-field couplings $\lambda_a$ and $\lambda_b$ for the forward (a-d) and backward (e-h) pumps. 
The gray dotted line and magenta dash-dotted curves denote the first-order [see Eq.\,(\ref{chic})] and second-order [see Eq.\,(\ref{2ndboundary})] phase transition boundaries, respectively. We consider $G/\kappa=1.5$ for both cases, and $\Delta_F/\Delta=0.5$ for the forward pump, and $\Delta_F/\Delta=-0.5$ for the backward pump, respectively. All other parameters are the same as in Fig.\,1 of the main text.}
	\label{figS1}
\end{figure}

In the case of tuned atom-field couplings where $\lambda_a=\lambda_b=\lambda$, the eigenvalues of $\Sigma_{\rm np}$ are given by 
 \begin{eqnarray}\label{Enp}
\!\!E_{\rm np,\pm}^{(1)}&=-\kappa\pm \frac{\Big\{16G^2+4\Delta_F^2p-\Delta^2(p^2+4)-\Big[(\Delta^4\lambda^4-32\Delta\Delta_FG^2-8\Delta^2\Delta_F^2p)(p-2)^2-16G^2\lambda^4\Delta_+\Delta_-+256G^4\Big]^{1/2}\Big\}^{1/2}}{2\sqrt{2}},\\
\!\!E_{\rm np,\pm}^{(2)}&=-\kappa\pm \frac{\Big\{16G^2+4\Delta_F^2p-\Delta^2(p^2+4)+\Big[(\Delta^4\lambda^4-32\Delta\Delta_FG^2-8\Delta^2\Delta_F^2p)(p-2)^2-16G^2\lambda^4\Delta_+\Delta_-+256G^4\Big]^{1/2}\Big\}^{1/2}}{2\sqrt{2}},
 \end{eqnarray}
with $p=\lambda^2-2$.
Setting the real part of the eigenvalues equal to zero, i.e., ${\rm Re}[E_{\rm np}]=0$, we obtain the critical pump strength for the second-order phase transition
\begin{align}\label{Gc2nd}
G_c^{\rm 2nd}=\frac{\Big\{\Delta^4(\lambda^2-2)^2+\big(2\kappa^2-\Delta_F ^2(\lambda^2-2)\big)^2+\Delta^2\Big[\kappa^2((\lambda^2-2)^2+4)-2\Delta_F^2(\lambda^2-2)^2\Big]\Big\}^{1/2}}{\Big[16\kappa^2+(\Delta-\Delta_F)^2(\lambda^2-4)^2\Big]^{1/2}},
\end{align}
which is Eq.\,(4) of the main text. {\color{black}In Fig.\,\ref{figS2}(a), we illustrate $G_c^{\rm 1st}$ and $G_c^{\rm 2nd}$ as a function of the Sagnac shift $\Delta_F$. The interplay of cavity rotation and directional pumping induces opposite Sagnac shifts, consequently causing the critical points moving towards larger (or smaller) values for the forward (or backward) pump.}


The tricritical point is the intersection point where the first- and second-order boundaries meet. By setting $G_{c}^{\rm 1st}=G_{c}^{\rm 2nd}$, we can determine the atom-field coupling of the tricritical point, 
 \begin{eqnarray}
\!\!\lambda_{\rm tric}\!=\!\Big(\frac{\Delta_-^2\!+\!\kappa^2}{\Delta_-^3(\Delta_+\Delta_-\!+\!\kappa^2)}\Big)^{{1}/{2}} \Big[
3\Delta_-^2\Delta_+\!-\!(\Delta_F\!-\!3\Delta)\kappa^2\!-\!\Big(\!\Delta_-^4\Delta_+^2\!+\!2\Delta_-^2\Delta_+(5\Delta\!+\!\Delta_F)\kappa^2\!+\!(\Delta_F\!-\!3\Delta)^2\kappa^4\Big)^{1/2}\Big]^{1/2}.
 \end{eqnarray}
If the fixed coupling $\lambda>\lambda_{\rm tric}$ (or $\lambda<\lambda_{\rm tric}$), surpassing the critical pump $G_{c}^{\rm 1st}$ (or $G_{c}^{\rm 2nd}$) can trigger a first- (or second-) order phase transition. 
In our considered parameter regime, where $\kappa/\Delta\sim 0.01$ and $-\Delta<\Delta_F<\Delta$, the equation for the atom-field coupling at the tricritical point can be simplified as:
 \begin{align}\label{lambdatric}
\lambda_{\rm tric}\approx  \sqrt{\frac{2(\Delta+\Delta_F)[(\Delta-\Delta_F)^2+\kappa^2]}{(\Delta-\Delta_F)[(\Delta+\Delta_F)(\Delta-\Delta_F)+\kappa^2]}}\approx \sqrt{2}.
 \end{align}
In Fig.\,\ref{figS2}, we plot the difference between $\lambda_{\rm tric}$ and $\sqrt{2}$. It can be seen that this difference is on the order of $10^{-3}$ in our considered parameter regime, which is consistent with the analytical result given in Eq.\,(\ref{lambdatric}). In other words, in our model, the atom-field coupling at the tricritical point is insensitive to the Sagnac shift. 

\begin{figure}
	\includegraphics[width=14.4cm]{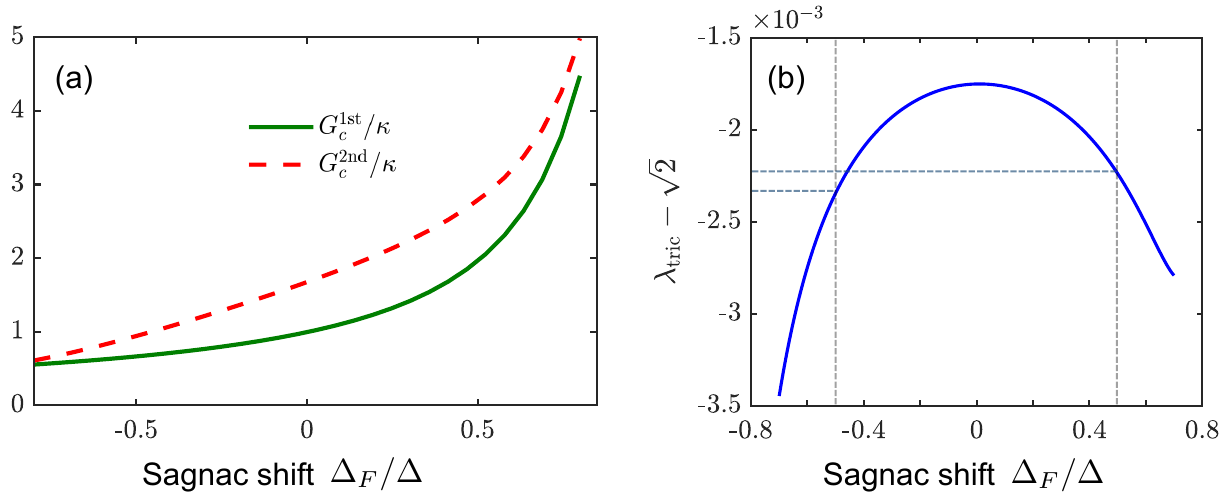}
	\caption{{{\color{black}(a) The first-order and second-order critical pumps $G_{c}^{\rm 1st}$ and $G_{c}^{\rm 2nd}$ versus the Sagnac shift $\Delta_F$}}. (b) The atom-field coupling strength $\lambda_{\rm tric}$ of the tricritical point minus $\sqrt{2}$ as a function of the Sagnac shift $\Delta_F$. The parameters used $\Delta_F/\Delta=\pm 0.5$ are labeled with the dotted curves.}
\label{figS2}
\end{figure}

\section{Dynamic evolution of two-operator correlators}


\subsection{Equations of motion in the normal phase}
To be consistent with the operator representation used in the main text, we perform a displacement on the cavity fields, i.e., $a\rightarrow c+\alpha\sqrt{\eta_+}$, $b\rightarrow d+\beta\sqrt{\eta_-}$, where $c$ and $d$ are cavity fluctuation operators. In the normal phase, $\alpha=\beta=0$. Using the equation $\partial \langle o\rangle/\partial t={\rm tr}\{o\partial \rho/\partial t\}$ for the expectation value $\langle o\rangle={\rm tr}\{o\rho\}$ of any observable $o$ of interest, and utilizing Eq.\,(\ref{mastereq:np}), we can obtain a set of closed equations of the motion for the two-operator fluctuation correlators. These equations take the following form: 
 \begin{align}\label{Wnp}
\frac{d{\textbf v}_{\rm np}}{dt}={W}_{\rm np}{ \textbf{v}_{\rm np}}+R_{\rm np},
\end{align} 
where ${\textbf v}_{\rm np}=[{\langle c^{\dag}c\rangle}, {\langle c^{\dag 2}\rangle}, {\langle c^{2}\rangle},{\langle cd\rangle}, {\langle c^{\dag}d^{\dag}\rangle},{\langle cd^{\dag}\rangle}, {\langle c^{\dag}d\rangle},{\langle d^{\dag}d\rangle},{\langle d^{2}\rangle},{\langle d^{\dag 2}\rangle}]^{\rm T}$,  $R_{\rm np}=[0,2iG,-2iG,0,0,0,0,0,0,0]^{T}$, 
and 
\begin{align}
W_{\rm np}=\left(\begin{array}{cccc}
W_{\rm np}^{(1,1)} &W_{\rm np}^{(1,2)}\\
W_{\rm np}^{(2,1)}&W_{\rm np}^{(2,2)}\\
\end{array}\right)
\end{align}
with \begin{eqnarray}
\!\!\!W_{\rm np}^{(1,1)}=\left(\begin{array}{ccccc}
\!-2\kappa\!&-\!2iG&\!2iG&\!0&\!0\\
\!4iG &2i\Delta_a\!-\!2\kappa&\!0&\!0&\!2iJ_s\\
\!-4iG&\!0&\!-2i\Delta_a\!-\!2\kappa &-2iJ_s&\!0\\
\!0&0&\!-iJ_s&-i(\Delta_a\!+\!\Delta_b)\!-\!2\kappa&\!0\\
0&iJ_s&0&0&i(\Delta_a\!+\!\Delta_b)\!-\!2\kappa\\
\end{array}\right),\nonumber
 \end{eqnarray}
\begin{eqnarray}
\!\!\!W_{\rm np}^{(1,2)}=\left(\begin{array}{ccccc}
iJ_s&-iJ_s&0&0&0\\
0&0&0&0&0\\
0 &0&0&0&0\\
0&-2iG&0 &-iJ_s&0\\
2iG&0&0&0&iJ_s\\
\end{array}\right),
W_{\rm np}^{(2,1)}=\left(\begin{array}{ccccc}
iJ_s&0&0&0&-2iG\\
-iJ_s &0&0&2iG&0\\
0&0&0 &0&0\\
0&0&0&-2iJ_s&0\\
0&0&0&0&2iJ_s\\
\end{array}\right),\nonumber
 \end{eqnarray}
\begin{eqnarray}
\!\!\!W_{\rm np}^{(2,2)}=\left(\begin{array}{ccccc}
\!-i(\Delta_a\!-\!\Delta_b)\!-\!2\kappa\!&\!0&\!-iJ_s&\!0&\!0\\
\!0 &i(\Delta_a\!-\!\Delta_b)\!-\!2\kappa&\!iJ_s&\!0&\!0\\
\!-iJ_s&\!iJ_s&\!-2\kappa &0&\!0\\
\!0&0&\!0&-2i\Delta_b\!-\!2\kappa&\!0\\
0&0&0&0&2i\Delta_b\!-\!2\kappa\\
\end{array}\right),\nonumber
 \end{eqnarray}
In the steady state, we can solve Equation (\ref{Wnp}) by setting the left-hand side to zero, yielding the following expression for the ten two-operator correlators in the steady state: $$\textbf{v}_{\rm np}=-W_{\rm np}^{-1}R_{\rm np}.$$ The analytical expressions of these correlators are complex and not presented here, but their numerical results are shown in Figs.\,3(a)-3(c) in the main text.

\subsection{Equations of motion in the superradiant phase}
To derive the dynamic equation for the two-operator fluctuation correlators in the superradiant phase, we start with the equation of motion for the reduced density matrix in the superradiant phase, given by \begin{align}\dot{{\rho}}_{\rm sp}=-i[H_{\rm sp},{\rho}_{\rm sp}]+\kappa(2c{\rho}_{\rm sp}c^{\dag}-c^{\dag}c{\rho}_{\rm sp}-{\rho}_{\rm sp}c^{\dag}c)+\kappa(2d{\rho}_{\rm sp}d^{\dag}-d^{\dag}d{\rho}_{\rm sp}-{\rho}_{\rm sp}d^{\dag}d).\end{align} Here, ${\rho}_{\rm sp}=\langle \tilde{\downarrow}\lvert \tilde{\rho}\rvert \tilde{\downarrow}\rangle$ represents the reduced density matrix in the superradiant phase.
Using the equation $\partial \langle o\rangle/\partial t={\rm tr}\{o\partial \rho/\partial t\}$ for the expectation value  $\langle o\rangle={\rm tr}\{o\rho\}$, we can express the time derivative of the two-operator fluctuation correlators as follows:
\begin{small}
\begin{align}
\!\!\!\!\!\!\!\frac{d\langle c^{\dag}c\rangle}{dt}=&\,2i(G+\Lambda_4e^{2i\phi})\langle c^2\rangle-2i(G+\Lambda_4e^{-2i\phi})\langle c^{\dag 2}\rangle+i\Lambda_3(\langle cd^{\dag}\rangle-\langle c^{\dag}d\rangle)+i\Lambda_6(e^{2i\phi}\langle cb\rangle-e^{-2i\phi}\langle c^{\dag}d^{\dag}\rangle)-2\kappa\langle c^{\dag}c\rangle,\\
\!\!\!\!\!\!\!\frac{d \langle c^{\dag 2}\rangle}{dt}=&\,2i(\Lambda_1+i\kappa)\langle c^{\dag 2}\rangle+2i(G+\Lambda_4e^{2i\phi})(2\langle c^{\dag}c\rangle+1)+2i\Lambda_3\langle c^{\dag}d^{\dag}\rangle+2i\Lambda_6e^{2i\phi}\langle c^{\dag}d\rangle,\\
\!\!\!\!\!\!\!\frac{d\langle c^2\rangle}{dt}=&-2i(\Lambda_1-i\kappa)\langle c^2\rangle-2i(G+\Lambda_4e^{-2i\phi})(2\langle c^{\dag}c\rangle+1)-2i\Lambda_3\langle cd\rangle-2i\Lambda_6e^{-2i\phi}\langle cd^{\dag}\rangle,\\
\!\!\!\!\!\!\!\frac{d\langle cd\rangle}{dt}=&\!-i(\Lambda_1\!+\!\Lambda_2\!-\!2i\kappa)\langle cd\rangle\!-\!2i(G\!+\!\Lambda_4e^{-2i\phi})\langle c^{\dag}d\rangle\!-\!i\Lambda_3(\langle c^2\rangle\!+\!\langle d^2\rangle)\rangle \!-\!2i\Lambda_5e^{-2i\phi}\langle cd^{\dag}\!-\!i\Lambda_6e^{-2i\phi}\langle d^{\dag}d\rangle\!-\!i\Lambda_6e^{-2i\phi}(\langle c^{\dag}c\rangle\!+\!1),\\
\!\!\!\!\!\!\!\frac{d\langle c^{\dag}d^{\dag}\rangle}{dt}=&\,i(\Lambda_1\!+\!\Lambda_2\!+\!2i\kappa)\langle c^{\dag}d^{\dag}\rangle\!+\!2i(G\!+\!\Lambda_4e^{2i\phi})\langle cd^{\dag}\rangle\!+\!i\Lambda_3(\langle c^{\dag 2}\rangle\!+\!\langle d^{\dag 2}\rangle)\rangle\!+\!2i\Lambda_5e^{2i\phi}\langle c^{\dag}d\!+\!i\Lambda_6e^{2i\phi}(\langle d^{\dag}d\rangle\!+\!1)\!+\!i\Lambda_6e^{2i\phi}\langle c^{\dag}c\rangle,\\
\!\!\!\!\!\!\!\frac{d\langle cd^{\dag}\rangle}{dt}=&\!-\!i(\Lambda_1\!-\!\Lambda_2\!-\!2i\kappa)\langle cd^{\dag}\rangle\!-\!2i(G\!+\!\Lambda_4e^{-2i\phi})\langle c^{\dag}d^{\dag}\rangle\!+\!i\Lambda_3(\langle c^{\dag}c\rangle\!-\!\langle d^{\dag}d\rangle)\!+\!2i\Lambda_5e^{2i\phi}\langle cd\rangle\!-\!i\Lambda_6(e^{-2i\phi}\langle d^{\dag 2}\rangle
\!-\!e^{2i\phi}\langle c^2\rangle),\\
\!\!\!\!\!\!\!\frac{d\langle c^{\dag}d\rangle}{dt}=&\,i(\Lambda_1\!-\!\Lambda_2\!+\!2i\kappa)\langle c^{\dag}d\rangle\!+\!2i(G\!+\!\Lambda_4e^{2i\phi})\langle cd\rangle\!-\!i\Lambda_3(\langle c^{\dag}c\rangle\!-\!\langle d^{\dag}d\rangle)\!-2i\Lambda_5e^{-2i\phi}\langle c^{\dag}d^{\dag}\rangle+i\Lambda_6(e^{2i\phi}\langle d^2\rangle-e^{-2i\phi}\langle c^{\dag 2}\rangle),\\
\!\!\!\!\!\!\!\frac{d\langle d^{\dag}d\rangle}{dt}=&-i\Lambda_3(\langle cd^{\dag}\rangle-\langle c^{\dag}d\rangle)+i\Lambda_6(e^{2i\phi}\langle cd\rangle-e^{-2i\phi}\langle c^{\dag}d^{\dag}\rangle)+2i\Lambda_5(e^{2i\phi}\langle d^2\rangle-e^{-2i\phi}\langle d^{\dag 2}\rangle)-2\kappa\langle d^{\dag}d\rangle,\\
\!\!\!\!\!\!\!\frac{d\langle d^2\rangle}{dt}=&-2i(\Lambda_2+i\kappa)\langle d^2\rangle-2i\Lambda_3\langle cd\rangle-2i\Lambda_6e^{-2i\phi}\langle c^{\dag}d\rangle-2i\Lambda_5e^{-2i\phi}(2\langle d^{\dag}d+1),\\
\!\!\!\!\!\!\!\frac{d\langle d^{\dag 2}\rangle}{dt}=&\,2i(\Lambda_2+i\kappa)\langle d^{\dag 2}\rangle+2i\Lambda_3\langle c^{\dag}d^{\dag}\rangle+2i\Lambda_6e^{2i\phi}\langle cd^{\dag}\rangle+2i\Lambda_5e^{2i\phi}(2\langle d^{\dag}d\rangle+1).
\end{align}
\end{small}
We solve these ten coupled equations to obtain the cavity fluctuation $\langle c^{\dag}c\rangle$ or $\langle d^{\dag}d\rangle$ in the steady state, as shown in Figs.\,3(a)-3(c) in the main text.

\section{Stability analysis}\label{section4}

In this section, we discuss the stability of the system. Starting with the Heisenberg equations (\ref{Heisenberg equations-1}-\ref{Heisenberg equations-6}), we perform a semiclassical analysis. Specifically, we set $\langle\sigma_x\rangle=X$, $\langle\sigma_y\rangle=Y$, $\langle \sigma_z\rangle =Z$, $\langle a\rangle=\alpha\sqrt{\eta_+}$, $\langle b\rangle=\beta\sqrt{\eta_-}$. Furthermore, we expand the order parameters as $\alpha\rightarrow \alpha+\delta\alpha, \beta\rightarrow\beta+\delta\beta$ and $X\rightarrow X+\delta X, Y\rightarrow Y+\delta Y, Z\rightarrow Z+\delta Z$, where $\alpha,\beta,X,Y,Z$ are the mean-field steady-state solutions shown in Sec.\,\ref{section1}, and $\delta \alpha, \delta \beta, \delta X, \delta Y,\delta Z$ are quantum fluctuations. These expansions are valid in the limits of $\Delta_q/(\Delta\pm\Delta_F)$ and $\Delta_q/G\rightarrow \infty$. The equations of motion for the quantum fluctuations satisfy $$\dot{\textbf{u}} = {\rm{M}}{\textbf u},$$ with ${\textbf u}=[\delta \alpha_{\rm re}, \delta\alpha_{\rm im}, \delta\beta_{\rm re}, \delta\beta_{\rm im},\delta X,\delta Y]^{\rm T}$ and the stability matrix $\rm {M}$ is given by
 \begin{figure}[h]
\includegraphics[width=13.4cm]{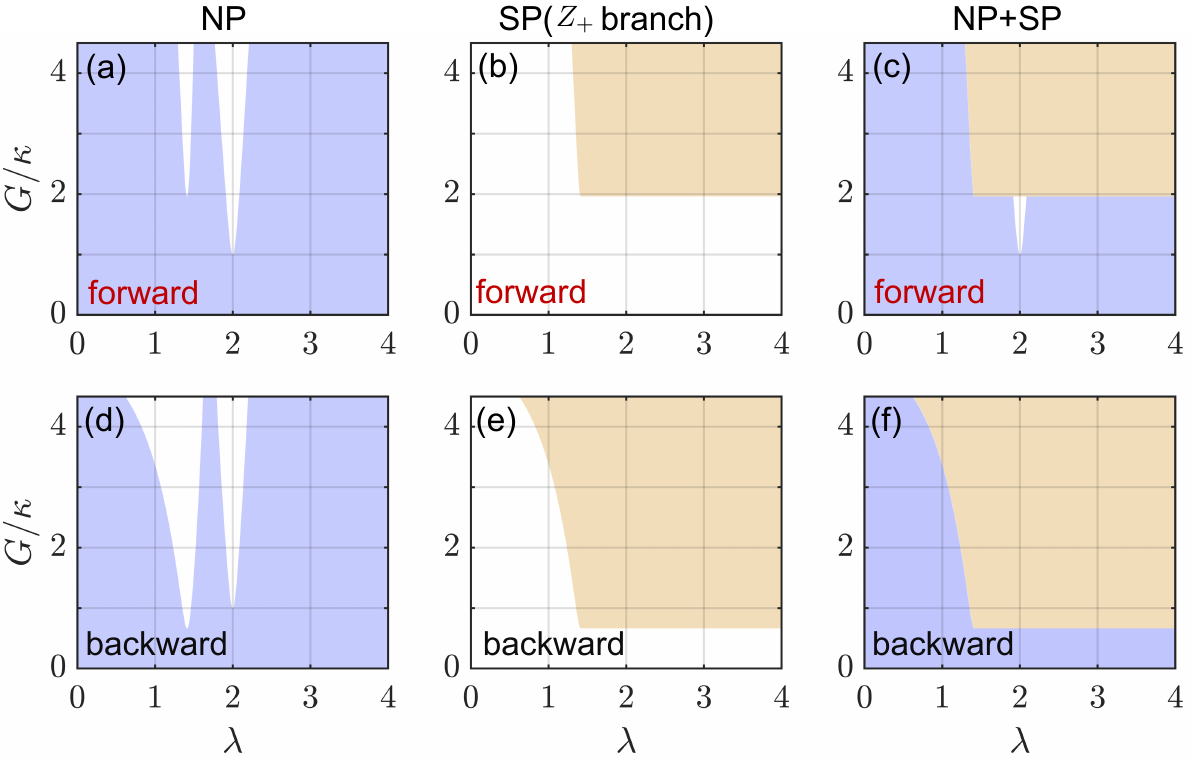}
\caption{Stability of the system for the forward and backward pumps in the normal phase (NP) and $Z_+$ branch of the superradiant phase (SP). The upper panels show the stability of the forward pump case, while the lower panels show the stability of the backward pump case. Panels (c,f) show the stability of the combination of the NP and SP. In the blank regions, the system is unstable. The parameter $\Delta_F/\Delta=0.5$ for the forward pump and $\Delta_F/\Delta=-0.5$ for the backward pump. The values of $\Delta_q$ is $10^4$, and the other parameters are the same as in Fig.\,3 of the main text.}
	\label{figS3}
\end{figure}
 \begin{eqnarray}
 \!\!\!{\rm M}=\left(\begin{array}{cccccc}\!-\kappa&\!\Delta_+\!-\!2G&\!0&\!J\sqrt{\frac{\Delta_+}{\Delta_-}}&\!0&\!-\frac{1}{4}\lambda_a\Delta_+\\
\!-(\Delta_+\!+\!2G)&\!-\kappa&\!-J\sqrt{\frac{\Delta_+}{\Delta_-}}&\!0&\!-\frac{1}{4}\lambda_a\Delta_+&\!0\\
\!0&\!J\sqrt{\frac{\Delta_-}{\Delta_+}}&\!-\kappa&\!\Delta_- &\!0&\!-\frac{1}{4}\lambda_b\Delta_-\\
\!-J\sqrt{\frac{\Delta_-}{\Delta_+}}&\!0&\!-\Delta_-&\!-\kappa&\!-\frac{1}{4}\lambda_b\Delta_- &\!0\\
\!0&\!-\lambda_a \Delta_q Z &\!0 &\!-\lambda_b \Delta_q Z&\!\frac{X}{Z}\Delta_q  (\lambda_a\alpha_{\rm im}\!+\!\lambda_b\beta_{\rm im})&\!\Delta_q[-1\!+\!\frac{Y}{Z}(\lambda_a\alpha_{\rm im}\!+\!\lambda_b\beta_{\rm im})]\\
\!-\lambda_a\Delta_q Z&\!0&\!-\lambda_b\Delta_q Z&\!0&\!\Delta_q[1\!+\!\frac{X}{Z}(\lambda_a\alpha_{\rm re}\!+\!\lambda_b\beta_{\rm re})]&\!\Delta_q\frac{Y}{Z}(\lambda_a\alpha_{\rm re}\!+\!\lambda_b\beta_{\rm re})
\end{array}\right),\nonumber\\
 \end{eqnarray}
where $\Delta_{\pm}=\Delta\pm\Delta_F$.
During the stability analysis, we separated the mean-field solutions and quantum fluctuations into real and imaginary parts, namely, $\alpha=\alpha_{\rm re}+i\alpha_{\rm im}$, $\beta=\beta_{\rm re}+i\beta_{\rm im}$ and $\delta\alpha=\delta \alpha_{\rm re}+i\delta\alpha_{\rm im}$, $\delta\beta=\delta \beta_{\rm re}+i\delta\beta_{\rm im}$. The stability of the system is determined by the eigenvalues of the matrix $\rm {M}$. If all eigenvalues have negative real parts, the system is stable and evolves into its steady state for $t\rightarrow \infty$. 

Figure\,\ref{figS3} shows the stable regions in the normal and superradiant phases for both forward and backward pump cases at $\Delta_q=10^4$. We find that for the forward pump, there is an unstable region located in the vicinity of $\lambda=2$ [see Fig.\,\ref{figS3}(c)]. However, for the backward pump case, this unstable region disappears, and both the normal and $Z_+$ branch of the superradiant phases are stable, as illustrated in Fig.\,\ref{figS3}(f). It is worth noting that the $Z_-$ branch solution in the superradiant phase is unstable (not shown here).

\section{The validity of the semiclassical approach}\label{section6}
\subsection{Overview}
{\color{black}In Sec. S1, we derive the mean-field solutions of cavity occupations by applying a semiclassical approach. Here, we offer a comprehensive review of the semiclassical approach, focusing on scenarios where the field interacts with a {\it single} atom. The semiclassical approach has proven to be effective in investigating various phase-transition-like behaviors within the Rabi model. These include the ground-state quantum phase transition (QPT)\,\cite{Ashhab2010,Ashhab2013,Hwang2015,Maoxin2017}, the excited-state QPT (ESQPT) indicated by the divergence of the semiclassical density of states\,\cite{Ricardo2016}, dissipative QPT\,\cite{Hwang2018}, and the breakdown of the photon blockade\,\cite{Carmichael2015}. It is worth noting that, although the semiclassical approach may not precisely capture the diverging quantum fluctuation of the cavity field in the limit of infinite detuning ($\Omega/\omega_0\rightarrow \infty$) in the Rabi model, it remains a reliable method for describing mean-field quantities such as the photon population and the atomic population of the ground state.

\subsection{Analytical derivations}

Here, we present an alternative yet intuitive derivation to illustrate that applying the semiclassical approximation solely to the cavity field leads to the semiclassical equations, aligning with the results presented in Sec.\,\ref{section1}. We initiate the derivation with the master equation (2) in the main text and utilize the equation $\partial \langle O\rangle/\partial t={\rm Tr}\{O\partial\rho/\partial t\}$, where the expectation value is denoted as $\langle O\rangle={\rm Tr}\{O\rho\}$. Subsequently, we obtain the equations of motion as follows:
\begin{align}\label{eq1}
\dot{\langle  a\rangle}&=-i(\Delta+\Delta_F)\langle a\rangle-\kappa\langle a\rangle-2iG\langle a^{\dagger}\rangle-iJ\langle b\rangle-\frac{i}{2}g_a\langle \sigma_x\rangle-\frac{1}{2}g_a\langle \sigma_y\rangle,\\
\dot{\langle b\rangle}&=-i(\Delta-\Delta_F)\langle b\rangle-\kappa\langle b\rangle-iJ\langle a\rangle-\frac{i}{2}g_b\langle \sigma_x\rangle-\frac{1}{2}g_b\langle \sigma_y\rangle, \label{eq3}\\
\dot{\langle \sigma_x\rangle}&=-\Delta_q\langle\sigma_y\rangle+ig_a(\langle a\sigma_z\rangle-\langle a^{\dagger}\sigma_z\rangle)+ig_b(\langle b\sigma_z\rangle-\langle b^{\dagger}\sigma_z\rangle),\label{eq3}\\
\dot{\langle \sigma_y \rangle}&=\Delta_q\langle \sigma_x\rangle-g_a(\langle a\sigma_z\rangle+\langle a^{\dagger}\sigma_z\rangle)-g_b(\langle b\sigma_z\rangle+\langle b^{\dagger}\sigma_z\rangle),\label{eq4}\\
\dot{\langle \sigma_z \rangle}&=ig_a(\langle a^{\dagger}\sigma_x\rangle-\langle a\sigma_x\rangle)+ig_b(\langle b^{\dagger}\sigma_x\rangle-\langle b\sigma_x\rangle)+g_a(\langle a\sigma_y\rangle+\langle a^{\dagger}\sigma_y\rangle)\nonumber\\&\,\,\,\,+g_b(\langle b\sigma_y\rangle+\langle b^{\dagger}\sigma_y\rangle).\label{eq5}
\end{align}
The semiclassical (or mean-field) equations can be derived from the Eqs. (\ref{eq1}-\ref{eq5}) by assuming the vanishing of second cumulants. For instance, this assumption implies $\langle a\sigma_z\rangle=\langle a\rangle \langle \sigma_z\rangle$. Specifically, we express the cavity field operators as $a=\langle a\rangle+\delta a$ and $b=\langle b\rangle+\delta b$, where the fluctuation $\delta a$ (or $\delta b$) is much smaller than the mean values $\langle a\rangle$ (or $\langle b \rangle$). When inserting the operator $a$ into terms involving the second cumulants, for example, $\langle a\sigma_z\rangle=\langle (\langle a\rangle+\delta a)\sigma_z \rangle=\langle a\rangle\langle\sigma_z\rangle+\langle\delta a\sigma_z\rangle$, neglecting the quantum fluctuation $\langle\delta a\sigma_z\rangle$, then we obtain $\langle a\sigma_z\rangle=\langle a\rangle \langle \sigma_z\rangle$. That is, regarding the atomic part, we retain the terms $\langle\sigma_{x,y,z}\rangle$ because they are finite constants and follow the spin-conservation law $\langle \sigma_x\rangle^2+\langle \sigma_y\rangle^2+\langle \sigma_z\rangle^2=1$. This approach allows us to derive the same semiclassical equations as discussed in Sec.\,\ref{section1}.

This approach can be validated through the low-energy Hamiltonian. As discussed in Sec.\,\ref{section2}.B, in the large detuning limit where the atomic energy scale significantly exceeds the field part, we can decouple the atomic subspaces and concentrate on the lowest atomic subspace. Based on Eq.\,(\ref{lowest-energy state}), we can calculate the mean value of the operator $\sigma_z$ in the lowest-energy state (or ground state) as follows:
\begin{align}
\langle \tilde{\downarrow}|\sigma_z|\tilde{\downarrow}\rangle=-\frac{1}{\sqrt{1+(\alpha_{\rm im}^2+\alpha_{\rm re}^2)\lambda_a ^2+2(\alpha_{\rm im}\beta_{\rm im}+\alpha_{\rm re}\beta_{\rm re})\lambda_a\lambda_b+(\beta_{\rm im}^2+\beta_{\rm re}^2)\lambda_b ^2}},
\end{align}
where $\alpha$ and $\beta$ represent the mean-field solutions of the cavity occupations, and $\lambda_{a,b}=2g_{a,b}/\sqrt{\Delta_q(\Delta\pm\Delta_F)}$ are dimensionless atom-field couplings of the Hamiltonian. When calculating $\langle \tilde{\downarrow}|\sigma_z|\tilde{\downarrow}\rangle$, the atomic part does not involve semiclassical approximation; only the cavity field is subjected to such an approximation.

In Fig. \,\ref{figR3}, we present the mean value of $\langle \sigma_z\rangle$ obtained from the mean-field approach (depicted by red circles) and $\langle \tilde{\downarrow}|\sigma_z|\tilde{\downarrow}\rangle$ derived from the full quantum description (illustrated by the black curve). Clearly, the mean values obtained from these two different methods are consistent, confirming the validity of applying semiclassical approximation solely to the cavity fields.

\begin{figure*}[h]
	\includegraphics[width=12cm]{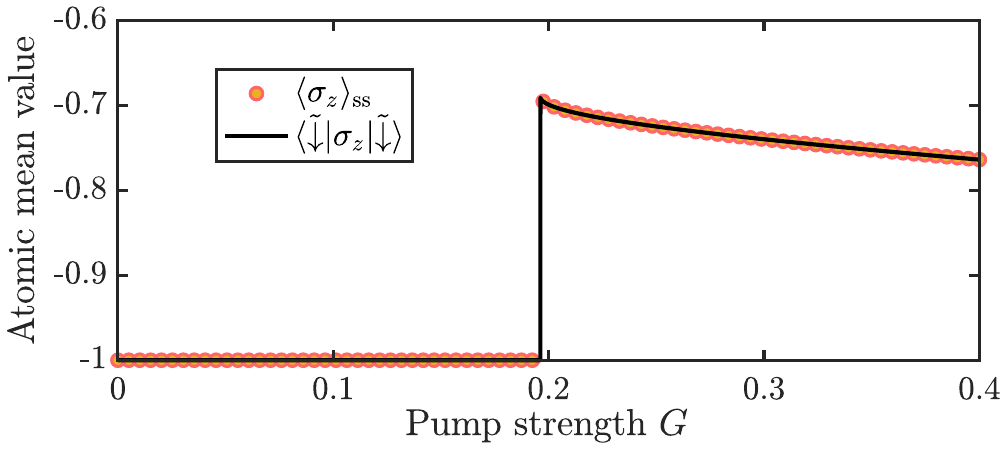}
	\caption{{\color{black}The atomic expectation values as a function of the pump strength $G$. The red circles represent the atomic expectation value in the steady state $\langle \sigma_z\rangle$, and the black curve illustrates the atomic expectation value in the lowest-energy state $\langle \tilde{\downarrow}|\sigma_z|\tilde{\downarrow}\rangle$. The parameters considered here are $\Delta=2, \Delta_F/\Delta=0.5, \lambda_a=\lambda_b=1.7, \kappa/\Delta =0.05$.}}	\label{figR3}
\end{figure*}
Based on the semiclassical equations (\ref{eq1}-\ref{eq5}), we can derive the steady-state mean amplitudes of the cavity fields $\langle a\rangle, \langle b\rangle$ and the atomic expectation values $\langle \sigma_x\rangle, \langle \sigma_y\rangle$ and $\langle \sigma_z\rangle$, see Sec.\,\ref{section1}. Interestingly, we find that all of these steady-state quantities exhibit a bifurcation when the system parameter, specifically the pump strength (or atom-field coupling strength), is increased. We refer to the point where bifurcation occurs as the critical point for the second-order superradiant phase transition, denoted as $G_c^{\rm 2nd}$. 
It is worth noting that the critical point obtained through the semiclassical calculation aligns well with those derived from the full quantum description of the low-energy Hamiltonian in the infinite-detuning limit $\Delta_q/(\Delta\pm\Delta_F)\rightarrow\infty$, as discussed in Sections \ref{section2} and \ref{section3}.B. 
\begin{figure*}[h]
	\includegraphics[width=16cm]{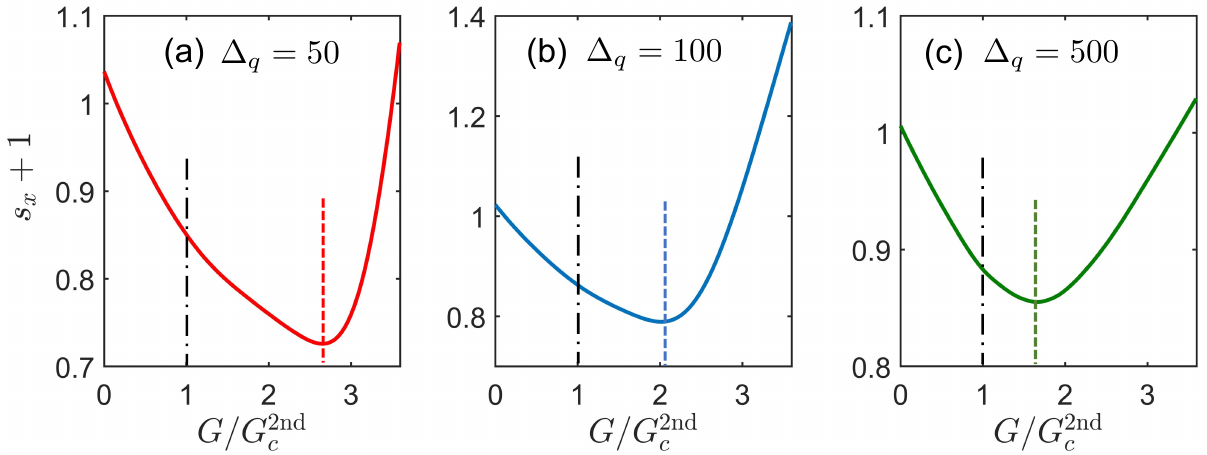}
	\caption{{\color{black}Numerical calculation of the position-squeezing quantifier $s_x+1$ as a function of $G/G_c^{\rm 2nd}$ for $\Delta_q=50$ (a), 100 (b), 500 (c). Here, the analytical value of critical pump strength provided by the semiclassical approach, i.e., $G_{c}^{\rm 2nd}$, is denoted by the black dash--dotted line. Other parameters are same as Fig. 2(a,b) in the main text.}}

	\label{figR1}
\end{figure*}
\subsection{Squeezing parameter}
To further illustrate the alignment of critical points obtained from the semiclassical approach and the full quantum approach, we begin with the master equation (2) in the main text and proceed to numerically calculate the position-squeezing parameter\,\cite{Ashhab2010,Ashhab2013},
 $$s_x =2\langle (X-\langle X\rangle)^2\rangle-1,$$
where the quadrature variable is defined as $X=(a+a^{\dagger})/\sqrt{2}$. In Figure \ref{figR1}, we plot the squeezing quantifier $s_x+1$ as a function of the pump strength $G$ (measured in comparison to the second-order critical pump strength $G_c^{\rm 2nd}$, which is the theoretical value obtained from the mean-field approach). As we increase the atomic detuning $\Delta_q$ (i.e., getting closer to the infinite detuning limit $\eta\rightarrow \infty$), the point where the squeezing quantifier reaches its minimum gradually approaches the theoretical critical pump $G_c^{\rm 2nd}$. These numerical results indicate that by increasing the atomic detuning $\Delta_q$, the critical pump converges towards the theoretical value provided by the mean-field approach\,\cite{Ashhab2010,Ashhab2013}.}

\section{Discussions on the dual-coupling JC model }
{\color{black}\subsection{Comparison between the dual-coupling JC model and the Rabi model}\label{section7}
Generally, the occurrence of the superradiant phase transition in the standard Rabi model, typically hinges on two critical conditions: (i) attaining an exceedingly large atomic frequency, such that $\omega_q/\omega\rightarrow \infty$, and (ii) achieving ultra-strong coupling between the atom and the field. Our model successfully overcomes the two critical challenging conditions present in the standard Rabi model. Specifically, for condition (i), in our model, the large atomic frequency was transformed into a large atomic detuning (i.e., $\Delta_q/\Delta\rightarrow \infty$, where \(\Delta_q=\omega_q-\omega_p/2\) and \(\Delta=\omega_0-\omega_p/2\)). By tuning the frequency of the pump field, one can readily achieve this large atomic detuning, which serves the same purpose as \(\omega_q/\omega\rightarrow \infty\) in the standard Rabi model. Thus,  condition (i) can be easily met in our model.
Simultaneously, the critical atom-field coupling strength required for the onset of superradiant phase transitions is determined by the detunings \(\Delta_q\) and \(\Delta\), rather than the atomic frequency \(\omega_q\) and resonator frequency \(\omega_0\). Consequently, our proposed scheme can alleviate constraint (ii) from ultra-strong coupling to strong coupling.  


\subsection{Comparison between all-optical controls and its magnetic or electronic counterpart}
In our model, the control over superradiant phase transitions and multicriticality is achieved through all-optical manipulation using external fields. This all-optical control presents several advantages in comparison to its magnetic and electronic counterparts. Firstly, in terms of integration, traditional methods for achieving optical nonreciprocity often rely on the magneto-optical Faraday effect\,\cite{wang2009observation,Khanikaev2010}. These devices tend to be bulky and necessitate large magnetic fields, making them inconvenient for integration. Issues such as crosstalk induced by the magnetic field and lattice mismatches between magneto-optic materials and silicon further complicate integration\,\cite{Dai2012}. In contrast, our all-optical method for breaking system reciprocity demonstrates {\it high compatibility} and {\it ease of integration} into photonic systems. This characteristic not only enhances the potential for diverse applications in quantum communication but also provides essential building blocks for a quantum network\,\cite{Kurizki2015}.
Secondly, all-optical systems are {\it more compact} than their magnetic and electronic counterparts. In our model, the microcavity is at the millimeter scale, and there have been reports of micrometer-scale optical cavities\,\cite{Ward2011}. This characteristic facilitates the development of on-chip nonreciprocal devices. 
Thirdly, all-optical system can be {\it easily reconfigured without need of the complex electronic components}. This characteristic makes opportunity to easily adjust the external pump strength for manipulating superradiant phase transitions and multicriticality. 
Fourthly, compared to the electronic counterparts, all-optical system, in general, can operate with {\it lower power consumption}\,\cite{jinno1990ultrafast}. 

On the other hand, these advantages may stimulate further theoretical and experimental explorations, potentially advancing on-chip nonreciprocal device development. Our research reveals nonreciprocal superradiant phase transitions and a rich phase diagram featured by controllable multicritical points. These findings not only hold fundamental research significance but also provide quantum resources for quantum metrology\,\cite{Hotter2024}.
Therefore, our work may inspire the development of integrated high-precision quantum sensing.
Moreover, extending our model to $N$- particle case may inspire many applications in optical field. Recently, steady-state superradiance has been demonstrated in a bad cavity regime, yielding lasing with a linewidth in the millihertz range\,\cite{bohnet2012steady}. From this perspective, the combination of superradiance with our all-optical system may open up  new avenues for designing on-chip unidirectional laser.
Additionally, our system presents opportunities to explore and manipulate a broader range of physical phenomena, including superradiant cooling\,\cite{Minghui2016} and atomic synchronization\,\cite{Minghui2014}. This broadened scope enhances the versatility of our platform, offering unique opportunities for advancing our understanding and practical utilization of these intriguing phenomena.}

\section{Discussion on the assumption of system parameters}\label{section5}
\subsection{The case of nonzero cavity hopping rate $J\ne 0 $}
\begin{figure}
\includegraphics[width=15.4cm]{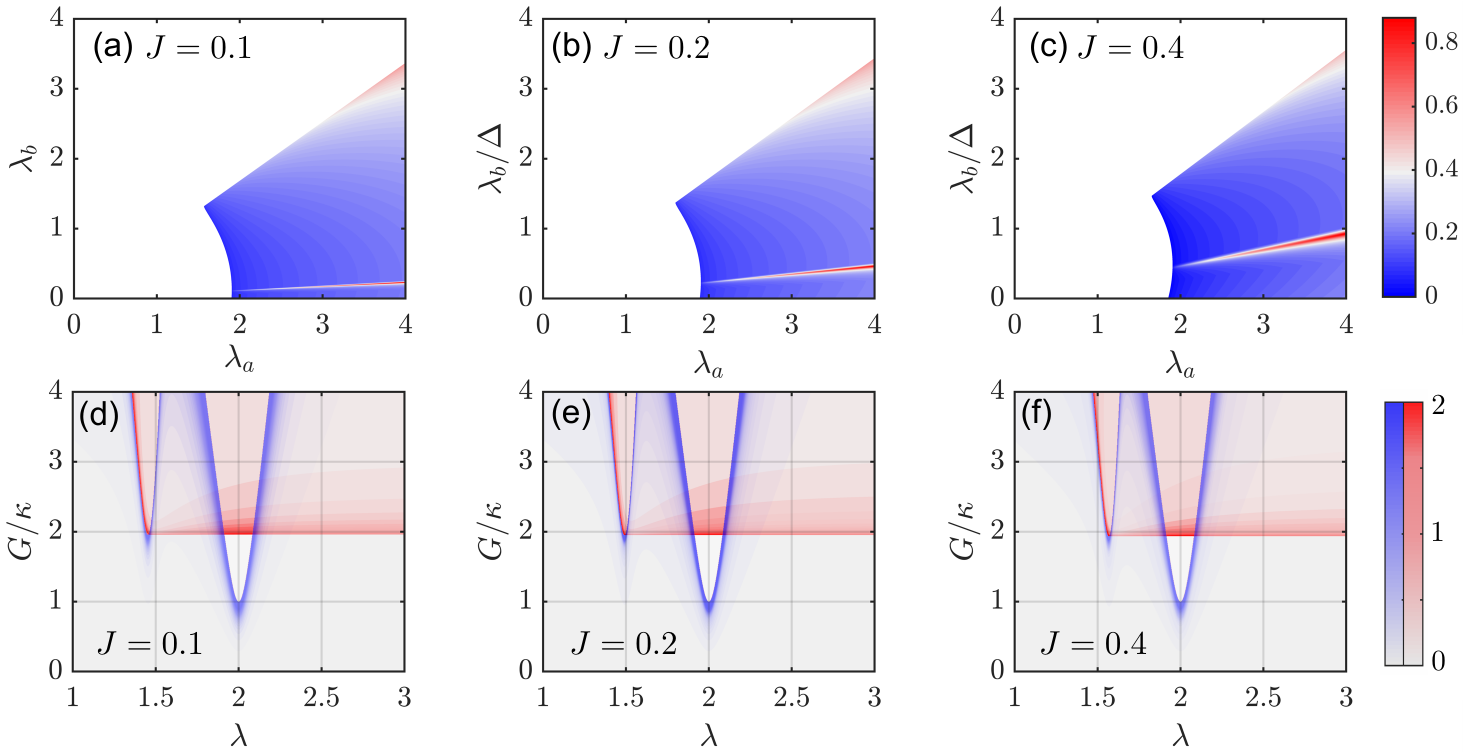}
\caption{(a-c) Phase diagram of the order parameter $\alpha_{\rm re}$ for different cavity hopping rate $J$. (d-e) Cavity fluctuation $\langle c^{\dag}c\rangle$ as functions of $\lambda$ and $G$. In all plots, we consider $\Delta_F/\Delta=0.5$.}
	\label{figS4}
\end{figure}

In this section, we investigate how the cavity hopping rate $J$ affects the superradiant transitions in the dual-coupling JC model. While the solutions for $J\ne0$ have complicated mathematical expressions, we present here only the numerical results. In Figs.\,\ref{figS4}(a)-\,\ref{figS4}(c) we plot the phase diagram of $\alpha_{\rm re}$ for different $J$. We observe that the non-zero value of $J$ induces a continuous increase of $\alpha_{\rm re}$ in the bulk of the superradiant phase. Comparing with Fig.\,\ref{figS1}, we find that $J$ causes only a small correction to the boundaries of the first- and second-order phase transitions. Moreover, from Figs.\,\ref{figS4}(d)-\,\ref{figS4}(f) we observe that the cone-shaped critical curve on the left shifts towards larger values of $\lambda$, which causes the critical point of the second-order transition to shift toward a larger pump strength $G$. Overall, we conclude that $J\ne 0$ does not alter the main physics discussed in our model.
{\color{black}
\subsection{Numerical simulations on steady-state mean photon number}

In Fig.\,\ref{figS5}, we conducted numerical simulations to analyze the steady-state mean photon number $\langle a^{\dagger}a\rangle$ as a function of the pump strength $G$. These simulations are based on the full Hamiltonian (1) and the master equation (2), carried through the Qutip. 
 The chosen system parameters for the simulations are as follows: cavity detuning $\Delta=2$, cavity decay rate $\kappa/\Delta=0.05$, atomic decay rate $\gamma=\kappa$, cavity hopping rate $J=\kappa$, and thermal photon number $\bar{n}=7.4\times 10^{-3}$.
It is observed in Fig.\,\ref{figS5} that for larger values of $\Delta_q$, the growth in $\langle a^{\dagger}a\rangle$ occurs at a faster rate. As $\Delta_q/\Delta$ increases, the tendency of $\langle a^{\dagger}a\rangle$ gradually approaches the case of infinite detuning limit, i.e., $\Delta_q/\Delta\rightarrow \infty$. In the infinite detuning limit, $\langle a^{\dagger}a\rangle$ diverges at the critical point. Note that due to the limited capabilities of computer, here, we have simulated only the case where $\Delta_q/\Delta$ is maximally set to 50, with the cavity modes $a$ and $b$ truncated to 18.
Even though, within reasonable experimental parameters, as we increase the atomic detuning value $\Delta_q/\Delta$, our numerical simulations, conducted without any assumptions and approximations, consistently converge towards the analytical results obtained for $\Delta_q/\Delta\rightarrow \infty$. This observation implies that our assumptions for the sake of analytical simplification are justified.

}

\begin{figure}
\includegraphics[width=14.4cm]{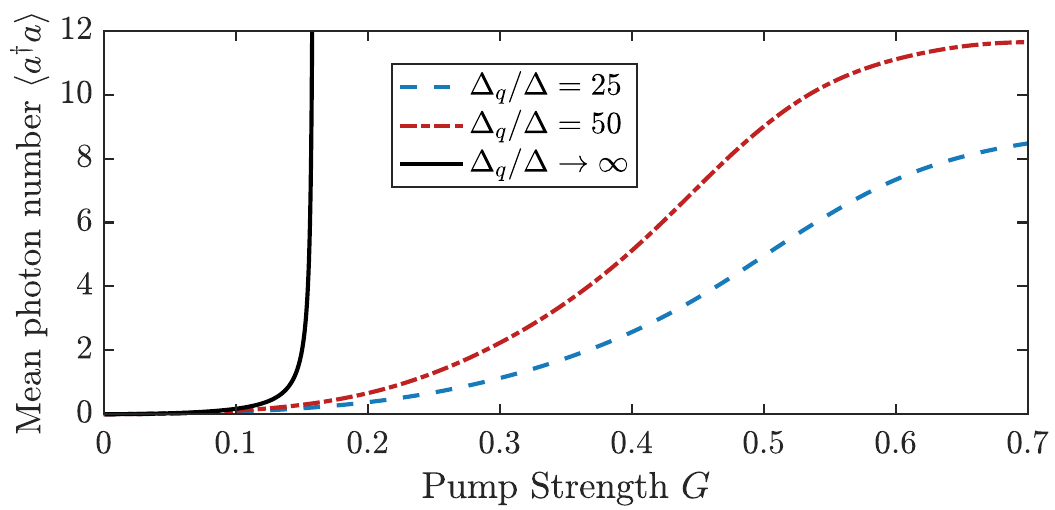}
\caption{{\color{black}Numerical simulation of the mean photon number $\langle a^{\dag}a\rangle$ as a function of the pump strength $G$ for finite detuning cases: $\Delta_q/\Delta=25$ (depicted by the blue dotted curve) and $\Delta_q/\Delta=50$ (illustrated by the red dash-dotted curve). The analytical result obtained for the infinite detuning case, $\Delta_q/\Delta\rightarrow\infty$, is shown by the black-solid curve. Here, we considered cavity decay $\kappa=0.1$, atomic decay $\gamma=0.1$, cavity hopping rate $J=0.1$, and thermal photon number $\bar{n}=7.4\times 10^{-3}$. Due to computational limitations, simulations were conducted only for the case where $\Delta_q/\Delta$ is maximally set to 50, with both cavity modes $a$ and $b$ truncated to 18. Additional parameters include $\Delta=2$, $\Delta_F=0$, $\lambda_a=\lambda_b=1.4$.}}
	\label{figS5}
\end{figure}
\section{Possible experimental implementations}\label{section9}
{\color{black}
To achieve nonreciprocal superradiant phase transitions and multicriticality, the possible experimental implementation relies on {\color{black}three} techniques: first, reaching the strong-coupling regime in the interaction between the atom and cavity fields; second, unidirectionally squeezing one of the cavity modes; third, the capability to rotate the resonator. Recently, strong-coupling cavity QED has been realized using WGM microcavities, involving trapped atoms\,\cite{aoki2006observation,dayan2008photon,Aoki2009PRL,alton2011strong,Junge2013,scheucher2016quantum,Will2021}, quantum dots\,\cite{kiraz2001cavity,Peter2005PRL,srinivasan2007linear,SrinivasanPhysRevA} and nitrogen-vacancy (NV) centers\,\cite{park2006cavity,barbour2010composite}. Combining these achievements with recent experiments on rotating resonator\,\cite{maayani2018flying} and optical squeezing\,\cite{Strekalov2011,lu2019periodically,lu2020toward}, we discuss the feasibility of implementing our model with two potential setups: (i) cold cesium atoms falling onto the surface of a WGM microdisk cavity\,\cite{aoki2006observation,dayan2008photon,Aoki2009PRL,alton2011strong} and (ii) a single trapped $^{85}$Rb atom interacting with a WGM microresonator\,\cite{Junge2013,scheucher2016quantum,Will2021}. Next, we provide detailed analysis of these experimental realization. 
\subsection{Strong coupling between an atom and microresonator}
\subsubsection*{(i) cold cesium atoms falling onto the surface of a WGM microdisk}
A possible experimental implementation platform could involve single cesium atoms radiatively coupled to a high-quality toroidal microresonator and in close proximity to the resonator's dielectric surface\,\cite{aoki2006observation,dayan2008photon,Aoki2009PRL,alton2011strong}. Specifically, a cloud of cold caesium atoms is located $\sim 800 \,\mu$m above the surface of the resonator. Several caesium atoms are released from an optical dipole-force trap and fall within the evanescent field of cavity mode, establishing strong coupling to the resonator's field. 
Due to the aerodynamic process caused by the cavity rotation, the falling atom hovers a short distance above the rapidly rotating resonator instead of crashing into its surface [see Sec. S7.\,B]. This is distinct from the scenario involving a static resonator, as described by Refs. \,\cite{aoki2006observation,alton2011strong}. 
A real-time detection scheme, relying on strong radiative interactions between individual atoms and the evanescent cavity field, can be employed to select atomic trajectories passing within 300 nm from the resonator’s surface. At this scale, the radiative atom-field interactions are characterized by strong coupling, evident in the observed vacuum-Rabi splitting. This experimental configuration can be theoretically modeled using the dual-coupling Jaynes-Cummings (JC) Hamiltonian, as demonstrated in Refs.\,\cite{aoki2006observation,dayan2008photon,alton2011strong}. In these schemes, the average atom-photon coupling has been achieved at approximately 40 MHz, surpassing the dissipative rates of both the atom and the cavity, thereby reaching the strong-coupling regime.

\subsubsection*{(ii) a single trapped atom interacting with a WGM microresonator}

Another possible experimental implementation platform could be trapping a single $^{85}$Rb atom at a small distance from the surface of the WGM bottle microresonator\,\cite{Junge2013,scheucher2016quantum,Will2021}. The stable and controlled interaction between a single atom and the resonator in the strong-coupling regime ($\sim 10$ MHz) is demonstrated in\,\cite{Will2021}. Specifically, a single $^{85}$Rb atom is trapped at a distance of about 200 nm from the resonator surface. This is achieved through a deep standing-wave optical dipole trap created by retroreflecting a focused trapping light field from the resonator surface. Additionally, a second, detuned compensation light field is employed to counteract the position-dependent detuning of the atomic resonance from the resonator mode. 
Resonant detection light is sent through the coupling fiber to detect the presence of a single atom in the resonator mode in real-time, and the transmitted power is monitored using a single-photon counting module. The transmission signal is detected using a field-programmable gate array-based system. Upon detecting an atom, the detection light is switched off, and the dipole trap is switched on, with the overall delay between detection and trapping being approximately 250 ns. This duration is significantly shorter than the average transit time of an atom through the evanescent field of the resonator mode, enabling us to capture a detected atom if it is located inside the trapping volume. 

\subsection{Directionally squeezing of the cavity mode}

With the advancement of nanofabrication techniques, a range of crystalline materials has been employed in the fabrication of WGM microresonators\,\cite{Ilchenko2004,Strekalov2011,Beckmann2011,fortsch2013versatile,Guo2016PRL,lu2019periodically,lu2020toward}. A notable advantage of crystalline cavities is their ability to support nonlinear optical processes, including optical parametric amplification (OPA). In a parametric amplifier, a pump beam interacting with a $\chi^{(2)}$ nonlinearity generates signal and idler beams. This OPA process is considered a crucial source of squeezed states of the radiative field. 

Inspired by recent experiments, here we consider a crystalline WGM microresonator with high second-order nonlinearity. To ensure phase matching for parametric down-conversion along the entire circumference of the resonator, an uniaxial crystal with the optical axis along the symmetry axis of the cavity should be used. The microdisk can be fabricated from a 5\% MgO-doped z-cut lithium niobate wafer, as demonstrated in Ref.\,\cite{Strekalov2011}, or periodically poled lithium niobate, as demonstrated in Refs.\,\cite{lu2019periodically,lu2020toward}. 
The pump field interacts with the $\chi^{(2)}$ nonlinearity through a tapered fiber, generating a squeezed cavity mode. This is an optical parametric amplification process.
The forward external light unidirectionally interacts with the clockwise $\chi^{(2)}$-nonlinearity, thereby directionally squeezing the clockwise cavity mode\,\cite{Tang2022}. In other words, the forward propagating mode through the waveguide selectively squeezes the copropagating mode in the resonator. Due to the rotation of the system, the light circulating in the resonator experiences a Sagnac-Fizeau shift, denoted by $\Delta_F$.


\subsection{Rotating resonator}
The rotating scheme can be implemented by mounting the WGM resonator on a turbine, causing the resonator to rotate with a certain angular velocity. This setup has been demonstrated in a recent experiment by Maayani {\it et al.}\,\cite{maayani2018flying}. 
Positioning the rotating microresonator near a single-mode telecommunications fiber allows light to be evanescently coupled into or out of the resonator through the tapered region. 
It is worth noting that in this rotating scheme, the aerodynamic process plays a crucial role: a rapidly rotating resonator can drag air into the region between the taper and the microdisk, forming a boundary layer of air. Due to the air pressure on the surface of the taper facing the resonator, the taper hovers a short distance above the rapidly spinning resonator, avoiding direct contact or adherence to the resonator. This distinguishes it from the situation with a stationary resonator. If any disturbance causes the taper to ascend beyond its stable equilibrium height, it naturally returns to its original position, a phenomenon referred to as `self-adjustment'.

\subsection{Discussion on the effect of unexpected noises and perturbations}

Based on the potential experimental implementations discussed above, we discuss how possible unexpected noises or perturbations might affect the performance of our system.
The unexpected noises and perturbations may be attributed to the thermal noise of the cavity and the perturbations caused by the rotation scheme. 

According to the experimentally feasible parameters, the cavity frequency is $\omega = 1.93 \times 10^{14}$ Hz. At room temperature $T = 300$ K, the thermal cavity number is $\bar{n} = 1/[\exp(\hbar\omega/k_BT)-1]\approx7.4\times 10^{-3}$, which is significantly less than 1 and has been omitted in our analytical considerations. However, in plotting Fig. \ref{figS5}, we have considered non-zero thermal cavity number  ($\bar{n}=7.4\times 10^{-3}$).

On the other hand, the rapidly rotating cavity drags a boundary layer of air around the resonator, which may induce the vibrations of the atom. This could alter the overlap between the atom and the mode volume of the resonator, modifying the atom-resonator coupling strength. However, as discussed in Sec. \ref{section9}.A, the coupling strength between the WGM resonator and an atom in current experiments is sufficient to achieve nonreciprocal phase transitions in our model. Moreover, in our model, we achieve phase transitions by effectively enhancing the pump strength of the external field, and the Sagnac shift remains robust against modifications in coupling. Therefore, it is safe to say that this perturbation would not change the nonreciprocal behaviors of our model.}


 %

\section{Extending the model to the case of $N$ particles ($N\gg 1$)}
{\color{black}
Our approach can be extended to scenarios involving an infinite number of atoms (\(N\rightarrow \infty\)), particularly within the context of the dual-coupling Tavis-Cummings (TC) model. Here, we give a brief discussion in this extension.

We consider $N$ two-level atoms interacting with two counter-propagating modes of WGM resonator. The resonator is made of materials with second-order nonlinearity. A classical field with frequency $\omega_p$ input from the forward (or backward) can directionally generate squeezing cavity modes $a$ (or $b$) through an optical parametric amplification process. The system Hamiltonian can be described by the dual-coupling TC model (in the forward pump):
 
\begin{align}\label{TChamiltonian}
H=\omega_0a^{\dag}a+\omega_0 b^{\dag}b+{\omega_q}J_z+\frac{g_a}{\sqrt{N}}(aJ_++a^{\dag}J_-)+\frac{g_b}{\sqrt{N}}(bJ_++b^{\dag}J_-)+G(a^{\dagger 2}e^{-i\omega_pt}+a^2e^{i\omega_pt}).
\end{align}
Here $J_z=(1/2)\sum_{i=1}^N\sigma_z$ and $J_{\pm}=\sum_{i=1}^N\sigma_{\pm}$ represent collective angular-momentum operators. There are $N$ atoms simultaneously coupled to cavity modes $a$ and $b$ with collective coupling strengths $g_a$ and $g_b$, respectively. The cavity field is unidirectionally pumped by an external field with the pump strength of $G$. Additionally, we consider the microresonator rotates counterclockwise with an angular velocity $\Omega$, causing the two cavity modes to experience Sagnac-Fizeau shifts with respect to their static resonance frequency $\omega_0$, i.e., $\omega_0\rightarrow \omega_0\pm \Delta_F$. In the frame rotating at $\omega_p/2$, Hamiltonian (\ref{TChamiltonian}) is transformed into
\begin{align}\label{rotatingTC}
H&=(\Delta+\Delta_F) a^{\dag}a+(\Delta-\Delta_F) b^{\dag}b+\Delta_qJ_z+\frac{g_a}{\sqrt{N}}(aJ_++a^{\dag}J_-)\nonumber\\
&+\frac{g_b}{\sqrt{N}}(bJ_++b^{\dag}J_-)+G(a^{\dagger 2}+a^2),
\end{align}
where the detunings are defined as $\Delta=\omega_0-\omega_p/2$ and $\Delta_q=\omega_q-\omega_p/2$. The form of Hamiltonian (\ref{rotatingTC}) closely resembles Hamiltonian (1) presented in the main text. Utilizing Hamiltonian (\ref{rotatingTC}), we can derive mean field equations for $\langle a\rangle, \langle b\rangle, \langle J_{x,y,z}\rangle$ and subsequently calculate the critical points for both first-order and second-order superradiant phase transitions. This process aligns with the method employed in the manuscript. 

The dual-coupling TC model maintains the advantages of the dual-coupling JC model. Firstly, from Hamiltonian (\ref{rotatingTC}), we infer that the critical atom-field coupling strength necessary for the occurrence of superradiant phase transitions corresponds to the detunings $\Delta_q$ and $\Delta$ rather than the atomic frequency $\omega_q$ and resonator frequency $\omega_0$. Thus, superradiant phase transitions in dual-coupling TC does not require ultra-strong atom-field coupling. Secondly, the dual-TC model, the condition of an extremely large detuning is transformed into the limit of an extremely large number of atoms, i.e., in the thermodynamic limit as $N\rightarrow\infty$. Thirdly, the control of superradiant phase transitions in the dual-TC model can also be achieved by adjusting the pump strength. }


\begin{thebibliography}{127}%
\makeatletter
\providecommand \@ifxundefined [1]{%
 \@ifx{#1\undefined}
}%
\providecommand \@ifnum [1]{%
 \ifnum #1\expandafter \@firstoftwo
 \else \expandafter \@secondoftwo
 \fi
}%
\providecommand \@ifx [1]{%
 \ifx #1\expandafter \@firstoftwo
 \else \expandafter \@secondoftwo
 \fi
}%
\providecommand \natexlab [1]{#1}%
\providecommand \enquote  [1]{``#1''}%
\providecommand \bibnamefont  [1]{#1}%
\providecommand \bibfnamefont [1]{#1}%
\providecommand \citenamefont [1]{#1}%
\providecommand \href@noop [0]{\@secondoftwo}%
\providecommand \href [0]{\begingroup \@sanitize@url \@href}%
\providecommand \@href[1]{\@@startlink{#1}\@@href}%
\providecommand \@@href[1]{\endgroup#1\@@endlink}%
\providecommand \@sanitize@url [0]{\catcode `\\12\catcode `\$12\catcode
  `\&12\catcode `\#12\catcode `\^12\catcode `\_12\catcode `\%12\relax}%
\providecommand \@@startlink[1]{}%
\providecommand \@@endlink[0]{}%
\providecommand \url  [0]{\begingroup\@sanitize@url \@url }%
\providecommand \@url [1]{\endgroup\@href {#1}{\urlprefix }}%
\providecommand \urlprefix  [0]{URL }%
\providecommand \Eprint [0]{\href }%
\providecommand \doibase [0]{https://doi.org/}%
\providecommand \selectlanguage [0]{\@gobble}%
\providecommand \bibinfo  [0]{\@secondoftwo}%
\providecommand \bibfield  [0]{\@secondoftwo}%
\providecommand \translation [1]{[#1]}%
\providecommand \BibitemOpen [0]{}%
\providecommand \bibitemStop [0]{}%
\providecommand \bibitemNoStop [0]{.\EOS\space}%
\providecommand \EOS [0]{\spacefactor3000\relax}%
\providecommand \BibitemShut  [1]{\csname bibitem#1\endcsname}%
\let\auto@bib@innerbib\@empty
\bibitem [{\citenamefont {Domb}(2000)}]{domb2000phase}%
  \BibitemOpen
  \bibfield  {author} {\bibinfo {author} {\bibfnamefont {C.}~\bibnamefont
  {Domb}},\ }\href@noop {} {\emph {\bibinfo {title} {Phase transitions and
  critical phenomena}}}\ (\bibinfo  {publisher} {Elsevier},\ \bibinfo {year}
  {2000})\BibitemShut {NoStop}%
\bibitem [{\citenamefont {Sachdev}()}]{sachdev1999quantum}%
  \BibitemOpen
  \bibfield  {author} {\bibinfo {author} {\bibfnamefont {S.}~\bibnamefont
  {Sachdev}},\ }\href@noop {} {\emph {\bibinfo {title} {Quantum phase
  transitions}}}\ (\bibinfo  {publisher} {Cambridge University Press,
  Cambridge, England, 2011})\BibitemShut {NoStop}%
\bibitem [{\citenamefont {Dicke}(1954)}]{Dicke1954}%
  \BibitemOpen
  \bibfield  {author} {\bibinfo {author} {\bibfnamefont {R.~H.}\ \bibnamefont
  {Dicke}},\ }\bibfield  {title} {\bibinfo {title} {Coherence in spontaneous
  radiation processes},\ }\href {https://doi.org/10.1103/PhysRev.93.99}
  {\bibfield  {journal} {\bibinfo  {journal} {Phys. Rev.}\ }\textbf {\bibinfo
  {volume} {93}},\ \bibinfo {pages} {99} (\bibinfo {year} {1954})}\BibitemShut
  {NoStop}%
\bibitem [{\citenamefont {Hepp}\ and\ \citenamefont
  {Lieb}(1973)}]{HEPP1973360}%
  \BibitemOpen
  \bibfield  {author} {\bibinfo {author} {\bibfnamefont {K.}~\bibnamefont
  {Hepp}}\ and\ \bibinfo {author} {\bibfnamefont {E.~H.}\ \bibnamefont
  {Lieb}},\ }\bibfield  {title} {\bibinfo {title} {On the superradiant phase
  transition for molecules in a quantized radiation field: the
  \uppercase{D}icke maser model},\ }\href
  {https://doi.org/https://doi.org/10.1016/0003-4916(73)90039-0} {\bibfield
  {journal} {\bibinfo  {journal} {Ann. Phys.}\ }\textbf {\bibinfo {volume}
  {76}},\ \bibinfo {pages} {360} (\bibinfo {year} {1973})}\BibitemShut
  {NoStop}%
\bibitem [{\citenamefont {Wang}\ and\ \citenamefont {Hioe}(1973)}]{Wang1973}%
  \BibitemOpen
  \bibfield  {author} {\bibinfo {author} {\bibfnamefont {Y.~K.}\ \bibnamefont
  {Wang}}\ and\ \bibinfo {author} {\bibfnamefont {F.~T.}\ \bibnamefont
  {Hioe}},\ }\bibfield  {title} {\bibinfo {title} {Phase transition in the
  \uppercase{D}icke model of superradiance},\ }\href
  {https://doi.org/10.1103/PhysRevA.7.831} {\bibfield  {journal} {\bibinfo
  {journal} {Phys. Rev. A}\ }\textbf {\bibinfo {volume} {7}},\ \bibinfo {pages}
  {831} (\bibinfo {year} {1973})}\BibitemShut {NoStop}%
\bibitem [{\citenamefont {Emary}\ and\ \citenamefont
  {Brandes}(2003)}]{EmaryClive2003}%
  \BibitemOpen
  \bibfield  {author} {\bibinfo {author} {\bibfnamefont {C.}~\bibnamefont
  {Emary}}\ and\ \bibinfo {author} {\bibfnamefont {T.}~\bibnamefont
  {Brandes}},\ }\bibfield  {title} {\bibinfo {title} {Quantum chaos triggered
  by precursors of a quantum phase transition: The \uppercase{D}icke model},\
  }\href {https://doi.org/10.1103/PhysRevLett.90.044101} {\bibfield  {journal}
  {\bibinfo  {journal} {Phys. Rev. Lett.}\ }\textbf {\bibinfo {volume} {90}},\
  \bibinfo {pages} {044101} (\bibinfo {year} {2003})}\BibitemShut {NoStop}%
\bibitem [{\citenamefont {Lambert}\ \emph {et~al.}(2004)\citenamefont
  {Lambert}, \citenamefont {Emary},\ and\ \citenamefont
  {Brandes}}]{Lambert2004}%
  \BibitemOpen
  \bibfield  {author} {\bibinfo {author} {\bibfnamefont {N.}~\bibnamefont
  {Lambert}}, \bibinfo {author} {\bibfnamefont {C.}~\bibnamefont {Emary}},\
  and\ \bibinfo {author} {\bibfnamefont {T.}~\bibnamefont {Brandes}},\
  }\bibfield  {title} {\bibinfo {title} {Entanglement and the phase transition
  in single-mode superradiance},\ }\href
  {https://doi.org/10.1103/PhysRevLett.92.073602} {\bibfield  {journal}
  {\bibinfo  {journal} {Phys. Rev. Lett.}\ }\textbf {\bibinfo {volume} {92}},\
  \bibinfo {pages} {073602} (\bibinfo {year} {2004})}\BibitemShut {NoStop}%
\bibitem [{\citenamefont {Dimer}\ \emph {et~al.}(2007)\citenamefont {Dimer},
  \citenamefont {Estienne}, \citenamefont {Parkins},\ and\ \citenamefont
  {Carmichael}}]{Dimer2007}%
  \BibitemOpen
  \bibfield  {author} {\bibinfo {author} {\bibfnamefont {F.}~\bibnamefont
  {Dimer}}, \bibinfo {author} {\bibfnamefont {B.}~\bibnamefont {Estienne}},
  \bibinfo {author} {\bibfnamefont {A.~S.}\ \bibnamefont {Parkins}},\ and\
  \bibinfo {author} {\bibfnamefont {H.~J.}\ \bibnamefont {Carmichael}},\
  }\bibfield  {title} {\bibinfo {title} {Proposed realization of the
  \uppercase{D}icke-model quantum phase transition in an optical cavity
  \uppercase{QED} system},\ }\href {https://doi.org/10.1103/PhysRevA.75.013804}
  {\bibfield  {journal} {\bibinfo  {journal} {Phys. Rev. A}\ }\textbf {\bibinfo
  {volume} {75}},\ \bibinfo {pages} {013804} (\bibinfo {year}
  {2007})}\BibitemShut {NoStop}%
\bibitem [{\citenamefont {Baumann}\ \emph {et~al.}(2010)\citenamefont
  {Baumann}, \citenamefont {Guerlin}, \citenamefont {Brennecke},\ and\
  \citenamefont {Esslinger}}]{Baumann2010}%
  \BibitemOpen
  \bibfield  {author} {\bibinfo {author} {\bibfnamefont {K.}~\bibnamefont
  {Baumann}}, \bibinfo {author} {\bibfnamefont {C.}~\bibnamefont {Guerlin}},
  \bibinfo {author} {\bibfnamefont {F.}~\bibnamefont {Brennecke}},\ and\
  \bibinfo {author} {\bibfnamefont {T.}~\bibnamefont {Esslinger}},\ }\bibfield
  {title} {\bibinfo {title} {\uppercase{D}icke quantum phase transition with a
  superfluid gas in an optical cavity},\ }\href
  {https://www.nature.com/articles/nature09009} {\bibfield  {journal} {\bibinfo
   {journal} {Nature (London)}\ }\textbf {\bibinfo {volume} {464}},\ \bibinfo
  {pages} {1301} (\bibinfo {year} {2010})}\BibitemShut {NoStop}%
\bibitem [{\citenamefont {Baden}\ \emph {et~al.}(2014)\citenamefont {Baden},
  \citenamefont {Arnold}, \citenamefont {Grimsmo}, \citenamefont {Parkins},\
  and\ \citenamefont {Barrett}}]{Baden2014}%
  \BibitemOpen
  \bibfield  {author} {\bibinfo {author} {\bibfnamefont {M.~P.}\ \bibnamefont
  {Baden}}, \bibinfo {author} {\bibfnamefont {K.~J.}\ \bibnamefont {Arnold}},
  \bibinfo {author} {\bibfnamefont {A.~L.}\ \bibnamefont {Grimsmo}}, \bibinfo
  {author} {\bibfnamefont {S.}~\bibnamefont {Parkins}},\ and\ \bibinfo {author}
  {\bibfnamefont {M.~D.}\ \bibnamefont {Barrett}},\ }\bibfield  {title}
  {\bibinfo {title} {Realization of the \uppercase{D}icke model using
  cavity-assisted raman transitions},\ }\href
  {https://doi.org/10.1103/PhysRevLett.113.020408} {\bibfield  {journal}
  {\bibinfo  {journal} {Phys. Rev. Lett.}\ }\textbf {\bibinfo {volume} {113}},\
  \bibinfo {pages} {020408} (\bibinfo {year} {2014})}\BibitemShut {NoStop}%
\bibitem [{\citenamefont {Nagy}\ \emph {et~al.}(2010)\citenamefont {Nagy},
  \citenamefont {K\'onya}, \citenamefont {Szirmai},\ and\ \citenamefont
  {Domokos}}]{Nagy2010}%
  \BibitemOpen
  \bibfield  {author} {\bibinfo {author} {\bibfnamefont {D.}~\bibnamefont
  {Nagy}}, \bibinfo {author} {\bibfnamefont {G.}~\bibnamefont {K\'onya}},
  \bibinfo {author} {\bibfnamefont {G.}~\bibnamefont {Szirmai}},\ and\ \bibinfo
  {author} {\bibfnamefont {P.}~\bibnamefont {Domokos}},\ }\bibfield  {title}
  {\bibinfo {title} {\uppercase{D}icke-model phase transition in the quantum
  motion of a \uppercase{B}ose-\uppercase{E}instein condensate in an optical
  cavity},\ }\href {https://doi.org/10.1103/PhysRevLett.104.130401} {\bibfield
  {journal} {\bibinfo  {journal} {Phys. Rev. Lett.}\ }\textbf {\bibinfo
  {volume} {104}},\ \bibinfo {pages} {130401} (\bibinfo {year}
  {2010})}\BibitemShut {NoStop}%
\bibitem [{\citenamefont {L\"u}\ \emph
  {et~al.}(2018{\natexlab{a}})\citenamefont {L\"u}, \citenamefont {Zheng},
  \citenamefont {Zhu},\ and\ \citenamefont {Wu}}]{Xin-You2018001}%
  \BibitemOpen
  \bibfield  {author} {\bibinfo {author} {\bibfnamefont {X.-Y.}\ \bibnamefont
  {L\"u}}, \bibinfo {author} {\bibfnamefont {L.-L.}\ \bibnamefont {Zheng}},
  \bibinfo {author} {\bibfnamefont {G.-L.}\ \bibnamefont {Zhu}},\ and\ \bibinfo
  {author} {\bibfnamefont {Y.}~\bibnamefont {Wu}},\ }\bibfield  {title}
  {\bibinfo {title} {Single-photon-triggered quantum phase transition},\ }\href
  {https://doi.org/10.1103/PhysRevApplied.9.064006} {\bibfield  {journal}
  {\bibinfo  {journal} {Phys. Rev. Appl.}\ }\textbf {\bibinfo {volume} {9}},\
  \bibinfo {pages} {064006} (\bibinfo {year} {2018}{\natexlab{a}})}\BibitemShut
  {NoStop}%
\bibitem [{\citenamefont {Kirton}\ \emph {et~al.}(2019)\citenamefont {Kirton},
  \citenamefont {Roses}, \citenamefont {Keeling},\ and\ \citenamefont
  {Dalla~Torre}}]{Kirton2019}%
  \BibitemOpen
  \bibfield  {author} {\bibinfo {author} {\bibfnamefont {P.}~\bibnamefont
  {Kirton}}, \bibinfo {author} {\bibfnamefont {M.~M.}\ \bibnamefont {Roses}},
  \bibinfo {author} {\bibfnamefont {J.}~\bibnamefont {Keeling}},\ and\ \bibinfo
  {author} {\bibfnamefont {E.~G.}\ \bibnamefont {Dalla~Torre}},\ }\bibfield
  {title} {\bibinfo {title} {Introduction to the \uppercase{D}icke model: From
  equilibrium to nonequilibrium, and vice versa},\ }\href
  {https://doi.org/https://doi.org/10.1002/qute.201800043} {\bibfield
  {journal} {\bibinfo  {journal} {Adv. Quantum Technol.}\ }\textbf {\bibinfo
  {volume} {2}},\ \bibinfo {pages} {1800043} (\bibinfo {year}
  {2019})}\BibitemShut {NoStop}%
\bibitem [{\citenamefont {Xu}\ \emph {et~al.}(2021)\citenamefont {Xu},
  \citenamefont {Fallas~Padilla},\ and\ \citenamefont {Pu}}]{Youjiang2021}%
  \BibitemOpen
  \bibfield  {author} {\bibinfo {author} {\bibfnamefont {Y.}~\bibnamefont
  {Xu}}, \bibinfo {author} {\bibfnamefont {D.}~\bibnamefont {Fallas~Padilla}},\
  and\ \bibinfo {author} {\bibfnamefont {H.}~\bibnamefont {Pu}},\ }\bibfield
  {title} {\bibinfo {title} {Multicriticality and quantum fluctuation in a
  generalized \uppercase{D}icke model},\ }\href
  {https://doi.org/10.1103/PhysRevA.104.043708} {\bibfield  {journal} {\bibinfo
   {journal} {Phys. Rev. A}\ }\textbf {\bibinfo {volume} {104}},\ \bibinfo
  {pages} {043708} (\bibinfo {year} {2021})}\BibitemShut {NoStop}%
\bibitem [{\citenamefont {Zhao}\ and\ \citenamefont
  {Hwang}(2022)}]{Jinchen2022}%
  \BibitemOpen
  \bibfield  {author} {\bibinfo {author} {\bibfnamefont {J.}~\bibnamefont
  {Zhao}}\ and\ \bibinfo {author} {\bibfnamefont {M.-J.}\ \bibnamefont
  {Hwang}},\ }\bibfield  {title} {\bibinfo {title} {Frustrated superradiant
  phase transition},\ }\href {https://doi.org/10.1103/PhysRevLett.128.163601}
  {\bibfield  {journal} {\bibinfo  {journal} {Phys. Rev. Lett.}\ }\textbf
  {\bibinfo {volume} {128}},\ \bibinfo {pages} {163601} (\bibinfo {year}
  {2022})}\BibitemShut {NoStop}%
\bibitem [{\citenamefont {Rabi}(1937)}]{Rabi1937}%
  \BibitemOpen
  \bibfield  {author} {\bibinfo {author} {\bibfnamefont {I.~I.}\ \bibnamefont
  {Rabi}},\ }\bibfield  {title} {\bibinfo {title} {Space quantization in a
  gyrating magnetic field},\ }\href {https://doi.org/10.1103/PhysRev.51.652}
  {\bibfield  {journal} {\bibinfo  {journal} {Phys. Rev.}\ }\textbf {\bibinfo
  {volume} {51}},\ \bibinfo {pages} {652} (\bibinfo {year} {1937})}\BibitemShut
  {NoStop}%
\bibitem [{\citenamefont {Braak}(2011)}]{Braak2011}%
  \BibitemOpen
  \bibfield  {author} {\bibinfo {author} {\bibfnamefont {D.}~\bibnamefont
  {Braak}},\ }\bibfield  {title} {\bibinfo {title} {Integrability of the
  \uppercase{R}abi model},\ }\href
  {https://doi.org/10.1103/PhysRevLett.107.100401} {\bibfield  {journal}
  {\bibinfo  {journal} {Phys. Rev. Lett.}\ }\textbf {\bibinfo {volume} {107}},\
  \bibinfo {pages} {100401} (\bibinfo {year} {2011})}\BibitemShut {NoStop}%
\bibitem [{\citenamefont {Xie}\ \emph {et~al.}(2017)\citenamefont {Xie},
  \citenamefont {Zhong}, \citenamefont {Batchelor},\ and\ \citenamefont
  {Lee}}]{xie2017quantum}%
  \BibitemOpen
  \bibfield  {author} {\bibinfo {author} {\bibfnamefont {Q.}~\bibnamefont
  {Xie}}, \bibinfo {author} {\bibfnamefont {H.}~\bibnamefont {Zhong}}, \bibinfo
  {author} {\bibfnamefont {M.~T.}\ \bibnamefont {Batchelor}},\ and\ \bibinfo
  {author} {\bibfnamefont {C.}~\bibnamefont {Lee}},\ }\bibfield  {title}
  {\bibinfo {title} {The quantum \uppercase{R}abi model: solution and
  dynamics},\ }\href
  {https://iopscience.iop.org/article/10.1088/1751-8121/aa5a65} {\bibfield
  {journal} {\bibinfo  {journal} {J. Phys. A: Math. Theor.}\ }\textbf {\bibinfo
  {volume} {50}},\ \bibinfo {pages} {113001} (\bibinfo {year}
  {2017})}\BibitemShut {NoStop}%
\bibitem [{\citenamefont {Hwang}\ \emph {et~al.}(2015)\citenamefont {Hwang},
  \citenamefont {Puebla},\ and\ \citenamefont {Plenio}}]{Hwang2015}%
  \BibitemOpen
  \bibfield  {author} {\bibinfo {author} {\bibfnamefont {M.-J.}\ \bibnamefont
  {Hwang}}, \bibinfo {author} {\bibfnamefont {R.}~\bibnamefont {Puebla}},\ and\
  \bibinfo {author} {\bibfnamefont {M.~B.}\ \bibnamefont {Plenio}},\ }\bibfield
   {title} {\bibinfo {title} {Quantum phase transition and universal dynamics
  in the \uppercase{R}abi model},\ }\href
  {https://doi.org/10.1103/PhysRevLett.115.180404} {\bibfield  {journal}
  {\bibinfo  {journal} {Phys. Rev. Lett.}\ }\textbf {\bibinfo {volume} {115}},\
  \bibinfo {pages} {180404} (\bibinfo {year} {2015})}\BibitemShut {NoStop}%
\bibitem [{\citenamefont {Liu}\ \emph {et~al.}(2017)\citenamefont {Liu},
  \citenamefont {Chesi}, \citenamefont {Ying}, \citenamefont {Chen},
  \citenamefont {Luo},\ and\ \citenamefont {Lin}}]{Maoxin2017}%
  \BibitemOpen
  \bibfield  {author} {\bibinfo {author} {\bibfnamefont {M.}~\bibnamefont
  {Liu}}, \bibinfo {author} {\bibfnamefont {S.}~\bibnamefont {Chesi}}, \bibinfo
  {author} {\bibfnamefont {Z.-J.}\ \bibnamefont {Ying}}, \bibinfo {author}
  {\bibfnamefont {X.}~\bibnamefont {Chen}}, \bibinfo {author} {\bibfnamefont
  {H.-G.}\ \bibnamefont {Luo}},\ and\ \bibinfo {author} {\bibfnamefont {H.-Q.}\
  \bibnamefont {Lin}},\ }\bibfield  {title} {\bibinfo {title} {Universal
  scaling and critical exponents of the anisotropic quantum \uppercase{R}abi
  model},\ }\href {https://doi.org/10.1103/PhysRevLett.119.220601} {\bibfield
  {journal} {\bibinfo  {journal} {Phys. Rev. Lett.}\ }\textbf {\bibinfo
  {volume} {119}},\ \bibinfo {pages} {220601} (\bibinfo {year}
  {2017})}\BibitemShut {NoStop}%
\bibitem [{\citenamefont {L\"u}\ \emph
  {et~al.}(2018{\natexlab{b}})\citenamefont {L\"u}, \citenamefont {Zhu},
  \citenamefont {Zheng},\ and\ \citenamefont {Wu}}]{Xin-You2018}%
  \BibitemOpen
  \bibfield  {author} {\bibinfo {author} {\bibfnamefont {X.-Y.}\ \bibnamefont
  {L\"u}}, \bibinfo {author} {\bibfnamefont {G.-L.}\ \bibnamefont {Zhu}},
  \bibinfo {author} {\bibfnamefont {L.-L.}\ \bibnamefont {Zheng}},\ and\
  \bibinfo {author} {\bibfnamefont {Y.}~\bibnamefont {Wu}},\ }\bibfield
  {title} {\bibinfo {title} {Entanglement and quantum superposition induced by
  a single photon},\ }\href {https://doi.org/10.1103/PhysRevA.97.033807}
  {\bibfield  {journal} {\bibinfo  {journal} {Phys. Rev. A}\ }\textbf {\bibinfo
  {volume} {97}},\ \bibinfo {pages} {033807} (\bibinfo {year}
  {2018}{\natexlab{b}})}\BibitemShut {NoStop}%
\bibitem [{\citenamefont {Hwang}\ \emph {et~al.}(2018)\citenamefont {Hwang},
  \citenamefont {Rabl},\ and\ \citenamefont {Plenio}}]{Hwang2018}%
  \BibitemOpen
  \bibfield  {author} {\bibinfo {author} {\bibfnamefont {M.-J.}\ \bibnamefont
  {Hwang}}, \bibinfo {author} {\bibfnamefont {P.}~\bibnamefont {Rabl}},\ and\
  \bibinfo {author} {\bibfnamefont {M.~B.}\ \bibnamefont {Plenio}},\ }\bibfield
   {title} {\bibinfo {title} {Dissipative phase transition in the open quantum
  \uppercase{R}abi model},\ }\href {https://doi.org/10.1103/PhysRevA.97.013825}
  {\bibfield  {journal} {\bibinfo  {journal} {Phys. Rev. A}\ }\textbf {\bibinfo
  {volume} {97}},\ \bibinfo {pages} {013825} (\bibinfo {year}
  {2018})}\BibitemShut {NoStop}%
\bibitem [{\citenamefont {Zhang}\ \emph {et~al.}(2021)\citenamefont {Zhang},
  \citenamefont {Hu}, \citenamefont {Fu}, \citenamefont {Luo}, \citenamefont
  {Pu},\ and\ \citenamefont {Zhang}}]{ZhangYu-Yu2021}%
  \BibitemOpen
  \bibfield  {author} {\bibinfo {author} {\bibfnamefont {Y.-Y.}\ \bibnamefont
  {Zhang}}, \bibinfo {author} {\bibfnamefont {Z.-X.}\ \bibnamefont {Hu}},
  \bibinfo {author} {\bibfnamefont {L.}~\bibnamefont {Fu}}, \bibinfo {author}
  {\bibfnamefont {H.-G.}\ \bibnamefont {Luo}}, \bibinfo {author} {\bibfnamefont
  {H.}~\bibnamefont {Pu}},\ and\ \bibinfo {author} {\bibfnamefont {X.-F.}\
  \bibnamefont {Zhang}},\ }\bibfield  {title} {\bibinfo {title} {Quantum phases
  in a quantum \uppercase{R}abi triangle},\ }\href
  {https://doi.org/10.1103/PhysRevLett.127.063602} {\bibfield  {journal}
  {\bibinfo  {journal} {Phys. Rev. Lett.}\ }\textbf {\bibinfo {volume} {127}},\
  \bibinfo {pages} {063602} (\bibinfo {year} {2021})}\BibitemShut {NoStop}%
\bibitem [{\citenamefont {Fallas~Padilla}\ \emph {et~al.}(2022)\citenamefont
  {Fallas~Padilla}, \citenamefont {Pu}, \citenamefont {Cheng},\ and\
  \citenamefont {Zhang}}]{Fallas2022}%
  \BibitemOpen
  \bibfield  {author} {\bibinfo {author} {\bibfnamefont {D.}~\bibnamefont
  {Fallas~Padilla}}, \bibinfo {author} {\bibfnamefont {H.}~\bibnamefont {Pu}},
  \bibinfo {author} {\bibfnamefont {G.-J.}\ \bibnamefont {Cheng}},\ and\
  \bibinfo {author} {\bibfnamefont {Y.-Y.}\ \bibnamefont {Zhang}},\ }\bibfield
  {title} {\bibinfo {title} {Understanding the quantum \uppercase{R}abi ring
  using analogies to quantum magnetism},\ }\href
  {https://doi.org/10.1103/PhysRevLett.129.183602} {\bibfield  {journal}
  {\bibinfo  {journal} {Phys. Rev. Lett.}\ }\textbf {\bibinfo {volume} {129}},\
  \bibinfo {pages} {183602} (\bibinfo {year} {2022})}\BibitemShut {NoStop}%
\bibitem [{\citenamefont {Liu}\ \emph {et~al.}(2022)\citenamefont {Liu},
  \citenamefont {Zhao}, \citenamefont {Yang},\ and\ \citenamefont
  {Luo}}]{liu2022process}%
  \BibitemOpen
  \bibfield  {author} {\bibinfo {author} {\bibfnamefont {J.}~\bibnamefont
  {Liu}}, \bibinfo {author} {\bibfnamefont {M.}~\bibnamefont {Zhao}}, \bibinfo
  {author} {\bibfnamefont {Y.-T.}\ \bibnamefont {Yang}},\ and\ \bibinfo
  {author} {\bibfnamefont {H.-G.}\ \bibnamefont {Luo}},\ }\bibfield  {title}
  {\bibinfo {title} {The process of superradiant phase transition for quantum
  \uppercase{R}abi model in view of nonclassical states},\ }\href
  {https://arxiv.org/abs/2211.16233} {\bibfield  {journal} {\bibinfo  {journal}
  {arXiv:2211.16233}\ } (\bibinfo {year} {2022})}\BibitemShut {NoStop}%
\bibitem [{\citenamefont {Shen}\ \emph {et~al.}(2022)\citenamefont {Shen},
  \citenamefont {Tang}, \citenamefont {Shi}, \citenamefont {Wu}, \citenamefont
  {Yang},\ and\ \citenamefont {Zheng}}]{Zhengshibiao2022}%
  \BibitemOpen
  \bibfield  {author} {\bibinfo {author} {\bibfnamefont {L.-T.}\ \bibnamefont
  {Shen}}, \bibinfo {author} {\bibfnamefont {C.-Q.}\ \bibnamefont {Tang}},
  \bibinfo {author} {\bibfnamefont {Z.}~\bibnamefont {Shi}}, \bibinfo {author}
  {\bibfnamefont {H.}~\bibnamefont {Wu}}, \bibinfo {author} {\bibfnamefont
  {Z.-B.}\ \bibnamefont {Yang}},\ and\ \bibinfo {author} {\bibfnamefont
  {S.-B.}\ \bibnamefont {Zheng}},\ }\bibfield  {title} {\bibinfo {title}
  {Squeezed-light-induced quantum phase transition in the \uppercase
  {J}aynes-\uppercase{C}ummings model},\ }\href
  {https://doi.org/10.1103/PhysRevA.106.023705} {\bibfield  {journal} {\bibinfo
   {journal} {Phys. Rev. A}\ }\textbf {\bibinfo {volume} {106}},\ \bibinfo
  {pages} {023705} (\bibinfo {year} {2022})}\BibitemShut {NoStop}%
\bibitem [{\citenamefont {Georgescu}\ \emph {et~al.}(2014)\citenamefont
  {Georgescu}, \citenamefont {Ashhab},\ and\ \citenamefont
  {Nori}}]{Georgescu2014}%
  \BibitemOpen
  \bibfield  {author} {\bibinfo {author} {\bibfnamefont {I.~M.}\ \bibnamefont
  {Georgescu}}, \bibinfo {author} {\bibfnamefont {S.}~\bibnamefont {Ashhab}},\
  and\ \bibinfo {author} {\bibfnamefont {F.}~\bibnamefont {Nori}},\ }\bibfield
  {title} {\bibinfo {title} {Quantum simulation},\ }\href
  {https://doi.org/10.1103/RevModPhys.86.153} {\bibfield  {journal} {\bibinfo
  {journal} {Rev. Mod. Phys.}\ }\textbf {\bibinfo {volume} {86}},\ \bibinfo
  {pages} {153} (\bibinfo {year} {2014})}\BibitemShut {NoStop}%
\bibitem [{\citenamefont {Chen}\ \emph {et~al.}(2021)\citenamefont {Chen},
  \citenamefont {Wu}, \citenamefont {Jiang}, \citenamefont {L{\"u}},
  \citenamefont {Peng},\ and\ \citenamefont {Du}}]{chen2021experimental}%
  \BibitemOpen
  \bibfield  {author} {\bibinfo {author} {\bibfnamefont {X.}~\bibnamefont
  {Chen}}, \bibinfo {author} {\bibfnamefont {Z.}~\bibnamefont {Wu}}, \bibinfo
  {author} {\bibfnamefont {M.}~\bibnamefont {Jiang}}, \bibinfo {author}
  {\bibfnamefont {X.-Y.}\ \bibnamefont {L{\"u}}}, \bibinfo {author}
  {\bibfnamefont {X.}~\bibnamefont {Peng}},\ and\ \bibinfo {author}
  {\bibfnamefont {J.}~\bibnamefont {Du}},\ }\bibfield  {title} {\bibinfo
  {title} {Experimental quantum simulation of superradiant phase transition
  beyond no-go theorem via antisqueezing},\ }\href
  {https://www.nature.com/articles/s41467-021-26573-5} {\bibfield  {journal}
  {\bibinfo  {journal} {Nat. Commun.}\ }\textbf {\bibinfo {volume} {12}},\
  \bibinfo {pages} {6281} (\bibinfo {year} {2021})}\BibitemShut {NoStop}%
\bibitem [{\citenamefont {Puebla}\ \emph {et~al.}(2017)\citenamefont {Puebla},
  \citenamefont {Hwang}, \citenamefont {Casanova},\ and\ \citenamefont
  {Plenio}}]{Puebla2017}%
  \BibitemOpen
  \bibfield  {author} {\bibinfo {author} {\bibfnamefont {R.}~\bibnamefont
  {Puebla}}, \bibinfo {author} {\bibfnamefont {M.-J.}\ \bibnamefont {Hwang}},
  \bibinfo {author} {\bibfnamefont {J.}~\bibnamefont {Casanova}},\ and\
  \bibinfo {author} {\bibfnamefont {M.~B.}\ \bibnamefont {Plenio}},\ }\bibfield
   {title} {\bibinfo {title} {Probing the dynamics of a superradiant quantum
  phase transition with a single trapped ion},\ }\href
  {https://doi.org/10.1103/PhysRevLett.118.073001} {\bibfield  {journal}
  {\bibinfo  {journal} {Phys. Rev. Lett.}\ }\textbf {\bibinfo {volume} {118}},\
  \bibinfo {pages} {073001} (\bibinfo {year} {2017})}\BibitemShut {NoStop}%
\bibitem [{\citenamefont {Cai}\ \emph {et~al.}(2021)\citenamefont {Cai},
  \citenamefont {Liu}, \citenamefont {Zhao}, \citenamefont {Wu}, \citenamefont
  {Mei}, \citenamefont {Jiang}, \citenamefont {He}, \citenamefont {Zhang},
  \citenamefont {Zhou},\ and\ \citenamefont {Duan}}]{cai2021observation}%
  \BibitemOpen
  \bibfield  {author} {\bibinfo {author} {\bibfnamefont {M.-L.}\ \bibnamefont
  {Cai}}, \bibinfo {author} {\bibfnamefont {Z.-D.}\ \bibnamefont {Liu}},
  \bibinfo {author} {\bibfnamefont {W.-D.}\ \bibnamefont {Zhao}}, \bibinfo
  {author} {\bibfnamefont {Y.-K.}\ \bibnamefont {Wu}}, \bibinfo {author}
  {\bibfnamefont {Q.-X.}\ \bibnamefont {Mei}}, \bibinfo {author} {\bibfnamefont
  {Y.}~\bibnamefont {Jiang}}, \bibinfo {author} {\bibfnamefont
  {L.}~\bibnamefont {He}}, \bibinfo {author} {\bibfnamefont {X.}~\bibnamefont
  {Zhang}}, \bibinfo {author} {\bibfnamefont {Z.-C.}\ \bibnamefont {Zhou}},\
  and\ \bibinfo {author} {\bibfnamefont {L.-M.}\ \bibnamefont {Duan}},\
  }\bibfield  {title} {\bibinfo {title} {Observation of a quantum phase
  transition in the quantum \uppercase{R}abi model with a single trapped ion},\
  }\href {https://www.nature.com/articles/s41467-021-21425-8} {\bibfield
  {journal} {\bibinfo  {journal} {Nat. Commun.}\ }\textbf {\bibinfo {volume}
  {12}},\ \bibinfo {pages} {1126} (\bibinfo {year} {2021})}\BibitemShut
  {NoStop}%
\bibitem [{\citenamefont {Zheng}\ \emph {et~al.}(2023)\citenamefont {Zheng},
  \citenamefont {Ning}, \citenamefont {Chen}, \citenamefont {L\"u},
  \citenamefont {Shen}, \citenamefont {Xu}, \citenamefont {Zhang},
  \citenamefont {Xu}, \citenamefont {Li}, \citenamefont {Xia}, \citenamefont
  {Wu}, \citenamefont {Yang}, \citenamefont {Miranowicz}, \citenamefont
  {Lambert}, \citenamefont {Zheng}, \citenamefont {Fan}, \citenamefont {Nori},\
  and\ \citenamefont {Zheng}}]{zheng2022emergent}%
  \BibitemOpen
  \bibfield  {author} {\bibinfo {author} {\bibfnamefont {R.-H.}\ \bibnamefont
  {Zheng}}, \bibinfo {author} {\bibfnamefont {W.}~\bibnamefont {Ning}},
  \bibinfo {author} {\bibfnamefont {Y.-H.}\ \bibnamefont {Chen}}, \bibinfo
  {author} {\bibfnamefont {J.-H.}\ \bibnamefont {L\"u}}, \bibinfo {author}
  {\bibfnamefont {L.-T.}\ \bibnamefont {Shen}}, \bibinfo {author}
  {\bibfnamefont {K.}~\bibnamefont {Xu}}, \bibinfo {author} {\bibfnamefont
  {Y.-R.}\ \bibnamefont {Zhang}}, \bibinfo {author} {\bibfnamefont
  {D.}~\bibnamefont {Xu}}, \bibinfo {author} {\bibfnamefont {H.}~\bibnamefont
  {Li}}, \bibinfo {author} {\bibfnamefont {Y.}~\bibnamefont {Xia}}, \bibinfo
  {author} {\bibfnamefont {F.}~\bibnamefont {Wu}}, \bibinfo {author}
  {\bibfnamefont {Z.-B.}\ \bibnamefont {Yang}}, \bibinfo {author}
  {\bibfnamefont {A.}~\bibnamefont {Miranowicz}}, \bibinfo {author}
  {\bibfnamefont {N.}~\bibnamefont {Lambert}}, \bibinfo {author} {\bibfnamefont
  {D.}~\bibnamefont {Zheng}}, \bibinfo {author} {\bibfnamefont
  {H.}~\bibnamefont {Fan}}, \bibinfo {author} {\bibfnamefont {F.}~\bibnamefont
  {Nori}},\ and\ \bibinfo {author} {\bibfnamefont {S.-B.}\ \bibnamefont
  {Zheng}},\ }\bibfield  {title} {\bibinfo {title} {Observation of a
  superradiant phase transition with emergent cat states},\ }\href
  {https://doi.org/10.1103/PhysRevLett.131.113601} {\bibfield  {journal}
  {\bibinfo  {journal} {Phys. Rev. Lett.}\ }\textbf {\bibinfo {volume} {131}},\
  \bibinfo {pages} {113601} (\bibinfo {year} {2023})}\BibitemShut {NoStop}%
\bibitem [{\citenamefont {Kirton}\ and\ \citenamefont
  {Keeling}(2017)}]{Kirton2017}%
  \BibitemOpen
  \bibfield  {author} {\bibinfo {author} {\bibfnamefont {P.}~\bibnamefont
  {Kirton}}\ and\ \bibinfo {author} {\bibfnamefont {J.}~\bibnamefont
  {Keeling}},\ }\bibfield  {title} {\bibinfo {title} {Suppressing and restoring
  the \uppercase{D}icke superradiance transition by dephasing and decay},\
  }\href {https://doi.org/10.1103/PhysRevLett.118.123602} {\bibfield  {journal}
  {\bibinfo  {journal} {Phys. Rev. Lett.}\ }\textbf {\bibinfo {volume} {118}},\
  \bibinfo {pages} {123602} (\bibinfo {year} {2017})}\BibitemShut {NoStop}%
\bibitem [{\citenamefont {Shammah}\ \emph {et~al.}(2018)\citenamefont
  {Shammah}, \citenamefont {Ahmed}, \citenamefont {Lambert}, \citenamefont
  {De~Liberato},\ and\ \citenamefont {Nori}}]{Shammah2018}%
  \BibitemOpen
  \bibfield  {author} {\bibinfo {author} {\bibfnamefont {N.}~\bibnamefont
  {Shammah}}, \bibinfo {author} {\bibfnamefont {S.}~\bibnamefont {Ahmed}},
  \bibinfo {author} {\bibfnamefont {N.}~\bibnamefont {Lambert}}, \bibinfo
  {author} {\bibfnamefont {S.}~\bibnamefont {De~Liberato}},\ and\ \bibinfo
  {author} {\bibfnamefont {F.}~\bibnamefont {Nori}},\ }\bibfield  {title}
  {\bibinfo {title} {Open quantum systems with local and collective incoherent
  processes: Efficient numerical simulations using permutational invariance},\
  }\href {https://doi.org/10.1103/PhysRevA.98.063815} {\bibfield  {journal}
  {\bibinfo  {journal} {Phys. Rev. A}\ }\textbf {\bibinfo {volume} {98}},\
  \bibinfo {pages} {063815} (\bibinfo {year} {2018})}\BibitemShut {NoStop}%
\bibitem [{\citenamefont {Soriente}\ \emph {et~al.}(2018)\citenamefont
  {Soriente}, \citenamefont {Donner}, \citenamefont {Chitra},\ and\
  \citenamefont {Zilberberg}}]{Soriente2018}%
  \BibitemOpen
  \bibfield  {author} {\bibinfo {author} {\bibfnamefont {M.}~\bibnamefont
  {Soriente}}, \bibinfo {author} {\bibfnamefont {T.}~\bibnamefont {Donner}},
  \bibinfo {author} {\bibfnamefont {R.}~\bibnamefont {Chitra}},\ and\ \bibinfo
  {author} {\bibfnamefont {O.}~\bibnamefont {Zilberberg}},\ }\bibfield  {title}
  {\bibinfo {title} {Dissipation-induced anomalous multicritical phenomena},\
  }\href {https://doi.org/10.1103/PhysRevLett.120.183603} {\bibfield  {journal}
  {\bibinfo  {journal} {Phys. Rev. Lett.}\ }\textbf {\bibinfo {volume} {120}},\
  \bibinfo {pages} {183603} (\bibinfo {year} {2018})}\BibitemShut {NoStop}%
\bibitem [{\citenamefont {Zhu}\ \emph {et~al.}(2020{\natexlab{a}})\citenamefont
  {Zhu}, \citenamefont {Ping}, \citenamefont {Yang},\ and\ \citenamefont
  {Agarwal}}]{ZhuCJ2020}%
  \BibitemOpen
  \bibfield  {author} {\bibinfo {author} {\bibfnamefont {C.~J.}\ \bibnamefont
  {Zhu}}, \bibinfo {author} {\bibfnamefont {L.~L.}\ \bibnamefont {Ping}},
  \bibinfo {author} {\bibfnamefont {Y.~P.}\ \bibnamefont {Yang}},\ and\
  \bibinfo {author} {\bibfnamefont {G.~S.}\ \bibnamefont {Agarwal}},\
  }\bibfield  {title} {\bibinfo {title} {Squeezed light induced symmetry
  breaking superradiant phase transition},\ }\href
  {https://doi.org/10.1103/PhysRevLett.124.073602} {\bibfield  {journal}
  {\bibinfo  {journal} {Phys. Rev. Lett.}\ }\textbf {\bibinfo {volume} {124}},\
  \bibinfo {pages} {073602} (\bibinfo {year} {2020}{\natexlab{a}})}\BibitemShut
  {NoStop}%
\bibitem [{\citenamefont {Lin}\ \emph {et~al.}(2022)\citenamefont {Lin},
  \citenamefont {Rosa-Medina}, \citenamefont {Ferri}, \citenamefont {Finger},
  \citenamefont {Kroeger}, \citenamefont {Donner}, \citenamefont {Esslinger},\
  and\ \citenamefont {Chitra}}]{Lin2022Rui}%
  \BibitemOpen
  \bibfield  {author} {\bibinfo {author} {\bibfnamefont {R.}~\bibnamefont
  {Lin}}, \bibinfo {author} {\bibfnamefont {R.}~\bibnamefont {Rosa-Medina}},
  \bibinfo {author} {\bibfnamefont {F.}~\bibnamefont {Ferri}}, \bibinfo
  {author} {\bibfnamefont {F.}~\bibnamefont {Finger}}, \bibinfo {author}
  {\bibfnamefont {K.}~\bibnamefont {Kroeger}}, \bibinfo {author} {\bibfnamefont
  {T.}~\bibnamefont {Donner}}, \bibinfo {author} {\bibfnamefont
  {T.}~\bibnamefont {Esslinger}},\ and\ \bibinfo {author} {\bibfnamefont
  {R.}~\bibnamefont {Chitra}},\ }\bibfield  {title} {\bibinfo {title}
  {Dissipation-engineered family of nearly dark states in many-body cavity-atom
  systems},\ }\href {https://doi.org/10.1103/PhysRevLett.128.153601} {\bibfield
   {journal} {\bibinfo  {journal} {Phys. Rev. Lett.}\ }\textbf {\bibinfo
  {volume} {128}},\ \bibinfo {pages} {153601} (\bibinfo {year}
  {2022})}\BibitemShut {NoStop}%
\bibitem [{\citenamefont {Xu}\ and\ \citenamefont {Pu}(2019)}]{Youjiang2019}%
  \BibitemOpen
  \bibfield  {author} {\bibinfo {author} {\bibfnamefont {Y.}~\bibnamefont
  {Xu}}\ and\ \bibinfo {author} {\bibfnamefont {H.}~\bibnamefont {Pu}},\
  }\bibfield  {title} {\bibinfo {title} {Emergent universality in a quantum
  tricritical \uppercase{D}icke model},\ }\href
  {https://doi.org/10.1103/PhysRevLett.122.193201} {\bibfield  {journal}
  {\bibinfo  {journal} {Phys. Rev. Lett.}\ }\textbf {\bibinfo {volume} {122}},\
  \bibinfo {pages} {193201} (\bibinfo {year} {2019})}\BibitemShut {NoStop}%
\bibitem [{\citenamefont {Zhu}\ \emph {et~al.}(2020{\natexlab{b}})\citenamefont
  {Zhu}, \citenamefont {Xu}, \citenamefont {Zhang},\ and\ \citenamefont
  {Liu}}]{Han-Jie2020}%
  \BibitemOpen
  \bibfield  {author} {\bibinfo {author} {\bibfnamefont {H.-J.}\ \bibnamefont
  {Zhu}}, \bibinfo {author} {\bibfnamefont {K.}~\bibnamefont {Xu}}, \bibinfo
  {author} {\bibfnamefont {G.-F.}\ \bibnamefont {Zhang}},\ and\ \bibinfo
  {author} {\bibfnamefont {W.-M.}\ \bibnamefont {Liu}},\ }\bibfield  {title}
  {\bibinfo {title} {Finite-component multicriticality at the superradiant
  quantum phase transition},\ }\href
  {https://doi.org/10.1103/PhysRevLett.125.050402} {\bibfield  {journal}
  {\bibinfo  {journal} {Phys. Rev. Lett.}\ }\textbf {\bibinfo {volume} {125}},\
  \bibinfo {pages} {050402} (\bibinfo {year} {2020}{\natexlab{b}})}\BibitemShut
  {NoStop}%
\bibitem [{\citenamefont {Larson}\ and\ \citenamefont
  {Irish}(2017)}]{Larson2017}%
  \BibitemOpen
  \bibfield  {author} {\bibinfo {author} {\bibfnamefont {J.}~\bibnamefont
  {Larson}}\ and\ \bibinfo {author} {\bibfnamefont {E.~K.}\ \bibnamefont
  {Irish}},\ }\bibfield  {title} {\bibinfo {title} {Some remarks on
  ‘superradiant’ phase transitions in light-matter systems},\ }\href
  {https://doi.org/10.1088/1751-8121/aa65dc} {\bibfield  {journal} {\bibinfo
  {journal} {J. Phys. A: Math. Theor.}\ }\textbf {\bibinfo {volume} {50}},\
  \bibinfo {pages} {174002} (\bibinfo {year} {2017})}\BibitemShut {NoStop}%
\bibitem [{\citenamefont {Carmichael}(2015)}]{Carmichael2015}%
  \BibitemOpen
  \bibfield  {author} {\bibinfo {author} {\bibfnamefont {H.~J.}\ \bibnamefont
  {Carmichael}},\ }\bibfield  {title} {\bibinfo {title} {Breakdown of photon
  blockade: A dissipative quantum phase transition in zero dimensions},\ }\href
  {https://doi.org/10.1103/PhysRevX.5.031028} {\bibfield  {journal} {\bibinfo
  {journal} {Phys. Rev. X}\ }\textbf {\bibinfo {volume} {5}},\ \bibinfo {pages}
  {031028} (\bibinfo {year} {2015})}\BibitemShut {NoStop}%
\bibitem [{\citenamefont {Fink}\ \emph {et~al.}(2017)\citenamefont {Fink},
  \citenamefont {Dombi}, \citenamefont {Vukics}, \citenamefont {Wallraff},\
  and\ \citenamefont {Domokos}}]{Fink2017}%
  \BibitemOpen
  \bibfield  {author} {\bibinfo {author} {\bibfnamefont {J.~M.}\ \bibnamefont
  {Fink}}, \bibinfo {author} {\bibfnamefont {A.}~\bibnamefont {Dombi}},
  \bibinfo {author} {\bibfnamefont {A.}~\bibnamefont {Vukics}}, \bibinfo
  {author} {\bibfnamefont {A.}~\bibnamefont {Wallraff}},\ and\ \bibinfo
  {author} {\bibfnamefont {P.}~\bibnamefont {Domokos}},\ }\bibfield  {title}
  {\bibinfo {title} {Observation of the photon-blockade breakdown phase
  transition},\ }\href {https://doi.org/10.1103/PhysRevX.7.011012} {\bibfield
  {journal} {\bibinfo  {journal} {Phys. Rev. X}\ }\textbf {\bibinfo {volume}
  {7}},\ \bibinfo {pages} {011012} (\bibinfo {year} {2017})}\BibitemShut
  {NoStop}%
\bibitem [{\citenamefont {Reiter}\ \emph {et~al.}(2020)\citenamefont {Reiter},
  \citenamefont {Nguyen}, \citenamefont {Home},\ and\ \citenamefont
  {Yelin}}]{Reiter2020}%
  \BibitemOpen
  \bibfield  {author} {\bibinfo {author} {\bibfnamefont {F.}~\bibnamefont
  {Reiter}}, \bibinfo {author} {\bibfnamefont {T.~L.}\ \bibnamefont {Nguyen}},
  \bibinfo {author} {\bibfnamefont {J.~P.}\ \bibnamefont {Home}},\ and\
  \bibinfo {author} {\bibfnamefont {S.~F.}\ \bibnamefont {Yelin}},\ }\bibfield
  {title} {\bibinfo {title} {Cooperative breakdown of the oscillator blockade
  in the \uppercase{D}icke model},\ }\href
  {https://doi.org/10.1103/PhysRevLett.125.233602} {\bibfield  {journal}
  {\bibinfo  {journal} {Phys. Rev. Lett.}\ }\textbf {\bibinfo {volume} {125}},\
  \bibinfo {pages} {233602} (\bibinfo {year} {2020})}\BibitemShut {NoStop}%
\bibitem [{\citenamefont {Meiser}\ \emph {et~al.}(2009)\citenamefont {Meiser},
  \citenamefont {Ye}, \citenamefont {Carlson},\ and\ \citenamefont
  {Holland}}]{Meiser2009}%
  \BibitemOpen
  \bibfield  {author} {\bibinfo {author} {\bibfnamefont {D.}~\bibnamefont
  {Meiser}}, \bibinfo {author} {\bibfnamefont {J.}~\bibnamefont {Ye}}, \bibinfo
  {author} {\bibfnamefont {D.~R.}\ \bibnamefont {Carlson}},\ and\ \bibinfo
  {author} {\bibfnamefont {M.~J.}\ \bibnamefont {Holland}},\ }\bibfield
  {title} {\bibinfo {title} {Prospects for a millihertz-linewidth laser},\
  }\href {https://doi.org/10.1103/PhysRevLett.102.163601} {\bibfield  {journal}
  {\bibinfo  {journal} {Phys. Rev. Lett.}\ }\textbf {\bibinfo {volume} {102}},\
  \bibinfo {pages} {163601} (\bibinfo {year} {2009})}\BibitemShut {NoStop}%
\bibitem [{\citenamefont {Bohnet}\ \emph {et~al.}(2012)\citenamefont {Bohnet},
  \citenamefont {Chen}, \citenamefont {Weiner}, \citenamefont {Meiser},
  \citenamefont {Holland},\ and\ \citenamefont {Thompson}}]{bohnet2012steady}%
  \BibitemOpen
  \bibfield  {author} {\bibinfo {author} {\bibfnamefont {J.~G.}\ \bibnamefont
  {Bohnet}}, \bibinfo {author} {\bibfnamefont {Z.}~\bibnamefont {Chen}},
  \bibinfo {author} {\bibfnamefont {J.~M.}\ \bibnamefont {Weiner}}, \bibinfo
  {author} {\bibfnamefont {D.}~\bibnamefont {Meiser}}, \bibinfo {author}
  {\bibfnamefont {M.~J.}\ \bibnamefont {Holland}},\ and\ \bibinfo {author}
  {\bibfnamefont {J.~K.}\ \bibnamefont {Thompson}},\ }\bibfield  {title}
  {\bibinfo {title} {A steady-state superradiant laser with less than one
  intracavity photon},\ }\href {https://www.nature.com/articles/nature10920}
  {\bibfield  {journal} {\bibinfo  {journal} {Nature (London)}\ }\textbf
  {\bibinfo {volume} {484}},\ \bibinfo {pages} {78} (\bibinfo {year}
  {2012})}\BibitemShut {NoStop}%
\bibitem [{\citenamefont {Norcia}\ and\ \citenamefont
  {Thompson}(2016)}]{Norcia2016}%
  \BibitemOpen
  \bibfield  {author} {\bibinfo {author} {\bibfnamefont {M.~A.}\ \bibnamefont
  {Norcia}}\ and\ \bibinfo {author} {\bibfnamefont {J.~K.}\ \bibnamefont
  {Thompson}},\ }\bibfield  {title} {\bibinfo {title} {Cold-strontium laser in
  the superradiant crossover regime},\ }\href
  {https://doi.org/10.1103/PhysRevX.6.011025} {\bibfield  {journal} {\bibinfo
  {journal} {Phys. Rev. X}\ }\textbf {\bibinfo {volume} {6}},\ \bibinfo {pages}
  {011025} (\bibinfo {year} {2016})}\BibitemShut {NoStop}%
\bibitem [{\citenamefont {Norcia}\ \emph {et~al.}(2016)\citenamefont {Norcia},
  \citenamefont {Winchester}, \citenamefont {Cline},\ and\ \citenamefont
  {Thompson}}]{norcia2016superradiance}%
  \BibitemOpen
  \bibfield  {author} {\bibinfo {author} {\bibfnamefont {M.~A.}\ \bibnamefont
  {Norcia}}, \bibinfo {author} {\bibfnamefont {M.~N.}\ \bibnamefont
  {Winchester}}, \bibinfo {author} {\bibfnamefont {J.~R.}\ \bibnamefont
  {Cline}},\ and\ \bibinfo {author} {\bibfnamefont {J.~K.}\ \bibnamefont
  {Thompson}},\ }\bibfield  {title} {\bibinfo {title} {Superradiance on the
  millihertz linewidth strontium clock transition},\ }\href
  {https://www.science.org/doi/10.1126/sciadv.1601231} {\bibfield  {journal}
  {\bibinfo  {journal} {Sci. Adv.}\ }\textbf {\bibinfo {volume} {2}},\ \bibinfo
  {pages} {e1601231} (\bibinfo {year} {2016})}\BibitemShut {NoStop}%
\bibitem [{\citenamefont {Gong}\ \emph {et~al.}(2018)\citenamefont {Gong},
  \citenamefont {Hamazaki},\ and\ \citenamefont {Ueda}}]{Gong2018}%
  \BibitemOpen
  \bibfield  {author} {\bibinfo {author} {\bibfnamefont {Z.}~\bibnamefont
  {Gong}}, \bibinfo {author} {\bibfnamefont {R.}~\bibnamefont {Hamazaki}},\
  and\ \bibinfo {author} {\bibfnamefont {M.}~\bibnamefont {Ueda}},\ }\bibfield
  {title} {\bibinfo {title} {Discrete time-crystalline order in cavity and
  circuit \uppercase{QED} systems},\ }\href
  {https://doi.org/10.1103/PhysRevLett.120.040404} {\bibfield  {journal}
  {\bibinfo  {journal} {Phys. Rev. Lett.}\ }\textbf {\bibinfo {volume} {120}},\
  \bibinfo {pages} {040404} (\bibinfo {year} {2018})}\BibitemShut {NoStop}%
\bibitem [{\citenamefont {Mattes}\ \emph {et~al.}(2023)\citenamefont {Mattes},
  \citenamefont {Lesanovsky},\ and\ \citenamefont
  {Carollo}}]{mattes2023entangled}%
  \BibitemOpen
  \bibfield  {author} {\bibinfo {author} {\bibfnamefont {R.}~\bibnamefont
  {Mattes}}, \bibinfo {author} {\bibfnamefont {I.}~\bibnamefont {Lesanovsky}},\
  and\ \bibinfo {author} {\bibfnamefont {F.}~\bibnamefont {Carollo}},\
  }\bibfield  {title} {\bibinfo {title} {Entangled time-crystal phase in an
  open quantum light-matter system},\ }\href {https://arxiv.org/abs/2303.07725}
  {\bibfield  {journal} {\bibinfo  {journal} {arXiv:2303.07725}\ } (\bibinfo
  {year} {2023})}\BibitemShut {NoStop}%
\bibitem [{\citenamefont {Ferioli}\ \emph {et~al.}(2023)\citenamefont
  {Ferioli}, \citenamefont {Glicenstein}, \citenamefont {Ferrier-Barbut},\ and\
  \citenamefont {Browaeys}}]{ferioli2023non}%
  \BibitemOpen
  \bibfield  {author} {\bibinfo {author} {\bibfnamefont {G.}~\bibnamefont
  {Ferioli}}, \bibinfo {author} {\bibfnamefont {A.}~\bibnamefont
  {Glicenstein}}, \bibinfo {author} {\bibfnamefont {I.}~\bibnamefont
  {Ferrier-Barbut}},\ and\ \bibinfo {author} {\bibfnamefont {A.}~\bibnamefont
  {Browaeys}},\ }\bibfield  {title} {\bibinfo {title} {A non-equilibrium
  superradiant phase transition in free space},\ }\href
  {https://www.nature.com/articles/s41567-023-02064-w} {\bibfield  {journal}
  {\bibinfo  {journal} {Nat. Phys.}\ ,\ \bibinfo {pages} {1}} (\bibinfo {year}
  {2023})}\BibitemShut {NoStop}%
\bibitem [{\citenamefont {Xu}\ \emph {et~al.}(2014)\citenamefont {Xu},
  \citenamefont {Tieri}, \citenamefont {Fine}, \citenamefont {Thompson},\ and\
  \citenamefont {Holland}}]{Minghui2014}%
  \BibitemOpen
  \bibfield  {author} {\bibinfo {author} {\bibfnamefont {M.}~\bibnamefont
  {Xu}}, \bibinfo {author} {\bibfnamefont {D.~A.}\ \bibnamefont {Tieri}},
  \bibinfo {author} {\bibfnamefont {E.~C.}\ \bibnamefont {Fine}}, \bibinfo
  {author} {\bibfnamefont {J.~K.}\ \bibnamefont {Thompson}},\ and\ \bibinfo
  {author} {\bibfnamefont {M.~J.}\ \bibnamefont {Holland}},\ }\bibfield
  {title} {\bibinfo {title} {Synchronization of two ensembles of atoms},\
  }\href {https://doi.org/10.1103/PhysRevLett.113.154101} {\bibfield  {journal}
  {\bibinfo  {journal} {Phys. Rev. Lett.}\ }\textbf {\bibinfo {volume} {113}},\
  \bibinfo {pages} {154101} (\bibinfo {year} {2014})}\BibitemShut {NoStop}%
\bibitem [{\citenamefont {Masson}\ and\ \citenamefont
  {Asenjo-Garcia}(2022)}]{masson2022universality}%
  \BibitemOpen
  \bibfield  {author} {\bibinfo {author} {\bibfnamefont {S.~J.}\ \bibnamefont
  {Masson}}\ and\ \bibinfo {author} {\bibfnamefont {A.}~\bibnamefont
  {Asenjo-Garcia}},\ }\bibfield  {title} {\bibinfo {title} {Universality of
  \uppercase{D}icke superradiance in arrays of quantum emitters},\ }\href
  {https://www.nature.com/articles/s41467-022-29805-4} {\bibfield  {journal}
  {\bibinfo  {journal} {Nat. Commun.}\ }\textbf {\bibinfo {volume} {13}},\
  \bibinfo {pages} {2285} (\bibinfo {year} {2022})}\BibitemShut {NoStop}%
\bibitem [{\citenamefont {Lin}\ \emph {et~al.}(2011)\citenamefont {Lin},
  \citenamefont {Ramezani}, \citenamefont {Eichelkraut}, \citenamefont
  {Kottos}, \citenamefont {Cao},\ and\ \citenamefont
  {Christodoulides}}]{Zin2011}%
  \BibitemOpen
  \bibfield  {author} {\bibinfo {author} {\bibfnamefont {Z.}~\bibnamefont
  {Lin}}, \bibinfo {author} {\bibfnamefont {H.}~\bibnamefont {Ramezani}},
  \bibinfo {author} {\bibfnamefont {T.}~\bibnamefont {Eichelkraut}}, \bibinfo
  {author} {\bibfnamefont {T.}~\bibnamefont {Kottos}}, \bibinfo {author}
  {\bibfnamefont {H.}~\bibnamefont {Cao}},\ and\ \bibinfo {author}
  {\bibfnamefont {D.~N.}\ \bibnamefont {Christodoulides}},\ }\bibfield  {title}
  {\bibinfo {title} {Unidirectional invisibility induced by
  $\mathcal{P}\mathcal{T}$-symmetric periodic structures},\ }\href
  {https://doi.org/10.1103/PhysRevLett.106.213901} {\bibfield  {journal}
  {\bibinfo  {journal} {Phys. Rev. Lett.}\ }\textbf {\bibinfo {volume} {106}},\
  \bibinfo {pages} {213901} (\bibinfo {year} {2011})}\BibitemShut {NoStop}%
\bibitem [{\citenamefont {Peng}\ \emph {et~al.}(2014)\citenamefont {Peng},
  \citenamefont {{\"O}zdemir}, \citenamefont {Lei}, \citenamefont {Monifi},
  \citenamefont {Gianfreda}, \citenamefont {Long}, \citenamefont {Fan},
  \citenamefont {Nori}, \citenamefont {Bender},\ and\ \citenamefont
  {Yang}}]{peng2014parity}%
  \BibitemOpen
  \bibfield  {author} {\bibinfo {author} {\bibfnamefont {B.}~\bibnamefont
  {Peng}}, \bibinfo {author} {\bibfnamefont {{\c{S}}.~K.}\ \bibnamefont
  {{\"O}zdemir}}, \bibinfo {author} {\bibfnamefont {F.}~\bibnamefont {Lei}},
  \bibinfo {author} {\bibfnamefont {F.}~\bibnamefont {Monifi}}, \bibinfo
  {author} {\bibfnamefont {M.}~\bibnamefont {Gianfreda}}, \bibinfo {author}
  {\bibfnamefont {G.~L.}\ \bibnamefont {Long}}, \bibinfo {author}
  {\bibfnamefont {S.}~\bibnamefont {Fan}}, \bibinfo {author} {\bibfnamefont
  {F.}~\bibnamefont {Nori}}, \bibinfo {author} {\bibfnamefont {C.~M.}\
  \bibnamefont {Bender}},\ and\ \bibinfo {author} {\bibfnamefont
  {L.}~\bibnamefont {Yang}},\ }\bibfield  {title} {\bibinfo {title}
  {Parity--time-symmetric whispering-gallery microcavities},\ }\href
  {https://www.nature.com/articles/nphys2927} {\bibfield  {journal} {\bibinfo
  {journal} {Nat. Phys.}\ }\textbf {\bibinfo {volume} {10}},\ \bibinfo {pages}
  {394} (\bibinfo {year} {2014})}\BibitemShut {NoStop}%
\bibitem [{\citenamefont {Chang}\ \emph {et~al.}(2014)\citenamefont {Chang},
  \citenamefont {Jiang}, \citenamefont {Hua}, \citenamefont {Yang},
  \citenamefont {Wen}, \citenamefont {Jiang}, \citenamefont {Li}, \citenamefont
  {Wang},\ and\ \citenamefont {Xiao}}]{chang2014parity}%
  \BibitemOpen
  \bibfield  {author} {\bibinfo {author} {\bibfnamefont {L.}~\bibnamefont
  {Chang}}, \bibinfo {author} {\bibfnamefont {X.}~\bibnamefont {Jiang}},
  \bibinfo {author} {\bibfnamefont {S.}~\bibnamefont {Hua}}, \bibinfo {author}
  {\bibfnamefont {C.}~\bibnamefont {Yang}}, \bibinfo {author} {\bibfnamefont
  {J.}~\bibnamefont {Wen}}, \bibinfo {author} {\bibfnamefont {L.}~\bibnamefont
  {Jiang}}, \bibinfo {author} {\bibfnamefont {G.}~\bibnamefont {Li}}, \bibinfo
  {author} {\bibfnamefont {G.}~\bibnamefont {Wang}},\ and\ \bibinfo {author}
  {\bibfnamefont {M.}~\bibnamefont {Xiao}},\ }\bibfield  {title} {\bibinfo
  {title} {Parity--time symmetry and variable optical isolation in
  active--passive-coupled microresonators},\ }\href
  {https://www.nature.com/articles/nphoton.2014.133} {\bibfield  {journal}
  {\bibinfo  {journal} {Nat. Photonics}\ }\textbf {\bibinfo {volume} {8}},\
  \bibinfo {pages} {524} (\bibinfo {year} {2014})}\BibitemShut {NoStop}%
\bibitem [{\citenamefont {Fan}\ \emph {et~al.}(2012)\citenamefont {Fan},
  \citenamefont {Wang}, \citenamefont {Varghese}, \citenamefont {Shen},
  \citenamefont {Niu}, \citenamefont {Xuan}, \citenamefont {Weiner},\ and\
  \citenamefont {Qi}}]{fan2012all}%
  \BibitemOpen
  \bibfield  {author} {\bibinfo {author} {\bibfnamefont {L.}~\bibnamefont
  {Fan}}, \bibinfo {author} {\bibfnamefont {J.}~\bibnamefont {Wang}}, \bibinfo
  {author} {\bibfnamefont {L.~T.}\ \bibnamefont {Varghese}}, \bibinfo {author}
  {\bibfnamefont {H.}~\bibnamefont {Shen}}, \bibinfo {author} {\bibfnamefont
  {B.}~\bibnamefont {Niu}}, \bibinfo {author} {\bibfnamefont {Y.}~\bibnamefont
  {Xuan}}, \bibinfo {author} {\bibfnamefont {A.~M.}\ \bibnamefont {Weiner}},\
  and\ \bibinfo {author} {\bibfnamefont {M.}~\bibnamefont {Qi}},\ }\bibfield
  {title} {\bibinfo {title} {An all-silicon passive optical diode},\ }\href
  {https://www.science.org/doi/10.1126/science.1214383} {\bibfield  {journal}
  {\bibinfo  {journal} {Science}\ }\textbf {\bibinfo {volume} {335}},\ \bibinfo
  {pages} {447} (\bibinfo {year} {2012})}\BibitemShut {NoStop}%
\bibitem [{\citenamefont {Cao}\ \emph {et~al.}(2017)\citenamefont {Cao},
  \citenamefont {Wang}, \citenamefont {Dong}, \citenamefont {Jing},
  \citenamefont {Liu}, \citenamefont {Chen}, \citenamefont {Ge}, \citenamefont
  {Gong},\ and\ \citenamefont {Xiao}}]{Qi-Tao2017}%
  \BibitemOpen
  \bibfield  {author} {\bibinfo {author} {\bibfnamefont {Q.-T.}\ \bibnamefont
  {Cao}}, \bibinfo {author} {\bibfnamefont {H.}~\bibnamefont {Wang}}, \bibinfo
  {author} {\bibfnamefont {C.-H.}\ \bibnamefont {Dong}}, \bibinfo {author}
  {\bibfnamefont {H.}~\bibnamefont {Jing}}, \bibinfo {author} {\bibfnamefont
  {R.-S.}\ \bibnamefont {Liu}}, \bibinfo {author} {\bibfnamefont
  {X.}~\bibnamefont {Chen}}, \bibinfo {author} {\bibfnamefont {L.}~\bibnamefont
  {Ge}}, \bibinfo {author} {\bibfnamefont {Q.}~\bibnamefont {Gong}},\ and\
  \bibinfo {author} {\bibfnamefont {Y.-F.}\ \bibnamefont {Xiao}},\ }\bibfield
  {title} {\bibinfo {title} {Experimental demonstration of spontaneous
  chirality in a nonlinear microresonator},\ }\href
  {https://doi.org/10.1103/PhysRevLett.118.033901} {\bibfield  {journal}
  {\bibinfo  {journal} {Phys. Rev. Lett.}\ }\textbf {\bibinfo {volume} {118}},\
  \bibinfo {pages} {033901} (\bibinfo {year} {2017})}\BibitemShut {NoStop}%
\bibitem [{\citenamefont {Manipatruni}\ \emph {et~al.}(2009)\citenamefont
  {Manipatruni}, \citenamefont {Robinson},\ and\ \citenamefont
  {Lipson}}]{Manipatruni2009}%
  \BibitemOpen
  \bibfield  {author} {\bibinfo {author} {\bibfnamefont {S.}~\bibnamefont
  {Manipatruni}}, \bibinfo {author} {\bibfnamefont {J.~T.}\ \bibnamefont
  {Robinson}},\ and\ \bibinfo {author} {\bibfnamefont {M.}~\bibnamefont
  {Lipson}},\ }\bibfield  {title} {\bibinfo {title} {Optical nonreciprocity in
  optomechanical structures},\ }\href
  {https://doi.org/10.1103/PhysRevLett.102.213903} {\bibfield  {journal}
  {\bibinfo  {journal} {Phys. Rev. Lett.}\ }\textbf {\bibinfo {volume} {102}},\
  \bibinfo {pages} {213903} (\bibinfo {year} {2009})}\BibitemShut {NoStop}%
\bibitem [{\citenamefont {Shen}\ \emph {et~al.}(2016)\citenamefont {Shen},
  \citenamefont {Zhang}, \citenamefont {Chen}, \citenamefont {Zou},
  \citenamefont {Xiao}, \citenamefont {Zou}, \citenamefont {Sun}, \citenamefont
  {Guo},\ and\ \citenamefont {Dong}}]{shen2016experimental}%
  \BibitemOpen
  \bibfield  {author} {\bibinfo {author} {\bibfnamefont {Z.}~\bibnamefont
  {Shen}}, \bibinfo {author} {\bibfnamefont {Y.-L.}\ \bibnamefont {Zhang}},
  \bibinfo {author} {\bibfnamefont {Y.}~\bibnamefont {Chen}}, \bibinfo {author}
  {\bibfnamefont {C.-L.}\ \bibnamefont {Zou}}, \bibinfo {author} {\bibfnamefont
  {Y.-F.}\ \bibnamefont {Xiao}}, \bibinfo {author} {\bibfnamefont {X.-B.}\
  \bibnamefont {Zou}}, \bibinfo {author} {\bibfnamefont {F.-W.}\ \bibnamefont
  {Sun}}, \bibinfo {author} {\bibfnamefont {G.-C.}\ \bibnamefont {Guo}},\ and\
  \bibinfo {author} {\bibfnamefont {C.-H.}\ \bibnamefont {Dong}},\ }\bibfield
  {title} {\bibinfo {title} {Experimental realization of optomechanically
  induced non-reciprocity},\ }\href
  {https://www.nature.com/articles/nphoton.2016.161} {\bibfield  {journal}
  {\bibinfo  {journal} {Nat. Photonics}\ }\textbf {\bibinfo {volume} {10}},\
  \bibinfo {pages} {657} (\bibinfo {year} {2016})}\BibitemShut {NoStop}%
\bibitem [{\citenamefont {Wang}\ \emph {et~al.}(2013)\citenamefont {Wang},
  \citenamefont {Zhou}, \citenamefont {Guo}, \citenamefont {Zhang},
  \citenamefont {Evers},\ and\ \citenamefont {Zhu}}]{Wang2013}%
  \BibitemOpen
  \bibfield  {author} {\bibinfo {author} {\bibfnamefont {D.-W.}\ \bibnamefont
  {Wang}}, \bibinfo {author} {\bibfnamefont {H.-T.}\ \bibnamefont {Zhou}},
  \bibinfo {author} {\bibfnamefont {M.-J.}\ \bibnamefont {Guo}}, \bibinfo
  {author} {\bibfnamefont {J.-X.}\ \bibnamefont {Zhang}}, \bibinfo {author}
  {\bibfnamefont {J.}~\bibnamefont {Evers}},\ and\ \bibinfo {author}
  {\bibfnamefont {S.-Y.}\ \bibnamefont {Zhu}},\ }\bibfield  {title} {\bibinfo
  {title} {Optical diode made from a moving photonic crystal},\ }\href
  {https://doi.org/10.1103/PhysRevLett.110.093901} {\bibfield  {journal}
  {\bibinfo  {journal} {Phys. Rev. Lett.}\ }\textbf {\bibinfo {volume} {110}},\
  \bibinfo {pages} {093901} (\bibinfo {year} {2013})}\BibitemShut {NoStop}%
\bibitem [{\citenamefont {Ramezani}\ \emph {et~al.}(2018)\citenamefont
  {Ramezani}, \citenamefont {Jha}, \citenamefont {Wang},\ and\ \citenamefont
  {Zhang}}]{Ramezani2018}%
  \BibitemOpen
  \bibfield  {author} {\bibinfo {author} {\bibfnamefont {H.}~\bibnamefont
  {Ramezani}}, \bibinfo {author} {\bibfnamefont {P.~K.}\ \bibnamefont {Jha}},
  \bibinfo {author} {\bibfnamefont {Y.}~\bibnamefont {Wang}},\ and\ \bibinfo
  {author} {\bibfnamefont {X.}~\bibnamefont {Zhang}},\ }\bibfield  {title}
  {\bibinfo {title} {Nonreciprocal localization of photons},\ }\href
  {https://doi.org/10.1103/PhysRevLett.120.043901} {\bibfield  {journal}
  {\bibinfo  {journal} {Phys. Rev. Lett.}\ }\textbf {\bibinfo {volume} {120}},\
  \bibinfo {pages} {043901} (\bibinfo {year} {2018})}\BibitemShut {NoStop}%
\bibitem [{\citenamefont {Zhang}\ \emph {et~al.}(2018)\citenamefont {Zhang},
  \citenamefont {Hu}, \citenamefont {Lin}, \citenamefont {Niu}, \citenamefont
  {Xia}, \citenamefont {Gong},\ and\ \citenamefont {Gong}}]{zhang2018thermal}%
  \BibitemOpen
  \bibfield  {author} {\bibinfo {author} {\bibfnamefont {S.}~\bibnamefont
  {Zhang}}, \bibinfo {author} {\bibfnamefont {Y.}~\bibnamefont {Hu}}, \bibinfo
  {author} {\bibfnamefont {G.}~\bibnamefont {Lin}}, \bibinfo {author}
  {\bibfnamefont {Y.}~\bibnamefont {Niu}}, \bibinfo {author} {\bibfnamefont
  {K.}~\bibnamefont {Xia}}, \bibinfo {author} {\bibfnamefont {J.}~\bibnamefont
  {Gong}},\ and\ \bibinfo {author} {\bibfnamefont {S.}~\bibnamefont {Gong}},\
  }\bibfield  {title} {\bibinfo {title} {Thermal-motion-induced non-reciprocal
  quantum optical system},\ }\href
  {https://www.nature.com/articles/s41566-018-0269-2} {\bibfield  {journal}
  {\bibinfo  {journal} {Nat. Photonics}\ }\textbf {\bibinfo {volume} {12}},\
  \bibinfo {pages} {744} (\bibinfo {year} {2018})}\BibitemShut {NoStop}%
\bibitem [{\citenamefont {Xia}\ \emph {et~al.}(2018)\citenamefont {Xia},
  \citenamefont {Nori},\ and\ \citenamefont {Xiao}}]{Keyu2018}%
  \BibitemOpen
  \bibfield  {author} {\bibinfo {author} {\bibfnamefont {K.}~\bibnamefont
  {Xia}}, \bibinfo {author} {\bibfnamefont {F.}~\bibnamefont {Nori}},\ and\
  \bibinfo {author} {\bibfnamefont {M.}~\bibnamefont {Xiao}},\ }\bibfield
  {title} {\bibinfo {title} {Cavity-free optical isolators and circulators
  using a chiral cross-\uppercase{K}err nonlinearity},\ }\href
  {https://doi.org/10.1103/PhysRevLett.121.203602} {\bibfield  {journal}
  {\bibinfo  {journal} {Phys. Rev. Lett.}\ }\textbf {\bibinfo {volume} {121}},\
  \bibinfo {pages} {203602} (\bibinfo {year} {2018})}\BibitemShut {NoStop}%
\bibitem [{\citenamefont {Tang}\ \emph {et~al.}(2022)\citenamefont {Tang},
  \citenamefont {Tang}, \citenamefont {Chen}, \citenamefont {Nori},
  \citenamefont {Xiao},\ and\ \citenamefont {Xia}}]{Tang2022}%
  \BibitemOpen
  \bibfield  {author} {\bibinfo {author} {\bibfnamefont {L.}~\bibnamefont
  {Tang}}, \bibinfo {author} {\bibfnamefont {J.}~\bibnamefont {Tang}}, \bibinfo
  {author} {\bibfnamefont {M.}~\bibnamefont {Chen}}, \bibinfo {author}
  {\bibfnamefont {F.}~\bibnamefont {Nori}}, \bibinfo {author} {\bibfnamefont
  {M.}~\bibnamefont {Xiao}},\ and\ \bibinfo {author} {\bibfnamefont
  {K.}~\bibnamefont {Xia}},\ }\bibfield  {title} {\bibinfo {title} {Quantum
  squeezing induced optical nonreciprocity},\ }\href
  {https://doi.org/10.1103/PhysRevLett.128.083604} {\bibfield  {journal}
  {\bibinfo  {journal} {Phys. Rev. Lett.}\ }\textbf {\bibinfo {volume} {128}},\
  \bibinfo {pages} {083604} (\bibinfo {year} {2022})}\BibitemShut {NoStop}%
\bibitem [{\citenamefont {Bennett}\ and\ \citenamefont
  {DiVincenzo}(2000)}]{bennett2000quantum}%
  \BibitemOpen
  \bibfield  {author} {\bibinfo {author} {\bibfnamefont {C.~H.}\ \bibnamefont
  {Bennett}}\ and\ \bibinfo {author} {\bibfnamefont {D.~P.}\ \bibnamefont
  {DiVincenzo}},\ }\bibfield  {title} {\bibinfo {title} {Quantum information
  and computation},\ }\href {https://www.nature.com/articles/35005001}
  {\bibfield  {journal} {\bibinfo  {journal} {Nature (London)}\ }\textbf
  {\bibinfo {volume} {404}},\ \bibinfo {pages} {247} (\bibinfo {year}
  {2000})}\BibitemShut {NoStop}%
\bibitem [{\citenamefont {Buluta}\ \emph {et~al.}(2011)\citenamefont {Buluta},
  \citenamefont {Ashhab},\ and\ \citenamefont {Nori}}]{Buluta_2011}%
  \BibitemOpen
  \bibfield  {author} {\bibinfo {author} {\bibfnamefont {I.}~\bibnamefont
  {Buluta}}, \bibinfo {author} {\bibfnamefont {S.}~\bibnamefont {Ashhab}},\
  and\ \bibinfo {author} {\bibfnamefont {F.}~\bibnamefont {Nori}},\ }\bibfield
  {title} {\bibinfo {title} {Natural and artificial atoms for quantum
  computation},\ }\href {https://doi.org/10.1088/0034-4885/74/10/104401}
  {\bibfield  {journal} {\bibinfo  {journal} {Rep. Prog. Phys.}\ }\textbf
  {\bibinfo {volume} {74}},\ \bibinfo {pages} {104401} (\bibinfo {year}
  {2011})}\BibitemShut {NoStop}%
\bibitem [{\citenamefont {Kimble}(2008)}]{kimble2008quantum}%
  \BibitemOpen
  \bibfield  {author} {\bibinfo {author} {\bibfnamefont {H.~J.}\ \bibnamefont
  {Kimble}},\ }\bibfield  {title} {\bibinfo {title} {The quantum internet},\
  }\href {https://www.nature.com/articles/nature07127} {\bibfield  {journal}
  {\bibinfo  {journal} {Nature (London)}\ }\textbf {\bibinfo {volume} {453}},\
  \bibinfo {pages} {1023} (\bibinfo {year} {2008})}\BibitemShut {NoStop}%
\bibitem [{\citenamefont {Fruchart}\ \emph {et~al.}(2021)\citenamefont
  {Fruchart}, \citenamefont {Hanai}, \citenamefont {Littlewood},\ and\
  \citenamefont {Vitelli}}]{fruchart2021non}%
  \BibitemOpen
  \bibfield  {author} {\bibinfo {author} {\bibfnamefont {M.}~\bibnamefont
  {Fruchart}}, \bibinfo {author} {\bibfnamefont {R.}~\bibnamefont {Hanai}},
  \bibinfo {author} {\bibfnamefont {P.~B.}\ \bibnamefont {Littlewood}},\ and\
  \bibinfo {author} {\bibfnamefont {V.}~\bibnamefont {Vitelli}},\ }\bibfield
  {title} {\bibinfo {title} {Non-reciprocal phase transitions},\ }\href
  {https://www.nature.com/articles/s41586-021-03375-9} {\bibfield  {journal}
  {\bibinfo  {journal} {Nature (London)}\ }\textbf {\bibinfo {volume} {592}},\
  \bibinfo {pages} {363} (\bibinfo {year} {2021})}\BibitemShut {NoStop}%
\bibitem [{\citenamefont {Chiacchio}\ \emph {et~al.}(2023)\citenamefont
  {Chiacchio}, \citenamefont {Nunnenkamp},\ and\ \citenamefont
  {Brunelli}}]{chiacchio2023}%
  \BibitemOpen
  \bibfield  {author} {\bibinfo {author} {\bibfnamefont {E.~I.~R.}\
  \bibnamefont {Chiacchio}}, \bibinfo {author} {\bibfnamefont {A.}~\bibnamefont
  {Nunnenkamp}},\ and\ \bibinfo {author} {\bibfnamefont {M.}~\bibnamefont
  {Brunelli}},\ }\bibfield  {title} {\bibinfo {title} {Nonreciprocal
  \uppercase{D}icke model},\ }\href
  {https://doi.org/10.1103/PhysRevLett.131.113602} {\bibfield  {journal}
  {\bibinfo  {journal} {Phys. Rev. Lett.}\ }\textbf {\bibinfo {volume} {131}},\
  \bibinfo {pages} {113602} (\bibinfo {year} {2023})}\BibitemShut {NoStop}%
\bibitem [{\citenamefont {Ferri}\ \emph {et~al.}(2021)\citenamefont {Ferri},
  \citenamefont {Rosa-Medina}, \citenamefont {Finger}, \citenamefont {Dogra},
  \citenamefont {Soriente}, \citenamefont {Zilberberg}, \citenamefont
  {Donner},\ and\ \citenamefont {Esslinger}}]{Ferri2021}%
  \BibitemOpen
  \bibfield  {author} {\bibinfo {author} {\bibfnamefont {F.}~\bibnamefont
  {Ferri}}, \bibinfo {author} {\bibfnamefont {R.}~\bibnamefont {Rosa-Medina}},
  \bibinfo {author} {\bibfnamefont {F.}~\bibnamefont {Finger}}, \bibinfo
  {author} {\bibfnamefont {N.}~\bibnamefont {Dogra}}, \bibinfo {author}
  {\bibfnamefont {M.}~\bibnamefont {Soriente}}, \bibinfo {author}
  {\bibfnamefont {O.}~\bibnamefont {Zilberberg}}, \bibinfo {author}
  {\bibfnamefont {T.}~\bibnamefont {Donner}},\ and\ \bibinfo {author}
  {\bibfnamefont {T.}~\bibnamefont {Esslinger}},\ }\bibfield  {title} {\bibinfo
  {title} {Emerging dissipative phases in a superradiant quantum gas with
  tunable decay},\ }\href {https://doi.org/10.1103/PhysRevX.11.041046}
  {\bibfield  {journal} {\bibinfo  {journal} {Phys. Rev. X}\ }\textbf {\bibinfo
  {volume} {11}},\ \bibinfo {pages} {041046} (\bibinfo {year}
  {2021})}\BibitemShut {NoStop}%
\bibitem [{\citenamefont {Klinder}\ \emph {et~al.}(2015)\citenamefont
  {Klinder}, \citenamefont {Ke{\ss}ler}, \citenamefont {Wolke}, \citenamefont
  {Mathey},\ and\ \citenamefont {Hemmerich}}]{klinder2015dynamical}%
  \BibitemOpen
  \bibfield  {author} {\bibinfo {author} {\bibfnamefont {J.}~\bibnamefont
  {Klinder}}, \bibinfo {author} {\bibfnamefont {H.}~\bibnamefont {Ke{\ss}ler}},
  \bibinfo {author} {\bibfnamefont {M.}~\bibnamefont {Wolke}}, \bibinfo
  {author} {\bibfnamefont {L.}~\bibnamefont {Mathey}},\ and\ \bibinfo {author}
  {\bibfnamefont {A.}~\bibnamefont {Hemmerich}},\ }\bibfield  {title} {\bibinfo
  {title} {Dynamical phase transition in the open \uppercase{D}icke model},\
  }\href {https://www.pnas.org/doi/10.1073/pnas.1417132112} {\bibfield
  {journal} {\bibinfo  {journal} {Proc. Natl. Acad. Sci. U. S. A.}\ }\textbf
  {\bibinfo {volume} {112}},\ \bibinfo {pages} {3290} (\bibinfo {year}
  {2015})}\BibitemShut {NoStop}%
\bibitem [{\citenamefont {Zhiqiang}\ \emph {et~al.}(2017)\citenamefont
  {Zhiqiang}, \citenamefont {Lee}, \citenamefont {Kumar}, \citenamefont
  {Arnold}, \citenamefont {Masson}, \citenamefont {Parkins},\ and\
  \citenamefont {Barrett}}]{zhiqiang2017nonequilibrium}%
  \BibitemOpen
  \bibfield  {author} {\bibinfo {author} {\bibfnamefont {Z.}~\bibnamefont
  {Zhiqiang}}, \bibinfo {author} {\bibfnamefont {C.~H.}\ \bibnamefont {Lee}},
  \bibinfo {author} {\bibfnamefont {R.}~\bibnamefont {Kumar}}, \bibinfo
  {author} {\bibfnamefont {K.}~\bibnamefont {Arnold}}, \bibinfo {author}
  {\bibfnamefont {S.~J.}\ \bibnamefont {Masson}}, \bibinfo {author}
  {\bibfnamefont {A.}~\bibnamefont {Parkins}},\ and\ \bibinfo {author}
  {\bibfnamefont {M.}~\bibnamefont {Barrett}},\ }\bibfield  {title} {\bibinfo
  {title} {Nonequilibrium phase transition in a spin-1 \uppercase{D}icke
  model},\ }\href
  {https://opg.optica.org/optica/fulltext.cfm?uri=optica-4-4-424&id=362715}
  {\bibfield  {journal} {\bibinfo  {journal} {Optica}\ }\textbf {\bibinfo
  {volume} {4}},\ \bibinfo {pages} {424} (\bibinfo {year} {2017})}\BibitemShut
  {NoStop}%
\bibitem [{\citenamefont {Feng}\ \emph {et~al.}(2015)\citenamefont {Feng},
  \citenamefont {Zhong}, \citenamefont {Liu}, \citenamefont {Yan},
  \citenamefont {Yang}, \citenamefont {Twamley},\ and\ \citenamefont
  {Wang}}]{feng2015exploring}%
  \BibitemOpen
  \bibfield  {author} {\bibinfo {author} {\bibfnamefont {M.}~\bibnamefont
  {Feng}}, \bibinfo {author} {\bibfnamefont {Y.}~\bibnamefont {Zhong}},
  \bibinfo {author} {\bibfnamefont {T.}~\bibnamefont {Liu}}, \bibinfo {author}
  {\bibfnamefont {L.}~\bibnamefont {Yan}}, \bibinfo {author} {\bibfnamefont
  {W.}~\bibnamefont {Yang}}, \bibinfo {author} {\bibfnamefont {J.}~\bibnamefont
  {Twamley}},\ and\ \bibinfo {author} {\bibfnamefont {H.}~\bibnamefont
  {Wang}},\ }\bibfield  {title} {\bibinfo {title} {Exploring the quantum
  critical behaviour in a driven \uppercase{T}avis--\uppercase{C}ummings
  circuit},\ }\href {https://www.nature.com/articles/ncomms8111} {\bibfield
  {journal} {\bibinfo  {journal} {Nat. Commun.}\ }\textbf {\bibinfo {volume}
  {6}},\ \bibinfo {pages} {7111} (\bibinfo {year} {2015})}\BibitemShut
  {NoStop}%
\bibitem [{\citenamefont {Garbe}\ \emph {et~al.}(2020)\citenamefont {Garbe},
  \citenamefont {Bina}, \citenamefont {Keller}, \citenamefont {Paris},\ and\
  \citenamefont {Felicetti}}]{Garbe2020}%
  \BibitemOpen
  \bibfield  {author} {\bibinfo {author} {\bibfnamefont {L.}~\bibnamefont
  {Garbe}}, \bibinfo {author} {\bibfnamefont {M.}~\bibnamefont {Bina}},
  \bibinfo {author} {\bibfnamefont {A.}~\bibnamefont {Keller}}, \bibinfo
  {author} {\bibfnamefont {M.~G.~A.}\ \bibnamefont {Paris}},\ and\ \bibinfo
  {author} {\bibfnamefont {S.}~\bibnamefont {Felicetti}},\ }\bibfield  {title}
  {\bibinfo {title} {Critical quantum metrology with a finite-component quantum
  phase transition},\ }\href {https://doi.org/10.1103/PhysRevLett.124.120504}
  {\bibfield  {journal} {\bibinfo  {journal} {Phys. Rev. Lett.}\ }\textbf
  {\bibinfo {volume} {124}},\ \bibinfo {pages} {120504} (\bibinfo {year}
  {2020})}\BibitemShut {NoStop}%
\bibitem [{\citenamefont {Chu}\ \emph {et~al.}(2021)\citenamefont {Chu},
  \citenamefont {Zhang}, \citenamefont {Yu},\ and\ \citenamefont
  {Cai}}]{Chu2021}%
  \BibitemOpen
  \bibfield  {author} {\bibinfo {author} {\bibfnamefont {Y.}~\bibnamefont
  {Chu}}, \bibinfo {author} {\bibfnamefont {S.}~\bibnamefont {Zhang}}, \bibinfo
  {author} {\bibfnamefont {B.}~\bibnamefont {Yu}},\ and\ \bibinfo {author}
  {\bibfnamefont {J.}~\bibnamefont {Cai}},\ }\bibfield  {title} {\bibinfo
  {title} {Dynamic framework for criticality-enhanced quantum sensing},\ }\href
  {https://doi.org/10.1103/PhysRevLett.126.010502} {\bibfield  {journal}
  {\bibinfo  {journal} {Phys. Rev. Lett.}\ }\textbf {\bibinfo {volume} {126}},\
  \bibinfo {pages} {010502} (\bibinfo {year} {2021})}\BibitemShut {NoStop}%
\bibitem [{\citenamefont {Ying}\ \emph {et~al.}(2022)\citenamefont {Ying},
  \citenamefont {Felicetti}, \citenamefont {Liu},\ and\ \citenamefont
  {Braak}}]{YingZu-Jian2022}%
  \BibitemOpen
  \bibfield  {author} {\bibinfo {author} {\bibfnamefont {Z.-J.}\ \bibnamefont
  {Ying}}, \bibinfo {author} {\bibfnamefont {S.}~\bibnamefont {Felicetti}},
  \bibinfo {author} {\bibfnamefont {G.}~\bibnamefont {Liu}},\ and\ \bibinfo
  {author} {\bibfnamefont {D.}~\bibnamefont {Braak}},\ }\bibfield  {title}
  {\bibinfo {title} {Critical quantum metrology in the non-linear quantum
  \uppercase{R}abi model},\ }\href {https://www.mdpi.com/1099-4300/24/8/1015}
  {\bibfield  {journal} {\bibinfo  {journal} {Entropy}\ }\textbf {\bibinfo
  {volume} {24}},\ \bibinfo {pages} {1015} (\bibinfo {year}
  {2022})}\BibitemShut {NoStop}%
\bibitem [{\citenamefont {He}\ \emph {et~al.}(2023)\citenamefont {He},
  \citenamefont {Lu}, \citenamefont {Yao}, \citenamefont {Zhu},\ and\
  \citenamefont {Ai}}]{he2023criticality}%
  \BibitemOpen
  \bibfield  {author} {\bibinfo {author} {\bibfnamefont {W.-T.}\ \bibnamefont
  {He}}, \bibinfo {author} {\bibfnamefont {C.-W.}\ \bibnamefont {Lu}}, \bibinfo
  {author} {\bibfnamefont {Y.-X.}\ \bibnamefont {Yao}}, \bibinfo {author}
  {\bibfnamefont {H.-Y.}\ \bibnamefont {Zhu}},\ and\ \bibinfo {author}
  {\bibfnamefont {Q.}~\bibnamefont {Ai}},\ }\bibfield  {title} {\bibinfo
  {title} {Criticality-based quantum metrology in the presence of
  decoherence},\ }\href
  {https://link.springer.com/article/10.1007/s11467-023-1278-2} {\bibfield
  {journal} {\bibinfo  {journal} {Front. Phys.}\ }\textbf {\bibinfo {volume}
  {18}},\ \bibinfo {pages} {31304} (\bibinfo {year} {2023})}\BibitemShut
  {NoStop}%
\bibitem [{\citenamefont {Aoki}\ \emph {et~al.}(2006)\citenamefont {Aoki},
  \citenamefont {Dayan}, \citenamefont {Wilcut}, \citenamefont {Bowen},
  \citenamefont {Parkins}, \citenamefont {Kippenberg}, \citenamefont {Vahala},\
  and\ \citenamefont {Kimble}}]{aoki2006observation}%
  \BibitemOpen
  \bibfield  {author} {\bibinfo {author} {\bibfnamefont {T.}~\bibnamefont
  {Aoki}}, \bibinfo {author} {\bibfnamefont {B.}~\bibnamefont {Dayan}},
  \bibinfo {author} {\bibfnamefont {E.}~\bibnamefont {Wilcut}}, \bibinfo
  {author} {\bibfnamefont {W.~P.}\ \bibnamefont {Bowen}}, \bibinfo {author}
  {\bibfnamefont {A.~S.}\ \bibnamefont {Parkins}}, \bibinfo {author}
  {\bibfnamefont {T.}~\bibnamefont {Kippenberg}}, \bibinfo {author}
  {\bibfnamefont {K.}~\bibnamefont {Vahala}},\ and\ \bibinfo {author}
  {\bibfnamefont {H.}~\bibnamefont {Kimble}},\ }\bibfield  {title} {\bibinfo
  {title} {Observation of strong coupling between one atom and a monolithic
  microresonator},\ }\href {https://www.nature.com/articles/nature05147}
  {\bibfield  {journal} {\bibinfo  {journal} {Nature}\ }\textbf {\bibinfo
  {volume} {443}},\ \bibinfo {pages} {671} (\bibinfo {year}
  {2006})}\BibitemShut {NoStop}%
\bibitem [{\citenamefont {Alton}\ \emph {et~al.}(2011)\citenamefont {Alton},
  \citenamefont {Stern}, \citenamefont {Aoki}, \citenamefont {Lee},
  \citenamefont {Ostby}, \citenamefont {Vahala},\ and\ \citenamefont
  {Kimble}}]{alton2011strong}%
  \BibitemOpen
  \bibfield  {author} {\bibinfo {author} {\bibfnamefont {D.}~\bibnamefont
  {Alton}}, \bibinfo {author} {\bibfnamefont {N.}~\bibnamefont {Stern}},
  \bibinfo {author} {\bibfnamefont {T.}~\bibnamefont {Aoki}}, \bibinfo {author}
  {\bibfnamefont {H.}~\bibnamefont {Lee}}, \bibinfo {author} {\bibfnamefont
  {E.}~\bibnamefont {Ostby}}, \bibinfo {author} {\bibfnamefont
  {K.}~\bibnamefont {Vahala}},\ and\ \bibinfo {author} {\bibfnamefont
  {H.}~\bibnamefont {Kimble}},\ }\bibfield  {title} {\bibinfo {title} {Strong
  interactions of single atoms and photons near a dielectric boundary},\ }\href
  {https://www.nature.com/articles/nphys1837} {\bibfield  {journal} {\bibinfo
  {journal} {Nat. Phys.}\ }\textbf {\bibinfo {volume} {7}},\ \bibinfo {pages}
  {159} (\bibinfo {year} {2011})}\BibitemShut {NoStop}%
\bibitem [{\citenamefont {Junge}\ \emph {et~al.}(2013)\citenamefont {Junge},
  \citenamefont {O'Shea}, \citenamefont {Volz},\ and\ \citenamefont
  {Rauschenbeutel}}]{Junge2013}%
  \BibitemOpen
  \bibfield  {author} {\bibinfo {author} {\bibfnamefont {C.}~\bibnamefont
  {Junge}}, \bibinfo {author} {\bibfnamefont {D.}~\bibnamefont {O'Shea}},
  \bibinfo {author} {\bibfnamefont {J.}~\bibnamefont {Volz}},\ and\ \bibinfo
  {author} {\bibfnamefont {A.}~\bibnamefont {Rauschenbeutel}},\ }\bibfield
  {title} {\bibinfo {title} {Strong coupling between single atoms and
  nontransversal photons},\ }\href
  {https://doi.org/10.1103/PhysRevLett.110.213604} {\bibfield  {journal}
  {\bibinfo  {journal} {Phys. Rev. Lett.}\ }\textbf {\bibinfo {volume} {110}},\
  \bibinfo {pages} {213604} (\bibinfo {year} {2013})}\BibitemShut {NoStop}%
\bibitem [{\citenamefont {Shomroni}\ \emph {et~al.}(2014)\citenamefont
  {Shomroni}, \citenamefont {Rosenblum}, \citenamefont {Lovsky}, \citenamefont
  {Bechler}, \citenamefont {Guendelman},\ and\ \citenamefont
  {Dayan}}]{shomroni2014all}%
  \BibitemOpen
  \bibfield  {author} {\bibinfo {author} {\bibfnamefont {I.}~\bibnamefont
  {Shomroni}}, \bibinfo {author} {\bibfnamefont {S.}~\bibnamefont {Rosenblum}},
  \bibinfo {author} {\bibfnamefont {Y.}~\bibnamefont {Lovsky}}, \bibinfo
  {author} {\bibfnamefont {O.}~\bibnamefont {Bechler}}, \bibinfo {author}
  {\bibfnamefont {G.}~\bibnamefont {Guendelman}},\ and\ \bibinfo {author}
  {\bibfnamefont {B.}~\bibnamefont {Dayan}},\ }\bibfield  {title} {\bibinfo
  {title} {All-optical routing of single photons by a one-atom switch
  controlled by a single photon},\ }\href
  {https://www.science.org/doi/10.1126/science.1254699} {\bibfield  {journal}
  {\bibinfo  {journal} {Science}\ }\textbf {\bibinfo {volume} {345}},\ \bibinfo
  {pages} {903} (\bibinfo {year} {2014})}\BibitemShut {NoStop}%
\bibitem [{\citenamefont {Scheucher}\ \emph {et~al.}(2016)\citenamefont
  {Scheucher}, \citenamefont {Hilico}, \citenamefont {Will}, \citenamefont
  {Volz},\ and\ \citenamefont {Rauschenbeutel}}]{scheucher2016quantum}%
  \BibitemOpen
  \bibfield  {author} {\bibinfo {author} {\bibfnamefont {M.}~\bibnamefont
  {Scheucher}}, \bibinfo {author} {\bibfnamefont {A.}~\bibnamefont {Hilico}},
  \bibinfo {author} {\bibfnamefont {E.}~\bibnamefont {Will}}, \bibinfo {author}
  {\bibfnamefont {J.}~\bibnamefont {Volz}},\ and\ \bibinfo {author}
  {\bibfnamefont {A.}~\bibnamefont {Rauschenbeutel}},\ }\bibfield  {title}
  {\bibinfo {title} {Quantum optical circulator controlled by a single chirally
  coupled atom},\ }\href {https://www.science.org/doi/10.1126/science.aaj2118}
  {\bibfield  {journal} {\bibinfo  {journal} {Science}\ }\textbf {\bibinfo
  {volume} {354}},\ \bibinfo {pages} {1577} (\bibinfo {year}
  {2016})}\BibitemShut {NoStop}%
\bibitem [{\citenamefont {Bechler}\ \emph {et~al.}(2018)\citenamefont
  {Bechler}, \citenamefont {Borne}, \citenamefont {Rosenblum}, \citenamefont
  {Guendelman}, \citenamefont {Mor}, \citenamefont {Netser}, \citenamefont
  {Ohana}, \citenamefont {Aqua}, \citenamefont {Drucker}, \citenamefont
  {Finkelstein}, \citenamefont {Lovsky}, \citenamefont {Bruch}, \citenamefont
  {Gurovich}, \citenamefont {Shafir},\ and\ \citenamefont
  {Dayan}}]{bechler2018passive}%
  \BibitemOpen
  \bibfield  {author} {\bibinfo {author} {\bibfnamefont {O.}~\bibnamefont
  {Bechler}}, \bibinfo {author} {\bibfnamefont {A.}~\bibnamefont {Borne}},
  \bibinfo {author} {\bibfnamefont {S.}~\bibnamefont {Rosenblum}}, \bibinfo
  {author} {\bibfnamefont {G.}~\bibnamefont {Guendelman}}, \bibinfo {author}
  {\bibfnamefont {O.~E.}\ \bibnamefont {Mor}}, \bibinfo {author} {\bibfnamefont
  {M.}~\bibnamefont {Netser}}, \bibinfo {author} {\bibfnamefont
  {T.}~\bibnamefont {Ohana}}, \bibinfo {author} {\bibfnamefont
  {Z.}~\bibnamefont {Aqua}}, \bibinfo {author} {\bibfnamefont {N.}~\bibnamefont
  {Drucker}}, \bibinfo {author} {\bibfnamefont {R.}~\bibnamefont
  {Finkelstein}}, \bibinfo {author} {\bibfnamefont {Y.}~\bibnamefont {Lovsky}},
  \bibinfo {author} {\bibfnamefont {R.}~\bibnamefont {Bruch}}, \bibinfo
  {author} {\bibfnamefont {D.}~\bibnamefont {Gurovich}}, \bibinfo {author}
  {\bibfnamefont {E.}~\bibnamefont {Shafir}},\ and\ \bibinfo {author}
  {\bibfnamefont {B.}~\bibnamefont {Dayan}},\ }\bibfield  {title} {\bibinfo
  {title} {A passive photon--atom qubit swap operation},\ }\href
  {https://www.nature.com/articles/s41567-018-0241-6} {\bibfield  {journal}
  {\bibinfo  {journal} {Nat. Phys.}\ }\textbf {\bibinfo {volume} {14}},\
  \bibinfo {pages} {996} (\bibinfo {year} {2018})}\BibitemShut {NoStop}%
\bibitem [{\citenamefont {Will}\ \emph {et~al.}(2021)\citenamefont {Will},
  \citenamefont {Masters}, \citenamefont {Rauschenbeutel}, \citenamefont
  {Scheucher},\ and\ \citenamefont {Volz}}]{Will2021}%
  \BibitemOpen
  \bibfield  {author} {\bibinfo {author} {\bibfnamefont {E.}~\bibnamefont
  {Will}}, \bibinfo {author} {\bibfnamefont {L.}~\bibnamefont {Masters}},
  \bibinfo {author} {\bibfnamefont {A.}~\bibnamefont {Rauschenbeutel}},
  \bibinfo {author} {\bibfnamefont {M.}~\bibnamefont {Scheucher}},\ and\
  \bibinfo {author} {\bibfnamefont {J.}~\bibnamefont {Volz}},\ }\bibfield
  {title} {\bibinfo {title} {Coupling a single trapped atom to a
  whispering-gallery-mode microresonator},\ }\href
  {https://doi.org/10.1103/PhysRevLett.126.233602} {\bibfield  {journal}
  {\bibinfo  {journal} {Phys. Rev. Lett.}\ }\textbf {\bibinfo {volume} {126}},\
  \bibinfo {pages} {233602} (\bibinfo {year} {2021})}\BibitemShut {NoStop}%
\bibitem [{\citenamefont {Maayani}\ \emph {et~al.}(2018)\citenamefont
  {Maayani}, \citenamefont {Dahan}, \citenamefont {Kligerman}, \citenamefont
  {Moses}, \citenamefont {Hassan}, \citenamefont {Jing}, \citenamefont {Nori},
  \citenamefont {Christodoulides},\ and\ \citenamefont
  {Carmon}}]{maayani2018flying}%
  \BibitemOpen
  \bibfield  {author} {\bibinfo {author} {\bibfnamefont {S.}~\bibnamefont
  {Maayani}}, \bibinfo {author} {\bibfnamefont {R.}~\bibnamefont {Dahan}},
  \bibinfo {author} {\bibfnamefont {Y.}~\bibnamefont {Kligerman}}, \bibinfo
  {author} {\bibfnamefont {E.}~\bibnamefont {Moses}}, \bibinfo {author}
  {\bibfnamefont {A.~U.}\ \bibnamefont {Hassan}}, \bibinfo {author}
  {\bibfnamefont {H.}~\bibnamefont {Jing}}, \bibinfo {author} {\bibfnamefont
  {F.}~\bibnamefont {Nori}}, \bibinfo {author} {\bibfnamefont {D.~N.}\
  \bibnamefont {Christodoulides}},\ and\ \bibinfo {author} {\bibfnamefont
  {T.}~\bibnamefont {Carmon}},\ }\bibfield  {title} {\bibinfo {title} {Flying
  couplers above spinning resonators generate irreversible refraction},\ }\href
  {https://www.nature.com/articles/s41586-018-0245-5} {\bibfield  {journal}
  {\bibinfo  {journal} {Nature (London)}\ }\textbf {\bibinfo {volume} {558}},\
  \bibinfo {pages} {569} (\bibinfo {year} {2018})}\BibitemShut {NoStop}%
\bibitem [{\citenamefont {Bliokh}\ and\ \citenamefont
  {Nori}(2015)}]{bliokh2015transverse}%
  \BibitemOpen
  \bibfield  {author} {\bibinfo {author} {\bibfnamefont {K.~Y.}\ \bibnamefont
  {Bliokh}}\ and\ \bibinfo {author} {\bibfnamefont {F.}~\bibnamefont {Nori}},\
  }\bibfield  {title} {\bibinfo {title} {Transverse and longitudinal angular
  momenta of light},\ }\href
  {https://www.sciencedirect.com/science/article/pii/S0370157315003336}
  {\bibfield  {journal} {\bibinfo  {journal} {Phys. Rep.}\ }\textbf {\bibinfo
  {volume} {592}},\ \bibinfo {pages} {1} (\bibinfo {year} {2015})}\BibitemShut
  {NoStop}%
\bibitem [{\citenamefont {Bliokh}\ \emph {et~al.}(2015)\citenamefont {Bliokh},
  \citenamefont {Smirnova},\ and\ \citenamefont {Nori}}]{bliokh2015quantum}%
  \BibitemOpen
  \bibfield  {author} {\bibinfo {author} {\bibfnamefont {K.~Y.}\ \bibnamefont
  {Bliokh}}, \bibinfo {author} {\bibfnamefont {D.}~\bibnamefont {Smirnova}},\
  and\ \bibinfo {author} {\bibfnamefont {F.}~\bibnamefont {Nori}},\ }\bibfield
  {title} {\bibinfo {title} {Quantum spin hall effect of light},\ }\href
  {https://www.science.org/doi/full/10.1126/science.aaa9519} {\bibfield
  {journal} {\bibinfo  {journal} {Science}\ }\textbf {\bibinfo {volume}
  {348}},\ \bibinfo {pages} {1448} (\bibinfo {year} {2015})}\BibitemShut
  {NoStop}%
\bibitem [{SM()}]{SM}%
  \BibitemOpen
  \href@noop {} {}\bibinfo {note} {See Supplemental Material at [url], which
  includes Refs. [88–109], for additional details on the discussions of phase
  transitions, numerical simulations, and potential experimental
  implementations.}\BibitemShut {Stop}%
\bibitem [{\citenamefont {Ashhab}\ and\ \citenamefont
  {Nori}(2010)}]{Ashhab2010}%
  \BibitemOpen
  \bibfield  {author} {\bibinfo {author} {\bibfnamefont {S.}~\bibnamefont
  {Ashhab}}\ and\ \bibinfo {author} {\bibfnamefont {F.}~\bibnamefont {Nori}},\
  }\bibfield  {title} {\bibinfo {title} {Qubit-oscillator systems in the
  ultrastrong-coupling regime and their potential for preparing nonclassical
  states},\ }\href {https://doi.org/10.1103/PhysRevA.81.042311} {\bibfield
  {journal} {\bibinfo  {journal} {Phys. Rev. A}\ }\textbf {\bibinfo {volume}
  {81}},\ \bibinfo {pages} {042311} (\bibinfo {year} {2010})}\BibitemShut
  {NoStop}%
\bibitem [{\citenamefont {Ashhab}(2013)}]{Ashhab2013}%
  \BibitemOpen
  \bibfield  {author} {\bibinfo {author} {\bibfnamefont {S.}~\bibnamefont
  {Ashhab}},\ }\bibfield  {title} {\bibinfo {title} {Superradiance transition
  in a system with a single qubit and a single oscillator},\ }\href
  {https://doi.org/10.1103/PhysRevA.87.013826} {\bibfield  {journal} {\bibinfo
  {journal} {Phys. Rev. A}\ }\textbf {\bibinfo {volume} {87}},\ \bibinfo
  {pages} {013826} (\bibinfo {year} {2013})}\BibitemShut {NoStop}%
\bibitem [{\citenamefont {Puebla}\ \emph {et~al.}(2016)\citenamefont {Puebla},
  \citenamefont {Hwang},\ and\ \citenamefont {Plenio}}]{Ricardo2016}%
  \BibitemOpen
  \bibfield  {author} {\bibinfo {author} {\bibfnamefont {R.}~\bibnamefont
  {Puebla}}, \bibinfo {author} {\bibfnamefont {M.-J.}\ \bibnamefont {Hwang}},\
  and\ \bibinfo {author} {\bibfnamefont {M.~B.}\ \bibnamefont {Plenio}},\
  }\bibfield  {title} {\bibinfo {title} {Excited-state quantum phase transition
  in the \uppercase{R}abi model},\ }\href
  {https://doi.org/10.1103/PhysRevA.94.023835} {\bibfield  {journal} {\bibinfo
  {journal} {Phys. Rev. A}\ }\textbf {\bibinfo {volume} {94}},\ \bibinfo
  {pages} {023835} (\bibinfo {year} {2016})}\BibitemShut {NoStop}%
\bibitem [{\citenamefont {Wang}\ \emph {et~al.}(2009)\citenamefont {Wang},
  \citenamefont {Chong}, \citenamefont {Joannopoulos},\ and\ \citenamefont
  {Solja{\v{c}}i{\'c}}}]{wang2009observation}%
  \BibitemOpen
  \bibfield  {author} {\bibinfo {author} {\bibfnamefont {Z.}~\bibnamefont
  {Wang}}, \bibinfo {author} {\bibfnamefont {Y.}~\bibnamefont {Chong}},
  \bibinfo {author} {\bibfnamefont {J.~D.}\ \bibnamefont {Joannopoulos}},\ and\
  \bibinfo {author} {\bibfnamefont {M.}~\bibnamefont {Solja{\v{c}}i{\'c}}},\
  }\bibfield  {title} {\bibinfo {title} {Observation of unidirectional
  backscattering-immune topological electromagnetic states},\ }\href
  {https://www.nature.com/articles/nature08293} {\bibfield  {journal} {\bibinfo
   {journal} {Nature}\ }\textbf {\bibinfo {volume} {461}},\ \bibinfo {pages}
  {772} (\bibinfo {year} {2009})}\BibitemShut {NoStop}%
\bibitem [{\citenamefont {Khanikaev}\ \emph {et~al.}(2010)\citenamefont
  {Khanikaev}, \citenamefont {Mousavi}, \citenamefont {Shvets},\ and\
  \citenamefont {Kivshar}}]{Khanikaev2010}%
  \BibitemOpen
  \bibfield  {author} {\bibinfo {author} {\bibfnamefont {A.~B.}\ \bibnamefont
  {Khanikaev}}, \bibinfo {author} {\bibfnamefont {S.~H.}\ \bibnamefont
  {Mousavi}}, \bibinfo {author} {\bibfnamefont {G.}~\bibnamefont {Shvets}},\
  and\ \bibinfo {author} {\bibfnamefont {Y.~S.}\ \bibnamefont {Kivshar}},\
  }\bibfield  {title} {\bibinfo {title} {One-way extraordinary optical
  transmission and nonreciprocal spoof plasmons},\ }\href
  {https://doi.org/10.1103/PhysRevLett.105.126804} {\bibfield  {journal}
  {\bibinfo  {journal} {Phys. Rev. Lett.}\ }\textbf {\bibinfo {volume} {105}},\
  \bibinfo {pages} {126804} (\bibinfo {year} {2010})}\BibitemShut {NoStop}%
\bibitem [{\citenamefont {Dai}\ \emph {et~al.}(2012)\citenamefont {Dai},
  \citenamefont {Bauters},\ and\ \citenamefont {Bowers}}]{Dai2012}%
  \BibitemOpen
  \bibfield  {author} {\bibinfo {author} {\bibfnamefont {D.}~\bibnamefont
  {Dai}}, \bibinfo {author} {\bibfnamefont {J.}~\bibnamefont {Bauters}},\ and\
  \bibinfo {author} {\bibfnamefont {J.~E.}\ \bibnamefont {Bowers}},\ }\bibfield
   {title} {\bibinfo {title} {Passive technologies for future large-scale
  photonic integrated circuits on silicon: polarization handling, light
  non-reciprocity and loss reduction},\ }\href
  {https://www.nature.com/articles/lsa20121} {\bibfield  {journal} {\bibinfo
  {journal} {Light: Science \& Applications}\ }\textbf {\bibinfo {volume}
  {1}},\ \bibinfo {pages} {e1} (\bibinfo {year} {2012})}\BibitemShut {NoStop}%
\bibitem [{\citenamefont {Kurizki}\ \emph {et~al.}(2015)\citenamefont
  {Kurizki}, \citenamefont {Bertet}, \citenamefont {Kubo}, \citenamefont
  {M{\o}lmer}, \citenamefont {Petrosyan}, \citenamefont {Rabl},\ and\
  \citenamefont {Schmiedmayer}}]{Kurizki2015}%
  \BibitemOpen
  \bibfield  {author} {\bibinfo {author} {\bibfnamefont {G.}~\bibnamefont
  {Kurizki}}, \bibinfo {author} {\bibfnamefont {P.}~\bibnamefont {Bertet}},
  \bibinfo {author} {\bibfnamefont {Y.}~\bibnamefont {Kubo}}, \bibinfo {author}
  {\bibfnamefont {K.}~\bibnamefont {M{\o}lmer}}, \bibinfo {author}
  {\bibfnamefont {D.}~\bibnamefont {Petrosyan}}, \bibinfo {author}
  {\bibfnamefont {P.}~\bibnamefont {Rabl}},\ and\ \bibinfo {author}
  {\bibfnamefont {J.}~\bibnamefont {Schmiedmayer}},\ }\bibfield  {title}
  {\bibinfo {title} {Quantum technologies with hybrid systems},\ }\href
  {https://www.pnas.org/doi/abs/10.1073/pnas.1419326112} {\bibfield  {journal}
  {\bibinfo  {journal} {Proc. Natl. Acad. Sci. U. S. A.}\ }\textbf {\bibinfo
  {volume} {112}},\ \bibinfo {pages} {3866} (\bibinfo {year}
  {2015})}\BibitemShut {NoStop}%
\bibitem [{\citenamefont {Ward}\ and\ \citenamefont {Benson}(2011)}]{Ward2011}%
  \BibitemOpen
  \bibfield  {author} {\bibinfo {author} {\bibfnamefont {J.}~\bibnamefont
  {Ward}}\ and\ \bibinfo {author} {\bibfnamefont {O.}~\bibnamefont {Benson}},\
  }\bibfield  {title} {\bibinfo {title} {W\uppercase{GM} microresonators:
  sensing, lasing and fundamental optics with microspheres},\ }\href
  {https://onlinelibrary.wiley.com/doi/abs/10.1002/lpor.201000025} {\bibfield
  {journal} {\bibinfo  {journal} {Laser \& Photonics Reviews}\ }\textbf
  {\bibinfo {volume} {5}},\ \bibinfo {pages} {553} (\bibinfo {year}
  {2011})}\BibitemShut {NoStop}%
\bibitem [{\citenamefont {Jinno}\ and\ \citenamefont
  {Matsumoto}(1990)}]{jinno1990ultrafast}%
  \BibitemOpen
  \bibfield  {author} {\bibinfo {author} {\bibfnamefont {M.}~\bibnamefont
  {Jinno}}\ and\ \bibinfo {author} {\bibfnamefont {T.}~\bibnamefont
  {Matsumoto}},\ }\bibfield  {title} {\bibinfo {title} {Ultrafast, low power,
  and highly stable all-optical switching in an all polarization maintaining
  fiber \uppercase{S}agnac interferometer},\ }\href
  {https://ieeexplore.ieee.org/document/54702} {\bibfield  {journal} {\bibinfo
  {journal} {IEEE Photonics Technol. Lett.}\ }\textbf {\bibinfo {volume} {2}},\
  \bibinfo {pages} {349} (\bibinfo {year} {1990})}\BibitemShut {NoStop}%
\bibitem [{\citenamefont {Dayan}\ \emph {et~al.}(2008)\citenamefont {Dayan},
  \citenamefont {Parkins}, \citenamefont {Aoki}, \citenamefont {Ostby},
  \citenamefont {Vahala},\ and\ \citenamefont {Kimble}}]{dayan2008photon}%
  \BibitemOpen
  \bibfield  {author} {\bibinfo {author} {\bibfnamefont {B.}~\bibnamefont
  {Dayan}}, \bibinfo {author} {\bibfnamefont {A.}~\bibnamefont {Parkins}},
  \bibinfo {author} {\bibfnamefont {T.}~\bibnamefont {Aoki}}, \bibinfo {author}
  {\bibfnamefont {E.}~\bibnamefont {Ostby}}, \bibinfo {author} {\bibfnamefont
  {K.}~\bibnamefont {Vahala}},\ and\ \bibinfo {author} {\bibfnamefont
  {H.}~\bibnamefont {Kimble}},\ }\bibfield  {title} {\bibinfo {title} {A photon
  turnstile dynamically regulated by one atom},\ }\href
  {https://www.science.org/doi/full/10.1126/science.1152261} {\bibfield
  {journal} {\bibinfo  {journal} {Science}\ }\textbf {\bibinfo {volume}
  {319}},\ \bibinfo {pages} {1062} (\bibinfo {year} {2008})}\BibitemShut
  {NoStop}%
\bibitem [{\citenamefont {Aoki}\ \emph {et~al.}(2009)\citenamefont {Aoki},
  \citenamefont {Parkins}, \citenamefont {Alton}, \citenamefont {Regal},
  \citenamefont {Dayan}, \citenamefont {Ostby}, \citenamefont {Vahala},\ and\
  \citenamefont {Kimble}}]{Aoki2009PRL}%
  \BibitemOpen
  \bibfield  {author} {\bibinfo {author} {\bibfnamefont {T.}~\bibnamefont
  {Aoki}}, \bibinfo {author} {\bibfnamefont {A.~S.}\ \bibnamefont {Parkins}},
  \bibinfo {author} {\bibfnamefont {D.~J.}\ \bibnamefont {Alton}}, \bibinfo
  {author} {\bibfnamefont {C.~A.}\ \bibnamefont {Regal}}, \bibinfo {author}
  {\bibfnamefont {B.}~\bibnamefont {Dayan}}, \bibinfo {author} {\bibfnamefont
  {E.}~\bibnamefont {Ostby}}, \bibinfo {author} {\bibfnamefont {K.~J.}\
  \bibnamefont {Vahala}},\ and\ \bibinfo {author} {\bibfnamefont {H.~J.}\
  \bibnamefont {Kimble}},\ }\bibfield  {title} {\bibinfo {title} {Efficient
  routing of single photons by one atom and a microtoroidal cavity},\ }\href
  {https://doi.org/10.1103/PhysRevLett.102.083601} {\bibfield  {journal}
  {\bibinfo  {journal} {Phys. Rev. Lett.}\ }\textbf {\bibinfo {volume} {102}},\
  \bibinfo {pages} {083601} (\bibinfo {year} {2009})}\BibitemShut {NoStop}%
\bibitem [{\citenamefont {Kiraz}\ \emph {et~al.}(2001)\citenamefont {Kiraz},
  \citenamefont {Michler}, \citenamefont {Becher}, \citenamefont {Gayral},
  \citenamefont {Imamo{\u{g}}lu}, \citenamefont {Zhang}, \citenamefont {Hu},
  \citenamefont {Schoenfeld},\ and\ \citenamefont {Petroff}}]{kiraz2001cavity}%
  \BibitemOpen
  \bibfield  {author} {\bibinfo {author} {\bibfnamefont {A.}~\bibnamefont
  {Kiraz}}, \bibinfo {author} {\bibfnamefont {P.}~\bibnamefont {Michler}},
  \bibinfo {author} {\bibfnamefont {C.}~\bibnamefont {Becher}}, \bibinfo
  {author} {\bibfnamefont {B.}~\bibnamefont {Gayral}}, \bibinfo {author}
  {\bibfnamefont {A.}~\bibnamefont {Imamo{\u{g}}lu}}, \bibinfo {author}
  {\bibfnamefont {L.}~\bibnamefont {Zhang}}, \bibinfo {author} {\bibfnamefont
  {E.}~\bibnamefont {Hu}}, \bibinfo {author} {\bibfnamefont {W.}~\bibnamefont
  {Schoenfeld}},\ and\ \bibinfo {author} {\bibfnamefont {P.}~\bibnamefont
  {Petroff}},\ }\bibfield  {title} {\bibinfo {title} {Cavity-quantum
  electrodynamics using a single \uppercase{I}n\uppercase{A}s quantum dot in a
  microdisk structure},\ }\href
  {https://pubs.aip.org/aip/apl/article/78/25/3932/112901/Cavity-quantum-electrodynamics-using-a-single-InAs}
  {\bibfield  {journal} {\bibinfo  {journal} {Appl. Phys. Lett.}\ }\textbf
  {\bibinfo {volume} {78}},\ \bibinfo {pages} {3932} (\bibinfo {year}
  {2001})}\BibitemShut {NoStop}%
\bibitem [{\citenamefont {Peter}\ \emph {et~al.}(2005)\citenamefont {Peter},
  \citenamefont {Senellart}, \citenamefont {Martrou}, \citenamefont
  {Lema\^{\i}tre}, \citenamefont {Hours}, \citenamefont {G\'erard},\ and\
  \citenamefont {Bloch}}]{Peter2005PRL}%
  \BibitemOpen
  \bibfield  {author} {\bibinfo {author} {\bibfnamefont {E.}~\bibnamefont
  {Peter}}, \bibinfo {author} {\bibfnamefont {P.}~\bibnamefont {Senellart}},
  \bibinfo {author} {\bibfnamefont {D.}~\bibnamefont {Martrou}}, \bibinfo
  {author} {\bibfnamefont {A.}~\bibnamefont {Lema\^{\i}tre}}, \bibinfo {author}
  {\bibfnamefont {J.}~\bibnamefont {Hours}}, \bibinfo {author} {\bibfnamefont
  {J.~M.}\ \bibnamefont {G\'erard}},\ and\ \bibinfo {author} {\bibfnamefont
  {J.}~\bibnamefont {Bloch}},\ }\bibfield  {title} {\bibinfo {title}
  {Exciton-photon strong-coupling regime for a single quantum dot embedded in a
  microcavity},\ }\href {https://doi.org/10.1103/PhysRevLett.95.067401}
  {\bibfield  {journal} {\bibinfo  {journal} {Phys. Rev. Lett.}\ }\textbf
  {\bibinfo {volume} {95}},\ \bibinfo {pages} {067401} (\bibinfo {year}
  {2005})}\BibitemShut {NoStop}%
\bibitem [{\citenamefont {Srinivasan}\ and\ \citenamefont
  {Painter}(2007{\natexlab{a}})}]{srinivasan2007linear}%
  \BibitemOpen
  \bibfield  {author} {\bibinfo {author} {\bibfnamefont {K.}~\bibnamefont
  {Srinivasan}}\ and\ \bibinfo {author} {\bibfnamefont {O.}~\bibnamefont
  {Painter}},\ }\bibfield  {title} {\bibinfo {title} {Linear and nonlinear
  optical spectroscopy of a strongly coupled microdisk--quantum dot system},\
  }\href {https://www.nature.com/articles/nature06274} {\bibfield  {journal}
  {\bibinfo  {journal} {Nature}\ }\textbf {\bibinfo {volume} {450}},\ \bibinfo
  {pages} {862} (\bibinfo {year} {2007}{\natexlab{a}})}\BibitemShut {NoStop}%
\bibitem [{\citenamefont {Srinivasan}\ and\ \citenamefont
  {Painter}(2007{\natexlab{b}})}]{SrinivasanPhysRevA}%
  \BibitemOpen
  \bibfield  {author} {\bibinfo {author} {\bibfnamefont {K.}~\bibnamefont
  {Srinivasan}}\ and\ \bibinfo {author} {\bibfnamefont {O.}~\bibnamefont
  {Painter}},\ }\bibfield  {title} {\bibinfo {title} {Mode coupling and
  cavity--quantum-dot interactions in a fiber-coupled microdisk cavity},\
  }\href {https://doi.org/10.1103/PhysRevA.75.023814} {\bibfield  {journal}
  {\bibinfo  {journal} {Phys. Rev. A}\ }\textbf {\bibinfo {volume} {75}},\
  \bibinfo {pages} {023814} (\bibinfo {year} {2007}{\natexlab{b}})}\BibitemShut
  {NoStop}%
\bibitem [{\citenamefont {Park}\ \emph {et~al.}(2006)\citenamefont {Park},
  \citenamefont {Cook},\ and\ \citenamefont {Wang}}]{park2006cavity}%
  \BibitemOpen
  \bibfield  {author} {\bibinfo {author} {\bibfnamefont {Y.-S.}\ \bibnamefont
  {Park}}, \bibinfo {author} {\bibfnamefont {A.~K.}\ \bibnamefont {Cook}},\
  and\ \bibinfo {author} {\bibfnamefont {H.}~\bibnamefont {Wang}},\ }\bibfield
  {title} {\bibinfo {title} {Cavity \uppercase{QED} with diamond nanocrystals
  and silica microspheres},\ }\href
  {https://pubs.acs.org/doi/full/10.1021/nl061342r} {\bibfield  {journal}
  {\bibinfo  {journal} {Nano Lett.}\ }\textbf {\bibinfo {volume} {6}},\
  \bibinfo {pages} {2075} (\bibinfo {year} {2006})}\BibitemShut {NoStop}%
\bibitem [{\citenamefont {Barbour}\ \emph {et~al.}(2010)\citenamefont
  {Barbour}, \citenamefont {Dinyari},\ and\ \citenamefont
  {Wang}}]{barbour2010composite}%
  \BibitemOpen
  \bibfield  {author} {\bibinfo {author} {\bibfnamefont {R.~J.}\ \bibnamefont
  {Barbour}}, \bibinfo {author} {\bibfnamefont {K.~N.}\ \bibnamefont
  {Dinyari}},\ and\ \bibinfo {author} {\bibfnamefont {H.}~\bibnamefont
  {Wang}},\ }\bibfield  {title} {\bibinfo {title} {A composite microcavity of
  diamond nanopillar and deformed silica microsphere with enhanced evanescent
  decay length},\ }\href
  {https://opg.optica.org/oe/fulltext.cfm?uri=oe-18-18-18968&id=205262}
  {\bibfield  {journal} {\bibinfo  {journal} {Optics Express}\ }\textbf
  {\bibinfo {volume} {18}},\ \bibinfo {pages} {18968} (\bibinfo {year}
  {2010})}\BibitemShut {NoStop}%
\bibitem [{\citenamefont {F\"urst}\ \emph {et~al.}(2011)\citenamefont
  {F\"urst}, \citenamefont {Strekalov}, \citenamefont {Elser}, \citenamefont
  {Aiello}, \citenamefont {Andersen}, \citenamefont {Marquardt},\ and\
  \citenamefont {Leuchs}}]{Strekalov2011}%
  \BibitemOpen
  \bibfield  {author} {\bibinfo {author} {\bibfnamefont {J.~U.}\ \bibnamefont
  {F\"urst}}, \bibinfo {author} {\bibfnamefont {D.~V.}\ \bibnamefont
  {Strekalov}}, \bibinfo {author} {\bibfnamefont {D.}~\bibnamefont {Elser}},
  \bibinfo {author} {\bibfnamefont {A.}~\bibnamefont {Aiello}}, \bibinfo
  {author} {\bibfnamefont {U.~L.}\ \bibnamefont {Andersen}}, \bibinfo {author}
  {\bibfnamefont {C.}~\bibnamefont {Marquardt}},\ and\ \bibinfo {author}
  {\bibfnamefont {G.}~\bibnamefont {Leuchs}},\ }\bibfield  {title} {\bibinfo
  {title} {Quantum light from a whispering-gallery-mode disk resonator},\
  }\href {https://doi.org/10.1103/PhysRevLett.106.113901} {\bibfield  {journal}
  {\bibinfo  {journal} {Phys. Rev. Lett.}\ }\textbf {\bibinfo {volume} {106}},\
  \bibinfo {pages} {113901} (\bibinfo {year} {2011})}\BibitemShut {NoStop}%
\bibitem [{\citenamefont {Ilchenko}\ \emph {et~al.}(2004)\citenamefont
  {Ilchenko}, \citenamefont {Savchenkov}, \citenamefont {Matsko},\ and\
  \citenamefont {Maleki}}]{Ilchenko2004}%
  \BibitemOpen
  \bibfield  {author} {\bibinfo {author} {\bibfnamefont {V.~S.}\ \bibnamefont
  {Ilchenko}}, \bibinfo {author} {\bibfnamefont {A.~A.}\ \bibnamefont
  {Savchenkov}}, \bibinfo {author} {\bibfnamefont {A.~B.}\ \bibnamefont
  {Matsko}},\ and\ \bibinfo {author} {\bibfnamefont {L.}~\bibnamefont
  {Maleki}},\ }\bibfield  {title} {\bibinfo {title} {Nonlinear optics and
  crystalline whispering gallery mode cavities},\ }\href
  {https://doi.org/10.1103/PhysRevLett.92.043903} {\bibfield  {journal}
  {\bibinfo  {journal} {Phys. Rev. Lett.}\ }\textbf {\bibinfo {volume} {92}},\
  \bibinfo {pages} {043903} (\bibinfo {year} {2004})}\BibitemShut {NoStop}%
\bibitem [{\citenamefont {Beckmann}\ \emph {et~al.}(2011)\citenamefont
  {Beckmann}, \citenamefont {Linnenbank}, \citenamefont {Steigerwald},
  \citenamefont {Sturman}, \citenamefont {Haertle}, \citenamefont {Buse},\ and\
  \citenamefont {Breunig}}]{Beckmann2011}%
  \BibitemOpen
  \bibfield  {author} {\bibinfo {author} {\bibfnamefont {T.}~\bibnamefont
  {Beckmann}}, \bibinfo {author} {\bibfnamefont {H.}~\bibnamefont
  {Linnenbank}}, \bibinfo {author} {\bibfnamefont {H.}~\bibnamefont
  {Steigerwald}}, \bibinfo {author} {\bibfnamefont {B.}~\bibnamefont
  {Sturman}}, \bibinfo {author} {\bibfnamefont {D.}~\bibnamefont {Haertle}},
  \bibinfo {author} {\bibfnamefont {K.}~\bibnamefont {Buse}},\ and\ \bibinfo
  {author} {\bibfnamefont {I.}~\bibnamefont {Breunig}},\ }\bibfield  {title}
  {\bibinfo {title} {Highly tunable low-threshold optical parametric
  oscillation in radially poled whispering gallery resonators},\ }\href
  {https://doi.org/10.1103/PhysRevLett.106.143903} {\bibfield  {journal}
  {\bibinfo  {journal} {Phys. Rev. Lett.}\ }\textbf {\bibinfo {volume} {106}},\
  \bibinfo {pages} {143903} (\bibinfo {year} {2011})}\BibitemShut {NoStop}%
\bibitem [{\citenamefont {F{\"o}rtsch}\ \emph {et~al.}(2013)\citenamefont
  {F{\"o}rtsch}, \citenamefont {F{\"u}rst}, \citenamefont {Wittmann},
  \citenamefont {Strekalov}, \citenamefont {Aiello}, \citenamefont {Chekhova},
  \citenamefont {Silberhorn}, \citenamefont {Leuchs},\ and\ \citenamefont
  {Marquardt}}]{fortsch2013versatile}%
  \BibitemOpen
  \bibfield  {author} {\bibinfo {author} {\bibfnamefont {M.}~\bibnamefont
  {F{\"o}rtsch}}, \bibinfo {author} {\bibfnamefont {J.~U.}\ \bibnamefont
  {F{\"u}rst}}, \bibinfo {author} {\bibfnamefont {C.}~\bibnamefont {Wittmann}},
  \bibinfo {author} {\bibfnamefont {D.}~\bibnamefont {Strekalov}}, \bibinfo
  {author} {\bibfnamefont {A.}~\bibnamefont {Aiello}}, \bibinfo {author}
  {\bibfnamefont {M.~V.}\ \bibnamefont {Chekhova}}, \bibinfo {author}
  {\bibfnamefont {C.}~\bibnamefont {Silberhorn}}, \bibinfo {author}
  {\bibfnamefont {G.}~\bibnamefont {Leuchs}},\ and\ \bibinfo {author}
  {\bibfnamefont {C.}~\bibnamefont {Marquardt}},\ }\bibfield  {title} {\bibinfo
  {title} {A versatile source of single photons for quantum information
  processing},\ }\href {https://www.nature.com/articles/ncomms2838} {\bibfield
  {journal} {\bibinfo  {journal} {Nat. Commun.}\ }\textbf {\bibinfo {volume}
  {4}},\ \bibinfo {pages} {1818} (\bibinfo {year} {2013})}\BibitemShut
  {NoStop}%
\bibitem [{\citenamefont {Guo}\ \emph {et~al.}(2016)\citenamefont {Guo},
  \citenamefont {Zou}, \citenamefont {Jung},\ and\ \citenamefont
  {Tang}}]{Guo2016PRL}%
  \BibitemOpen
  \bibfield  {author} {\bibinfo {author} {\bibfnamefont {X.}~\bibnamefont
  {Guo}}, \bibinfo {author} {\bibfnamefont {C.-L.}\ \bibnamefont {Zou}},
  \bibinfo {author} {\bibfnamefont {H.}~\bibnamefont {Jung}},\ and\ \bibinfo
  {author} {\bibfnamefont {H.~X.}\ \bibnamefont {Tang}},\ }\bibfield  {title}
  {\bibinfo {title} {On-chip strong coupling and efficient frequency conversion
  between telecom and visible optical modes},\ }\href
  {https://doi.org/10.1103/PhysRevLett.117.123902} {\bibfield  {journal}
  {\bibinfo  {journal} {Phys. Rev. Lett.}\ }\textbf {\bibinfo {volume} {117}},\
  \bibinfo {pages} {123902} (\bibinfo {year} {2016})}\BibitemShut {NoStop}%
\bibitem [{\citenamefont {Jing}\ \emph {et~al.}(2018)\citenamefont {Jing},
  \citenamefont {L\"{u}}, \citenamefont {\"{O}zdemir}, \citenamefont {Carmon},\
  and\ \citenamefont {Nori}}]{Jing2018}%
  \BibitemOpen
  \bibfield  {author} {\bibinfo {author} {\bibfnamefont {H.}~\bibnamefont
  {Jing}}, \bibinfo {author} {\bibfnamefont {H.}~\bibnamefont {L\"{u}}},
  \bibinfo {author} {\bibfnamefont {S.~K.}\ \bibnamefont {\"{O}zdemir}},
  \bibinfo {author} {\bibfnamefont {T.}~\bibnamefont {Carmon}},\ and\ \bibinfo
  {author} {\bibfnamefont {F.}~\bibnamefont {Nori}},\ }\bibfield  {title}
  {\bibinfo {title} {Nanoparticle sensing with a spinning resonator},\ }\href
  {https://doi.org/10.1364/OPTICA.5.001424} {\bibfield  {journal} {\bibinfo
  {journal} {Optica}\ }\textbf {\bibinfo {volume} {5}},\ \bibinfo {pages}
  {1424} (\bibinfo {year} {2018})}\BibitemShut {NoStop}%
\bibitem [{\citenamefont {Huang}\ \emph {et~al.}(2018)\citenamefont {Huang},
  \citenamefont {Miranowicz}, \citenamefont {Liao}, \citenamefont {Nori},\ and\
  \citenamefont {Jing}}]{Huang2018}%
  \BibitemOpen
  \bibfield  {author} {\bibinfo {author} {\bibfnamefont {R.}~\bibnamefont
  {Huang}}, \bibinfo {author} {\bibfnamefont {A.}~\bibnamefont {Miranowicz}},
  \bibinfo {author} {\bibfnamefont {J.-Q.}\ \bibnamefont {Liao}}, \bibinfo
  {author} {\bibfnamefont {F.}~\bibnamefont {Nori}},\ and\ \bibinfo {author}
  {\bibfnamefont {H.}~\bibnamefont {Jing}},\ }\bibfield  {title} {\bibinfo
  {title} {Nonreciprocal photon blockade},\ }\href
  {https://doi.org/10.1103/PhysRevLett.121.153601} {\bibfield  {journal}
  {\bibinfo  {journal} {Phys. Rev. Lett.}\ }\textbf {\bibinfo {volume} {121}},\
  \bibinfo {pages} {153601} (\bibinfo {year} {2018})}\BibitemShut {NoStop}%
\bibitem [{\citenamefont {Jiao}\ \emph {et~al.}(2020)\citenamefont {Jiao},
  \citenamefont {Zhang}, \citenamefont {Zhang}, \citenamefont {Miranowicz},
  \citenamefont {Kuang},\ and\ \citenamefont {Jing}}]{Jiao2020}%
  \BibitemOpen
  \bibfield  {author} {\bibinfo {author} {\bibfnamefont {Y.-F.}\ \bibnamefont
  {Jiao}}, \bibinfo {author} {\bibfnamefont {S.-D.}\ \bibnamefont {Zhang}},
  \bibinfo {author} {\bibfnamefont {Y.-L.}\ \bibnamefont {Zhang}}, \bibinfo
  {author} {\bibfnamefont {A.}~\bibnamefont {Miranowicz}}, \bibinfo {author}
  {\bibfnamefont {L.-M.}\ \bibnamefont {Kuang}},\ and\ \bibinfo {author}
  {\bibfnamefont {H.}~\bibnamefont {Jing}},\ }\bibfield  {title} {\bibinfo
  {title} {Nonreciprocal optomechanical entanglement against backscattering
  losses},\ }\href {https://doi.org/10.1103/PhysRevLett.125.143605} {\bibfield
  {journal} {\bibinfo  {journal} {Phys. Rev. Lett.}\ }\textbf {\bibinfo
  {volume} {125}},\ \bibinfo {pages} {143605} (\bibinfo {year}
  {2020})}\BibitemShut {NoStop}%
\bibitem [{\citenamefont {Hotter}\ \emph {et~al.}(2024)\citenamefont {Hotter},
  \citenamefont {Ritsch},\ and\ \citenamefont {Gietka}}]{Hotter2024}%
  \BibitemOpen
  \bibfield  {author} {\bibinfo {author} {\bibfnamefont {C.}~\bibnamefont
  {Hotter}}, \bibinfo {author} {\bibfnamefont {H.}~\bibnamefont {Ritsch}},\
  and\ \bibinfo {author} {\bibfnamefont {K.}~\bibnamefont {Gietka}},\
  }\bibfield  {title} {\bibinfo {title} {Combining critical and quantum
  metrology},\ }\href {https://doi.org/10.1103/PhysRevLett.132.060801}
  {\bibfield  {journal} {\bibinfo  {journal} {Phys. Rev. Lett.}\ }\textbf
  {\bibinfo {volume} {132}},\ \bibinfo {pages} {060801} (\bibinfo {year}
  {2024})}\BibitemShut {NoStop}%
\bibitem [{\citenamefont {Lu}\ \emph {et~al.}(2019)\citenamefont {Lu},
  \citenamefont {Surya}, \citenamefont {Liu}, \citenamefont {Bruch},
  \citenamefont {Gong}, \citenamefont {Xu},\ and\ \citenamefont
  {Tang}}]{lu2019periodically}%
  \BibitemOpen
  \bibfield  {author} {\bibinfo {author} {\bibfnamefont {J.}~\bibnamefont
  {Lu}}, \bibinfo {author} {\bibfnamefont {J.~B.}\ \bibnamefont {Surya}},
  \bibinfo {author} {\bibfnamefont {X.}~\bibnamefont {Liu}}, \bibinfo {author}
  {\bibfnamefont {A.~W.}\ \bibnamefont {Bruch}}, \bibinfo {author}
  {\bibfnamefont {Z.}~\bibnamefont {Gong}}, \bibinfo {author} {\bibfnamefont
  {Y.}~\bibnamefont {Xu}},\ and\ \bibinfo {author} {\bibfnamefont {H.~X.}\
  \bibnamefont {Tang}},\ }\bibfield  {title} {\bibinfo {title} {Periodically
  poled thin-film lithium niobate microring resonators with a second-harmonic
  generation efficiency of 250,000\%/w},\ }\href
  {https://opg.optica.org/optica/fulltext.cfm?uri=optica-6-12-1455&id=423402}
  {\bibfield  {journal} {\bibinfo  {journal} {Optica}\ }\textbf {\bibinfo
  {volume} {6}},\ \bibinfo {pages} {1455} (\bibinfo {year} {2019})}\BibitemShut
  {NoStop}%
\bibitem [{\citenamefont {Lu}\ \emph {et~al.}(2020)\citenamefont {Lu},
  \citenamefont {Li}, \citenamefont {Zou}, \citenamefont {Al~Sayem},\ and\
  \citenamefont {Tang}}]{lu2020toward}%
  \BibitemOpen
  \bibfield  {author} {\bibinfo {author} {\bibfnamefont {J.}~\bibnamefont
  {Lu}}, \bibinfo {author} {\bibfnamefont {M.}~\bibnamefont {Li}}, \bibinfo
  {author} {\bibfnamefont {C.-L.}\ \bibnamefont {Zou}}, \bibinfo {author}
  {\bibfnamefont {A.}~\bibnamefont {Al~Sayem}},\ and\ \bibinfo {author}
  {\bibfnamefont {H.~X.}\ \bibnamefont {Tang}},\ }\bibfield  {title} {\bibinfo
  {title} {Toward 1\% single-photon anharmonicity with periodically poled
  lithium niobate microring resonators},\ }\href
  {https://opg.optica.org/optica/fulltext.cfm?uri=optica-7-12-1654&id=442855}
  {\bibfield  {journal} {\bibinfo  {journal} {Optica}\ }\textbf {\bibinfo
  {volume} {7}},\ \bibinfo {pages} {1654} (\bibinfo {year} {2020})}\BibitemShut
  {NoStop}%
\bibitem [{\citenamefont {Scully}\ and\ \citenamefont
  {Zubairy}(1999)}]{scully1999quantum}%
  \BibitemOpen
  \bibfield  {author} {\bibinfo {author} {\bibfnamefont {M.~O.}\ \bibnamefont
  {Scully}}\ and\ \bibinfo {author} {\bibfnamefont {M.~S.}\ \bibnamefont
  {Zubairy}},\ }\href@noop {} {\emph {\bibinfo {title} {Quantum optics}}}\
  (\bibinfo  {publisher} {Cambridge},\ \bibinfo {year} {1999})\BibitemShut
  {NoStop}%
\bibitem [{\citenamefont {Agarwal}(2012)}]{agarwal2012quantum}%
  \BibitemOpen
  \bibfield  {author} {\bibinfo {author} {\bibfnamefont {G.~S.}\ \bibnamefont
  {Agarwal}},\ }\href@noop {} {\emph {\bibinfo {title} {Quantum optics}}}\
  (\bibinfo  {publisher} {Cambridge University Press},\ \bibinfo {year}
  {2012})\BibitemShut {NoStop}%
\bibitem [{\citenamefont {Huang}\ and\ \citenamefont
  {Agarwal}(2009)}]{Huang2009}%
  \BibitemOpen
  \bibfield  {author} {\bibinfo {author} {\bibfnamefont {S.}~\bibnamefont
  {Huang}}\ and\ \bibinfo {author} {\bibfnamefont {G.~S.}\ \bibnamefont
  {Agarwal}},\ }\bibfield  {title} {\bibinfo {title} {Normal-mode splitting in
  a coupled system of a nanomechanical oscillator and a parametric amplifier
  cavity},\ }\href {https://doi.org/10.1103/PhysRevA.80.033807} {\bibfield
  {journal} {\bibinfo  {journal} {Phys. Rev. A}\ }\textbf {\bibinfo {volume}
  {80}},\ \bibinfo {pages} {033807} (\bibinfo {year} {2009})}\BibitemShut
  {NoStop}%
\bibitem [{\citenamefont {L\"u}\ \emph {et~al.}(2015)\citenamefont {L\"u},
  \citenamefont {Wu}, \citenamefont {Johansson}, \citenamefont {Jing},
  \citenamefont {Zhang},\ and\ \citenamefont {Nori}}]{Xin-You2015}%
  \BibitemOpen
  \bibfield  {author} {\bibinfo {author} {\bibfnamefont {X.-Y.}\ \bibnamefont
  {L\"u}}, \bibinfo {author} {\bibfnamefont {Y.}~\bibnamefont {Wu}}, \bibinfo
  {author} {\bibfnamefont {J.~R.}\ \bibnamefont {Johansson}}, \bibinfo {author}
  {\bibfnamefont {H.}~\bibnamefont {Jing}}, \bibinfo {author} {\bibfnamefont
  {J.}~\bibnamefont {Zhang}},\ and\ \bibinfo {author} {\bibfnamefont
  {F.}~\bibnamefont {Nori}},\ }\bibfield  {title} {\bibinfo {title} {Squeezed
  optomechanics with phase-matched amplification and dissipation},\ }\href
  {https://doi.org/10.1103/PhysRevLett.114.093602} {\bibfield  {journal}
  {\bibinfo  {journal} {Phys. Rev. Lett.}\ }\textbf {\bibinfo {volume} {114}},\
  \bibinfo {pages} {093602} (\bibinfo {year} {2015})}\BibitemShut {NoStop}%
\bibitem [{\citenamefont {Qin}\ \emph {et~al.}(2018)\citenamefont {Qin},
  \citenamefont {Miranowicz}, \citenamefont {Li}, \citenamefont {L\"u},
  \citenamefont {You},\ and\ \citenamefont {Nori}}]{Qin2018}%
  \BibitemOpen
  \bibfield  {author} {\bibinfo {author} {\bibfnamefont {W.}~\bibnamefont
  {Qin}}, \bibinfo {author} {\bibfnamefont {A.}~\bibnamefont {Miranowicz}},
  \bibinfo {author} {\bibfnamefont {P.-B.}\ \bibnamefont {Li}}, \bibinfo
  {author} {\bibfnamefont {X.-Y.}\ \bibnamefont {L\"u}}, \bibinfo {author}
  {\bibfnamefont {J.~Q.}\ \bibnamefont {You}},\ and\ \bibinfo {author}
  {\bibfnamefont {F.}~\bibnamefont {Nori}},\ }\bibfield  {title} {\bibinfo
  {title} {Exponentially enhanced light-matter interaction, cooperativities,
  and steady-state entanglement using parametric amplification},\ }\href
  {https://doi.org/10.1103/PhysRevLett.120.093601} {\bibfield  {journal}
  {\bibinfo  {journal} {Phys. Rev. Lett.}\ }\textbf {\bibinfo {volume} {120}},\
  \bibinfo {pages} {093601} (\bibinfo {year} {2018})}\BibitemShut {NoStop}%
\bibitem [{\citenamefont {Qin}\ \emph {et~al.}(2022)\citenamefont {Qin},
  \citenamefont {Miranowicz},\ and\ \citenamefont {Nori}}]{QinWei2022}%
  \BibitemOpen
  \bibfield  {author} {\bibinfo {author} {\bibfnamefont {W.}~\bibnamefont
  {Qin}}, \bibinfo {author} {\bibfnamefont {A.}~\bibnamefont {Miranowicz}},\
  and\ \bibinfo {author} {\bibfnamefont {F.}~\bibnamefont {Nori}},\ }\bibfield
  {title} {\bibinfo {title} {Beating the 3 db limit for intracavity squeezing
  and its application to nondemolition qubit readout},\ }\href
  {https://doi.org/10.1103/PhysRevLett.129.123602} {\bibfield  {journal}
  {\bibinfo  {journal} {Phys. Rev. Lett.}\ }\textbf {\bibinfo {volume} {129}},\
  \bibinfo {pages} {123602} (\bibinfo {year} {2022})}\BibitemShut {NoStop}%
\bibitem [{\citenamefont {Malykin}(2000)}]{malykin2000sagnac}%
  \BibitemOpen
  \bibfield  {author} {\bibinfo {author} {\bibfnamefont {G.~B.}\ \bibnamefont
  {Malykin}},\ }\bibfield  {title} {\bibinfo {title} {The \uppercase{S}agnac
  effect: correct and incorrect explanations},\ }\href
  {https://iopscience.iop.org/article/10.1070/PU2000v043n12ABEH000830/meta}
  {\bibfield  {journal} {\bibinfo  {journal} {Phys.-Usp.}\ }\textbf {\bibinfo
  {volume} {43}},\ \bibinfo {pages} {1229} (\bibinfo {year}
  {2000})}\BibitemShut {NoStop}%
\bibitem [{\citenamefont {Nagy}\ \emph {et~al.}(2011)\citenamefont {Nagy},
  \citenamefont {Szirmai},\ and\ \citenamefont {Domokos}}]{Nagy2011}%
  \BibitemOpen
  \bibfield  {author} {\bibinfo {author} {\bibfnamefont {D.}~\bibnamefont
  {Nagy}}, \bibinfo {author} {\bibfnamefont {G.}~\bibnamefont {Szirmai}},\ and\
  \bibinfo {author} {\bibfnamefont {P.}~\bibnamefont {Domokos}},\ }\bibfield
  {title} {\bibinfo {title} {Critical exponent of a quantum-noise-driven phase
  transition: The open-system dicke model},\ }\href
  {https://doi.org/10.1103/PhysRevA.84.043637} {\bibfield  {journal} {\bibinfo
  {journal} {Phys. Rev. A}\ }\textbf {\bibinfo {volume} {84}},\ \bibinfo
  {pages} {043637} (\bibinfo {year} {2011})}\BibitemShut {NoStop}%
\bibitem [{\citenamefont {Johansson}\ \emph {et~al.}(2012)\citenamefont
  {Johansson}, \citenamefont {Nation},\ and\ \citenamefont
  {Nori}}]{JOHANSSON20121760}%
  \BibitemOpen
  \bibfield  {author} {\bibinfo {author} {\bibfnamefont {J.}~\bibnamefont
  {Johansson}}, \bibinfo {author} {\bibfnamefont {P.}~\bibnamefont {Nation}},\
  and\ \bibinfo {author} {\bibfnamefont {F.}~\bibnamefont {Nori}},\ }\bibfield
  {title} {\bibinfo {title} {Qutip: An open-source python framework for the
  dynamics of open quantum systems},\ }\href
  {https://doi.org/https://doi.org/10.1016/j.cpc.2012.02.021} {\bibfield
  {journal} {\bibinfo  {journal} {Comput. Phys. Commun.}\ }\textbf {\bibinfo
  {volume} {183}},\ \bibinfo {pages} {1760} (\bibinfo {year}
  {2012})}\BibitemShut {NoStop}%
\bibitem [{\citenamefont {Johansson}\ \emph {et~al.}(2013)\citenamefont
  {Johansson}, \citenamefont {Nation},\ and\ \citenamefont
  {Nori}}]{JOHANSSON20131234}%
  \BibitemOpen
  \bibfield  {author} {\bibinfo {author} {\bibfnamefont {J.}~\bibnamefont
  {Johansson}}, \bibinfo {author} {\bibfnamefont {P.}~\bibnamefont {Nation}},\
  and\ \bibinfo {author} {\bibfnamefont {F.}~\bibnamefont {Nori}},\ }\bibfield
  {title} {\bibinfo {title} {Qutip 2: A python framework for the dynamics of
  open quantum systems},\ }\href
  {https://doi.org/https://doi.org/10.1016/j.cpc.2012.11.019} {\bibfield
  {journal} {\bibinfo  {journal} {Comput. Phys. Commun.}\ }\textbf {\bibinfo
  {volume} {184}},\ \bibinfo {pages} {1234} (\bibinfo {year}
  {2013})}\BibitemShut {NoStop}%
\bibitem [{\citenamefont {Xu}\ \emph {et~al.}(2016)\citenamefont {Xu},
  \citenamefont {J\"ager}, \citenamefont {Sch\"utz}, \citenamefont {Cooper},
  \citenamefont {Morigi},\ and\ \citenamefont {Holland}}]{Minghui2016}%
  \BibitemOpen
  \bibfield  {author} {\bibinfo {author} {\bibfnamefont {M.}~\bibnamefont
  {Xu}}, \bibinfo {author} {\bibfnamefont {S.~B.}\ \bibnamefont {J\"ager}},
  \bibinfo {author} {\bibfnamefont {S.}~\bibnamefont {Sch\"utz}}, \bibinfo
  {author} {\bibfnamefont {J.}~\bibnamefont {Cooper}}, \bibinfo {author}
  {\bibfnamefont {G.}~\bibnamefont {Morigi}},\ and\ \bibinfo {author}
  {\bibfnamefont {M.~J.}\ \bibnamefont {Holland}},\ }\bibfield  {title}
  {\bibinfo {title} {Supercooling of atoms in an optical resonator},\ }\href
  {https://doi.org/10.1103/PhysRevLett.116.153002} {\bibfield  {journal}
  {\bibinfo  {journal} {Phys. Rev. Lett.}\ }\textbf {\bibinfo {volume} {116}},\
  \bibinfo {pages} {153002} (\bibinfo {year} {2016})}\BibitemShut {NoStop}%
\bibitem [{\citenamefont {Braunstein}\ and\ \citenamefont {van
  Loock}(2005)}]{Braunstein2005}%
  \BibitemOpen
  \bibfield  {author} {\bibinfo {author} {\bibfnamefont {S.~L.}\ \bibnamefont
  {Braunstein}}\ and\ \bibinfo {author} {\bibfnamefont {P.}~\bibnamefont {van
  Loock}},\ }\bibfield  {title} {\bibinfo {title} {Quantum information with
  continuous variables},\ }\href {https://doi.org/10.1103/RevModPhys.77.513}
  {\bibfield  {journal} {\bibinfo  {journal} {Rev. Mod. Phys.}\ }\textbf
  {\bibinfo {volume} {77}},\ \bibinfo {pages} {513} (\bibinfo {year}
  {2005})}\BibitemShut {NoStop}%
\end{thebibliography}
\end{document}